%% file: paper.tex
\documentclass{lmcs}

\usepackage[utf8]{inputenc}
\usepackage{amsmath,
	    amsfonts,
	    amstext,
	    amssymb,
	    mathrsfs,
	    stmaryrd,
	    latexsym,
	    enumitem,
	    proof,
	    graphicx,
	    color,
	    hyperref,
	    comment,
	    semantic, 
	    tikz}

\usetikzlibrary{arrows, positioning}
\tikzset{modal/.style = {>= stealth', shorten >= 0pt, shorten <= 0pt, auto,
			 node distance = 1cm, semithick}, 
	 point/.style = {circle, draw, fill = black, inner sep = 0.5mm}}

\input{Commands}

\begin{document}

\title[Noninterference Analysis of Systems with Nondeterminism and Probabilities]
      {Noninterference Analysis of \\ 
       Irreversible \blue{Systems and} Reversible Systems \\ 
       \blue{Featuring both} Nondeterminism and Probabilities}

\author[A.~Esposito]{Andrea Esposito\lmcsorcid{0009-0009-2259-902X}}
\author[A.~Aldini]{Alessandro Aldini\lmcsorcid{0000-0002-7250-5011}}
\author[M.~Bernardo]{Marco Bernardo\lmcsorcid{0000-0003-0267-6170}}
\address{Dipartimento di Scienze Pure e Applicate, Universit\`a di Urbino, Italy}
\email{andrea.esposito@uniurb.it, alessandro.aldini@uniurb.it, marco.bernardo@uniurb.it}



\begin{abstract}
\blue{The theory of noninterference} supports the analysis of secure computations in multi-level security
systems. Classical equivalence-based approaches to noninterference mainly rely on bisimilarity. In a
nondeterministic setting, assessing noninterference through weak bisimilarity is adequate for irreversible
systems, whereas for reversible ones branching bisimilarity has been recently proven to be more appropriate.
In this paper we address the same two families of systems with the difference that probabilities come into
play in addition to nondeterminism according to the alternating model of Hansson and Jonsson. For
irreversible systems we extend the results of Aldini, Bravetti, and Gorrieri developed in a
generative-reactive probabilistic setting, while for reversible systems we extend the results of Esposito,
Aldini, Bernardo, and Rossi developed in a purely nondeterministic setting. We recast noninterference
properties by adopting probabilistic variants of weak and branching bisimilarities for irreversible and
reversible systems\blue{,} respectively. Then we investigate a taxonomy of those properties as well as their
preservation and compositionality aspects, along with a comparison with \blue{earlier taxonomies}. The
adequacy of the extended noninterference theory is illustrated via a probabilistic smart contract lottery.
\end{abstract}

\maketitle

%
%
\section{Introduction}
\label{sec:intro}
%
%

The notion of noninterference was introduced in~\cite{GM82} to reason about the way in which illegitimate
information flows can occur in multi-level security systems due to covert channels from high-level agents to
low-level ones. Since the first definition, conceived for deterministic systems, in the last four decades a
\blue{vast amount} of work has been done to extend the approach to a variety of more expressive domains,
such as nondeterministic systems, systems in which quantitative aspects like time and probability play a
central role, and reversible systems; see, e.g.,
\cite{FG01,Ald06,Man11,HS12,VS98,SabSan00,BT03,ABG04,AB09,HMPR21,EABR25} and the references therein.
Likewise, to verify information-flow security properties based on noninterference, several different
approaches have been proposed ranging from the application of type theory~\cite{ZM04} and abstract
interpretation~\cite{GM18} to control flow and equivalence or model checking~\cite{FPR02,Mar03,AB11}.

Noninterference guarantees that low-level agents cannot infer from their observations what \blue{happens at
the} high level. Regardless of its specific definition, noninterference is closely tied to the notion of
behavioral equivalence~\cite{Gla01} because, given a multi-level security system, the idea is to compare the
system behavior with high-level actions being prevented and the system behavior with the same actions being
hidden. A natural framework in which to study system behavior is given by process
algebra~\cite{Mil89a,BHR84,BW90}. In this setting, weak bisimilarity~\cite{Mil89a} has been employed
in~\cite{FG01} to reason formally about covert channels and illegitimate information flows as well as to
study a classification of noninterference properties for irreversible nondeterministic systems.

In~\cite{EABR25} noninterference analysis has been extended to reversible systems, which feature both
forward and backward computations. Reversibility has started to gain attention in computing since it has
been shown that it may achieve lower levels of energy consumption~\cite{Lan61,Ben73}. Its applications range
from biochemical reaction modeling~\cite{PUY12,Pin17} and parallel discrete-event
simulation~\cite{PP14,SOJB18} to robotics~\cite{LES18}, wireless communications~\cite{SPP19}, fault-tolerant
systems~\cite{DK05,VKH10,LLMSS13,VS18}, program debugging~\cite{GLM14,LNPV18a}, and distributed
algorithms~\cite{Yca93,BLMMRS23}.

As shown in~\cite{EABR25}, noninterference properties based on weak bisimilarity are not adequate in a
reversible context because they fail to detect information flows emerging when backward computations are
triggered. A more appropriate semantics turns out to be branching bisimilarity~\cite{GW96} because it
coincides with weak back-and-forth bisimilarity~\cite{DMV90}. The latter behavioral equivalence requires
systems to be able to mimic each other's behavior stepwise not only when performing actions in the standard
forward direction, but also when undoing those actions in the backward direction. Formally, weak
back-and-forth bisimilarity is defined over computation paths instead of states thus preserving not only
causality but also history, as backward moves are constrained to take place along the same path followed in
the forward direction even in the presence of concurrency.

In this paper we extend the approach of~\cite{EABR25} to a probabilistic setting, so as to address
noninterference properties in a framework featuring nondeterminism, probabilities, and reversibility. The
starting point for our study is given by the probabilistic noninterference properties developed
in~\cite{ABG04} over a probabilistic process calculus based on a combination of the generative and reactive
probabilistic models of~\cite{GSS95}. In addition to probabilistic choice, in~\cite{ABG04} other operators
such as parallel composition and hiding are decorated with a probabilistic parameter, so that the selection
among all the actions that are executable by a process is fully probabilistic. Moreover, the behavioral
equivalence considered in~\cite{ABG04} is akin to the weak probabilistic bisimilarity of~\cite{BH97}, where
the latter equivalence is shown to coincide with probabilistic branching bisimilarity over fully
probabilistic processes.

Here we adopt a more expressive model combining nondeterminism and probabilities, called the strictly
alternating model~\cite{HJ90}. In this model, states are divided into nondeterministic and probabilistic
states, while transitions are divided into action transitions -- each labeled with an action and going from
a nondeterministic state to a probabilistic one -- and probabilistic transitions -- each labeled with a
probability and going from a probabilistic state to a nondeterministic one. A more flexible variant, called
the non-strictly alternating model~\cite{PLS00}, allows for action transitions also between two
nondeterministic states. 

Following~\cite{HJ90} we build a process calculus that, unlike the one in~\cite{ABG04}, supports
nondeterminism and decorates with probabilistic parameters only probabilistic choices. As for behavioral
equivalences, we resort to the weak probabilistic bisimilarity of~\cite{PLS00} and the probabilistic
branching bisimilarity of~\cite{AGT12}, originally developed in the non-strictly alternating model.  By
using these two equivalences we recast the noninterference properties of~\cite{FG01,FR06} for irreversible
systems and the noninterference properties of~\cite{EABR25} \linebreak for reversible systems, respectively,
to study their preservation and compositionality aspects as well as to provide a taxonomy similar to those
in~\cite{FG01,EABR25}.

Unlike~\cite{ABG04}, the resulting noninterference properties are lighter as they do not need additional
universal quantifications over probabilistic parameters. Furthermore, reversibility comes into play by
extending one of the results of~\cite{DMV90} to the strictly alternating model; we show that a probabilistic
variant of weak back-and-forth bisimilarity coincides with the probabilistic branching bisimilarity
of~\cite{AGT12}. Finally, we point out that for proving some of our results we have to resort to the
bisimulation-up-to technique~\cite{SM92} and therefore introduce probabilistic variants of up-to
weak~\cite{Mil89a} and branching~\cite{Gla93} bisimulations.

This paper, which is a revised and extended version of~\cite{EAB24}, is organized as follows. In
Section~\ref{sec:prob_basic_def_res} we recall the strictly alternating model of~\cite{HJ90} along with
various definitions of strong and weak bisimilarities for it -- with the weak ones being those
in~\cite{PLS00,AGT12} -- and a process calculus interpreted on it. In Section~\ref{sec:prob_bisim_sec_prop}
we recast in our probabilistic framework a selection of noninterference properties taken
from~\cite{FG01,FR06,EABR25}. In Section~\ref{sec:prob_bisim_sec_prop_char} we study their preservation and
compositionality characteristics as well as their taxonomy and relate it to the nondeterministic taxonomy
of~\cite{EABR25}. In Section~\ref{sec:prob_branching_is_bf} we establish a connection with reversibility by
introducing a weak probabilistic back-and-forth bisimilarity and proving that it coincides with
probabilistic branching bisimilarity. In Section~\ref{sec:prob_example} we present an example of a lottery
implemented through a probabilistic smart contract to show the adequacy of our approach when dealing with
information flows in systems featuring nondeterminism and probabilities, both in the irreversible case and
in the reversible one. Finally, in Section~\ref{sec:concl} we provide some concluding remarks and directions
for future work.

%
%
\section{Background Definitions and Results}
\label{sec:prob_basic_def_res}
%
%

In this section we recall the strictly alternating model of~\cite{HJ90} (Section~\ref{sec:plts}) along with
strong and weak probabilistic bisimilarities~\cite{PLS00} and probabilistic branching
bisimilarity~\cite{AGT12} (Section~\ref{sec:prob_bisim}). Then we introduce a probabilistic process language
that is inspired by~\cite{HJ90} (Section~\ref{sec:prob_proc_lang}), through which we will express
bisimulation-based information-flow security properties accounting for nondeterminism and probabilities.

%
\subsection{Probabilistic Labeled Transition Systems}
\label{sec:plts}
%

To represent the behavior of a process featuring nondeterminism and probabilities, we use a probabilistic
labeled transition system. This is a variant of a labeled transition system~\cite{Kel76} whose transitions
are labeled with actions or probabilities. Since we adopt the strictly alternating model of~\cite{HJ90}, we
distinguish between nondeterministic and probabilistic states. The transitions of the former are labeled
only with actions, while the transitions of the latter are labeled only with probabilities. Every action
transition leads from a nondeterministic state to a probabilistic one, while every probabilistic transition
leads from a probabilistic state to a nondeterministic one. We denote by $\cals_{\rm n}$ (resp.\ $\cals_{\rm
p}$) the set of nondeterministic (resp.\ probabilistic) states and by $\cala_{\tau}$ the action set
containing a set $\cala$ of observable actions and one unobservable action $\tau \notin \cala$.

	\begin{defi}\label{def:plts}

A \emph{probabilistic labeled transition system (PLTS)} is a triple $(\cals, \cala_{\tau}, \! \arrow{}{}
\!)$ where $\cals = \cals_{\rm n} \cup \cals_{\rm p}$ with $\cals_{\rm n} \cap \cals_{\rm p} = \emptyset$ is
a \blue{finite or} countable set of states, $\cala_{\tau} = \cala \cup \{ \tau \}$ is a countable set of
actions, and $\! \arrow{}{} \! = \! \arrow{}{\rm a} \cup \arrow{}{\rm p}$ is the transition relation, with
$\arrow{}{\rm a} \subseteq \cals_{\rm n} \times \cala_{\tau} \times \cals_{\rm p}$ being the action
transition relation whilst $\arrow{}{\rm p} \subseteq \cals_{\rm p} \times \realns_{]0, 1]} \times
\cals_{\rm n}$ being the probabilistic transition relation satisfying $\sum_{(s, p, s') \in \! \arrow{}{\rm
p}} p \in \{ 0, 1 \}$ for all $s \in \cals_{\rm p}$.
\fullbox

	\end{defi}

An action transition $(s, a, s')$ is written $s \arrow{a}{\rm a} s'$ while a probabilistic transition $(s,
p, s')$ is written $s \arrow{p}{\rm p} s'$, where $s$ is the source state and $s'$ is the target state.
\blue{A} computation of length $n$ from $s_{0}$ to $s_{n}$ \blue{is} a sequence $\theta = s_{0} \, \ell_{1}
\, s_{1} \, \ell_{2} \, s_{2} \dots s_{n - 1} \, \ell_{n} \, s_{n}$ where each $(s_{i - 1}, \ell_{i},
s_{i})$ is a transition with $\ell_{i} \in \cala_{\tau} \cup \realns_{]0, 1]}$; we denote by
$\trace(\theta)$ the subsequence of $\ell_{1} \, \ell_{2} \dots \ell_{n}$ composed of all occurring action
labels and by $\varepsilon_{s}$ the empty computation from $s$. We say that $s'$ is reachable from $s$,
written $s' \in \reach(s)$, iff there exists a computation from $s$ to $s'$.

%
\subsection{Probabilistic Bisimulation Equivalences}
\label{sec:prob_bisim}
%

We introduce four variants of probabilistic bisimilarity~\cite{LS91} over PLTSs. Each of them can be shown
to be the largest probabilistic \blue{bisimulation} of its kind by first proving that the transitive closure
of the union of probabilistic bisimulations of that kind is a probabilistic bisimulation of that kind
too~\cite{GSS95}.

Bisimilarity~\cite{Par81,Mil89a} identifies processes that are able to mimic each other's behavior stepwise,
hence having the same branching structure. In the strictly alternating model, this extends to probabilistic
behavior~\cite{HJ90}. Let $\pi(s, C) = \sum_{s \arrow{p}{\rm p} s', s' \in C} p$ be the cumulative
probability with which state $s \, \in \cals$ reaches a state in $C \, \subseteq \cals$; note that $\pi(s,
C) = 0$ \linebreak when $s$ is not a probabilistic state or $C$ does not contain any nondeterministic state.

	\begin{defi}\label{def:strong_prob_bisim}

Let $(\cals, \cala_{\tau}, \! \arrow{}{} \!)$ be a PLTS. We say that $s_{1}, s_{2} \in \cals$ are
\emph{strongly probabilistic bisimilar}, written $s_{1} \sbis{\rm p} s_{2}$, iff $(s_{1}, s_{2}) \in \calb$
for some strong probabilistic bisimulation $\calb$. An equivalence relation $\calb \subseteq (\cals_{\rm n}
\times \cals_{\rm n}) \cup (\cals_{\rm p} \times \cals_{\rm p})$ is a \emph{strong probabilistic
bisimulation} iff, whenever $(s_{1}, s_{2}) \in \calb$, then:

		\begin{itemize}

\item For each $s_{1} \arrow{a}{\rm a} s'_{1}$ there exists $s_{2} \arrow{a}{\rm a} s'_{2}$ such that
$(s'_{1}, s'_{2}) \in \calb$.

\item $\pi(s_{1}, C) = \pi(s_{2}, C)$ for all equivalence classes $C$ in the quotient set $\cals_{\rm n} /
\calb$.
\fullbox

		\end{itemize}

	\end{defi}

In~\cite{PLS00} a strong probabilistic bisimilarity more liberal than the one in~\cite{HJ90} allows a
nondeterministic state and a probabilistic state to be identified when the latter concentrates all of its
probabilistic mass in reaching the former. Think, e.g., of a probabilistic state whose outgoing transitions
all reach the same nondeterministic state. To this purpose the following function of $s, s' \in \cals$ is
introduced in~\cite{PLS00}:
\cws{0}{\proba(s, s') \: = \:
\begin{cases}
p & \text{if } s \in \cals_{\rm p} \land p = \sum_{s \arrow{p'}{\rm p} s'} p' > 0 \\
1 & \text{if } s \in \cals_{\rm n} \land s' = s \\
0 & \text{otherwise} \\
\end{cases}}
and is then lifted to $C \, \subseteq \cals$ by letting $\proba(s, C) = \sum_{s' \in C} \proba(s, s')$.

	\begin{defi}\label{def:strong_mix_prob_bisim}

Let $(\cals, \cala_{\tau}, \! \arrow{}{} \!)$ be a PLTS. We say that $s_{1}, s_{2} \in \cals$ are
\emph{strongly mix-probabilistic bisimilar}, written $s_{1} \sbis{\rm pm} s_{2}$, iff $(s_{1}, s_{2}) \in
\calb$ for some strong mix-probabilistic bisimulation $\calb$. An equivalence relation $\calb$ over $\cals$
is a \emph{strong mix-probabilistic bisimulation} iff, whenever $(s_{1}, s_{2}) \in \calb$, then:

		\begin{itemize}

\item If $s_{1}, s_{2} \in \cals_{\tt n}$, for each $s_{1} \arrow{a}{\rm a} s'_{1}$ there exists $s_{2}
\arrow{a}{\rm a} s'_{2}$ such that $(s'_{1}, s'_{2}) \in \calb$.

\item $\proba(s_{1}, C) = \proba(s_{2}, C)$ for all equivalence classes $C \in \cals / \calb$.
\fullbox

		\end{itemize}

	\end{defi}

Weak bisimilarity~\cite{Mil89a} is additionally capable of abstracting from unobservable actions. In a
probabilistic setting, it may be the case that probabilistic transitions are abstracted away too. Let $s
\warrow{}{} s'$ mean that $s' \in \reach(s)$ and, when $s' \neq s$, there exists a finite sequence of
transitions from $s$ to $s'$ in which there are only $\tau$-transitions alternating with probabilistic
transitions. Moreover let $\warrow{a}{}$ stand for $\warrow{}{} \! \arrow{a}{\rm a} \! \warrow{}{}$ and
$\warrow{\hat{a}}{}$ stand for $\warrow{}{}$ if $a = \tau$ or $\warrow{a}{}$ if $a \neq \tau$.

The weak probabilistic bisimilarity below is the one of~\cite{PLS00} redefined over the strictly alternating
model without using weighted probabilities and by imposing a pure $\proba$-based equality check like in the
previous two strong bisimilarities. As for the clause about nondeterministic behavior matching, if either
state has an outgoing $a$-transition, then the other state is the root of a tree in which $(i)$ all
\blue{paths in the tree} are of the form $\warrow{\hat{a}}{} \!$, $(ii)$~nondeterministic branching is
resolved via a deterministic scheduler, and $(iii)$ all the leaves in the tree are in the same equivalence
class as the target state of the $a$-transition and the cumulative probability of reaching them is $1$.
Formally, a deterministic scheduler is a function $\sigma$ from the set of computations to $\arrow{}{\rm a}
\cup \{ \bot \}$ such that, whenever $\sigma(\theta) = s \arrow{a}{\rm a} s'$, then $s$ is the last state of
$\theta$; symbol $\bot$ denotes scheduler halting. We denote by $\sched(s)$ the set of deterministic
schedulers for the computations starting from $s$.
Given $s \in \cals$, $\Phi \subseteq \cala_{\tau}^{*}$, $C \subseteq \cals$, a~computation $\theta$, and a
scheduler $\sigma$ for the initial state of $\theta$, we define the probability $\proba_{\sigma}(s, \Phi, C,
\theta)$ as the smallest solution to the following set of equations:
\cws{0}{X_{\sigma}(s, \Phi, C, \theta) \: = \:
\begin{cases}
\sum\limits_{s \arrow{p}{\rm p} s'} p \cdot X_{\sigma}(s', \Phi, C, \theta \, p \, s') & \text{if } s \in
\cals_{\rm p} \\
X_{\sigma}(s', \Phi - a, C, \theta \, a \, s') & \text{if } s \in \cals_{\rm n} \land \sigma(\theta) = s
\arrow{a}{\rm a} s' \\
1 & \text{if } \blue{s \in \cals_{\rm n} \, \land \,} \sigma(\theta) = \bot \land \varepsilon \in \Phi \land
s \in C \\
0 & \text{if } \blue{s \in \cals_{\rm n} \land \sigma(\theta) = \bot \land (\varepsilon \notin \Phi \lor s
\notin C)} \\
\end{cases}}
where $\Phi - a = \{ \alpha \in \cala_{\tau}^{*} \mid a \, \alpha \in \Phi \}$. In the definition below, we
use the regular expression $\tau^{*} \, \hat{a} \, \tau^{*}$ to denote the set of traces of \blue{paths} of
the form $\warrow{\hat{a}}{} \!$ and $[s'_{1}]_{\calb}$ to express the equivalence class of state $s'_{1}$
with respect to the equivalence relation $\calb$ over states.

	\begin{defi}\label{def:weak_prob_bisim}

Let $(\cals, \cala_{\tau}, \! \arrow{}{} \!)$ be a PLTS. We say that $s_{1}, s_{2} \in \cals$ are
\emph{weakly probabilistic bisimilar}, written $s_{1} \wbis{\rm pw} s_{2}$, iff $(s_{1}, s_{2}) \in \calb$
for some weak probabilistic bisimulation $\calb$. \linebreak An equivalence relation $\calb$ over $\cals$ is
a \emph{weak probabilistic bisimulation} iff, whenever $(s_{1}, s_{2}) \in \calb$, then:

		\begin{itemize}

\item For each $s_{1} \arrow{a}{\rm a} s'_{1}$ there exists $\sigma \in \sched(s_{2})$ such that
$\proba_{\sigma}(s_{2}, \tau^{*} \, \hat{a} \, \tau^{*}, [s'_{1}]_{\calb}, \varepsilon_{s_{2}}) = 1$.

\item $\proba(s_{1}, C) = \proba(s_{2}, C)$ for all equivalence classes $C \in \cals / \calb$.
\fullbox

		\end{itemize}

	\end{defi}


Branching bisimilarity~\cite{GW96} is finer than weak bisimilarity as it preserves the branching structure
of processes even when abstracting from $\tau$-actions -- see condition $(s_{1}, \bar{s}_{2}) \in \calb$ in
the definition below. The probabilistic branching bisimilarity that follows is the one of~\cite{AGT12}
redefined over the strictly alternating model in a style closer to~\cite{GW96}.

	\begin{defi}\label{def:prob_branching_bisim}

Let $(\cals, \cala_{\tau}, \! \arrow{}{} \!)$ be a PLTS. We say that $s_{1}, s_{2} \in \cals$ are
\emph{probabilistic branching bisimilar}, written $s_{1} \wbis{\rm pb} s_{2}$, iff $(s_{1}, s_{2}) \in
\calb$ for some probabilistic branching bisimulation $\calb$. An equivalence relation $\calb$ over $\cals$
is a \emph{probabilistic branching bisimulation} iff, whenever $(s_{1}, s_{2}) \in \calb$, then: 

		\begin{itemize}

\item For each $s_{1} \arrow{a}{\rm a} s'_{1}$:

			\begin{itemize}

\item either $a = \tau$ and $(s'_{1}, s_{2}) \in \calb$;

\item or there exists $s_{2} \warrow{}{} \bar{s}_{2} \arrow{a}{\rm a} s'_{2}$ such that $(s_{1},
\bar{s}_{2}) \in \calb$ and $(s'_{1}, s'_{2}) \in \calb$.

			\end{itemize}

\item $\proba(s_{1}, C) = \proba(s_{2}, C)$ for all equivalence classes $C \in \cals / \calb$.
\fullbox

		\end{itemize}

	\end{defi}

An example that highlights the higher distinguishing power of probabilistic branching bisimilarity is given
in Figure~\ref{fig:prob_wb_brb_cex}, where every PLTS is depicted as a directed graph in which vertices
represent states and action- or probability-labeled edges represent transitions. The initial states $s_{1}$
and $s_{2}$ of the two PLTSs are weakly probabilistic bisimilar but not probabilistic branching bisimilar.
On the one hand, each of the two states reachable from $s_{1}$ with probability $0.5$ and the state
reachable from $s_{2}$ with probability $1$ are all weakly probabilistic bisimilar and hence the cumulative
probability to reach them is the same from both initial states. On the other hand, the two states reachable
from $s_{1}$ are not probabilistic branching bisimilar, because if the one on the right performs $a$ then
the one on the left cannot respond by performing $\tau$, $1$, and $a$ because the state executing $a$ no
longer enables $b$. Thus, with respect to probabilistic branching bisimilarity, $s_{1}$ reaches with
probability $0.5$ two different equivalence classes, while $s_{2}$ reaches with probability $1$ only one of
them.


	\begin{figure}[t]

\begin{center}
\begin{tikzpicture}[modal]

\node[point] (root) [label = 90: {$s_{1}$}]                                                 {};
\node[point] (p1)   [below left = of root, xshift = -10pt, yshift = -10pt, label = 180: {}] {};
\node[point] (p2)   [below right = of root, xshift = 10pt, yshift = -10pt, label = 180: {}] {};
\node[point] (p3)   [below left = of p1, label = 0: {}]                                     {};
\node[point] (p4)   [below right = of p1, label = 180: {}]                                  {};
\node[point] (p5)   [below left = of p2, label = 180: {}]                                   {};
\node[point] (p6)   [below = of p2, label = 180: {}]                                        {};
\node[point] (p7)   [below right = of p2, label = 180: {}]                                  {};
\node[point] (p8)   [below = of p3, label = 0: {}]                                          {};
\node[point] (p9)   [below = of p5, label = 0: {}]                                          {};
\node[point] (p10)  [right = of p8, label = 0: {}]                                          {};
\node[point] (p11)  [right = of p9, label = 0: {}]                                          {};

\path[->] (root) edge node[above left]  {$0.5$}               (p1);
\path[->] (root) edge node[above right] {$0.5$}               (p2);
\path[->] (p1)   edge node[above left]  {$\tau$}              (p3);
\path[->] (p1)   edge node[above]       {$b \hspace{-0.1cm}$} (p4);
\path[->] (p2)   edge node[above left]  {$\tau$}              (p5);
\path[->] (p2)   edge node[left]        {$a$}                 (p6);
\path[->] (p2)   edge node[above right] {$b$}                 (p7);
\path[->] (p3)   edge node[left]        {$1$}                 (p8);
\path[->] (p5)   edge node[left]        {$1$}                 (p9);
\path[->] (p8)   edge node[above]       {$a$}                 (p10);
\path[->] (p9)   edge node[above]       {$a$}                 (p11);

\node[point] (qroot) [right = 5.5cm of root, label = 90: {$s_{2}$}] {};
\node[point] (q1)    [below = of qroot, label = 0: {}]              {};
\node[point] (q2)    [below left = of q1, label = 0: {}]            {};
\node[point] (q3)    [below right = of q1, label = 180: {}]         {};
\node[point] (q4)    [below = of q2, label = 180: {}]               {};
\node[point] (q5)    [right = of q4, label = 180: {}]               {};

\path[->] (qroot) edge node[right]       {$1$}    (q1);
\path[->] (q1)    edge node[above left]  {$\tau$} (q2);
\path[->] (q1)    edge node[above right] {$b$}    (q3);
\path[->] (q2)    edge node[left]        {$1$}    (q4);
\path[->] (q4)    edge node[above]       {$a$}    (q5);

\end{tikzpicture}
\end{center}

\caption{States $s_{1}$ and $s_{2}$ are related by $\wbis{\rm pw}$ but distinguished by $\wbis{\rm pb}$}
\label{fig:prob_wb_brb_cex}

	\end{figure}

%
\subsection{A Probabilistic Process Calculus with High and Low Actions}
\label{sec:prob_proc_lang}
%

We now introduce a probabilistic process calculus to formalize the security properties of interest. To
address two security levels, we partition the set $\cala$ of observable actions into $\cala_{\calh} \cup
\cala_{\call}$, with $\cala_{\calh} \cap \cala_{\call} = \emptyset$, where $\cala_{\calh}$ is the set of
high-level actions, ranged over by $h$, and $\cala_{\call}$ is the set of low-level actions, ranged over by
$l$. Note that $\tau \notin \cala_{\calh} \cup \cala_{\call}$.

The overall set of process terms is given by $\procs = \procs_{\rm n} \cup \procs_{\rm p}$ and ranged over
by $E$. The set \linebreak $\procs_{\rm n}$ of nondeterministic process terms, ranged over by $N$, is
obtained by considering typical operators from CCS~\cite{Mil89a} and CSP~\cite{BHR84}. The set $\procs_{\rm
p}$ of probabilistic process terms, ranged over by $P$, is obtained by taking a probabilistic choice
operator similar to the one of~\cite{HJ90}. In addition to action prefix, nondeterministic choice,
probabilistic choice, and \linebreak parallel composition -- which is taken from CSP so as not to hide
synchronizations among high-level actions by turning them into $\tau$ as would happen with the CCS parallel
composition -- we \linebreak include restriction and hiding, as they are necessary to formalize
noninterference properties, and recursion (which was not considered in~\cite{EAB24}). The syntax for
$\procs$ is: \\[-0.2cm]
\cws{0}{\begin{array}{rcl}
N & \!\! ::= \!\! & \nil \mid a \, .\, P \mid N + N \mid N \pco{L} N \mid N \setminus L \mid N \, / \, L
\mid \mathit{N\!K} \\
P & \!\! ::= \!\! & \bigoplus_{i \in I} [p_{i}] N_{i} \mid P \pco{L} P \mid P \setminus L \mid P \, / \, L
\mid \mathit{P\!K} \\
\end{array}}
where:

	\begin{itemize}

\item $\nil$ is the terminated process.

\item $a \, . \, \_$, for $a \in \cala_{\tau}$, is the action prefix operator describing a process that can
initially perform action $a$.

\item $\_ + \_$ is the alternative composition operator expressing a nondeterministic choice between two
processes based on their initially executable actions.

\item $\bigoplus_{i \in I} [p_{i}] \, \_$, for $I$ finite and \blue{non}empty, is the generalized
probabilistic composition operator expressing a probabilistic choice among finitely many processes each with
probability $p_{i} \in \realns_{]0, 1]}$ and such that $\sum_{i \in I} p_{i} = 1$. We will use $[p_{1}]
N_{1} \oplus [p_{2}] N_{2}$ as a shorthand for $\bigoplus_{i \in \{ 1, 2 \}} [p_{i}] N_{i}$ and $a \, . \,
N$ as a shorthand for $a \, . \, [1] N$ especially when $N$ is $\nil$.

\item $\_ \pco{L} \_$, for $L \subseteq \cala$, is the parallel composition operator allowing two processes
to proceed independently on any action not in $L$ and forcing them to synchronize on every action in $L$
\linebreak as well as on probabilities (which are multiplied)~\cite{HJ90}.

\item $\_ \setminus L$, for $L \subseteq \cala$, is the restriction operator, which prevents the execution
of all actions belonging to $L$.

\item $\_ \, / \, L$, for $L \subseteq \cala$, is the hiding operator, which turns all the executed actions
belonging to $L$ \linebreak into the unobservable action $\tau$. 

\item $\mathit{N\!K}$ (resp.\ $\mathit{P\!K}$) is a process constant equipped with a defining equation of
the form $\mathit{N\!K} \eqdef N$ (resp.\ $\mathit{P\!K} \eqdef P$), where every constant possibly occurring
in $N$ (resp.\ $P$) -- including $\mathit{N\!K}$ (resp.\ $\mathit{P\!K}$) itself thus allowing for recursion
-- must be in the scope of an action prefix.

	\end{itemize}

	\begin{table}[t]

\[\begin{array}{|rc|}
\hline 
& \\[-0.3cm]
\emph{Prefix} & a \, . \, P \arrow{a}{\rm a} P \\
\emph{Choice} & \inference[]{N_{1} \arrow{a}{\rm a} P_{1}}
{N_{1} + N_{2} \arrow{a}{\rm a} P_{1}} \qquad\qquad\qquad
\inference[]{N_{2} \arrow{a}{\rm a} P_{2}}
{N_{1} + N_{2} \arrow{a}{\rm a} P_{2}} \\[0.4cm]
\emph{Parallel} & \inference[]{N_{1} \arrow{a}{\rm a} P_{1} \quad a \notin L}
{N_{1} \pco{L} N_{2} \arrow{a}{\rm a} P_{1} \pco{L} [1] N_{2}} \qquad\qquad\qquad 
\inference[]{N_{2} \arrow{a}{\rm a} P_{2} \quad a \notin L}
{N_{1} \pco{L} N_{2} \arrow{a}{\rm a} [1] N_{1} \pco{L} P_{2}} \\[0.4cm]
\emph{Synch} & \inference[]{N_{1} \arrow{a}{\rm a} P_{1} \quad N_{2} \arrow{a}{\rm a} P_{2} \quad a \in L}
{N_{1} \pco{L} N_{2} \arrow{a}{\rm a} P_{1} \pco{L} P_{2}} \\[0.4cm]
\emph{Restriction} & \inference[]{N \arrow{a}{\rm a} P \quad a \notin L}
{N \setminus L \arrow{a}{\rm a} P \setminus L} \\[0.4cm]
\emph{Hiding} & \inference[]{N \arrow{a}{\rm a} P \quad a \in L}
{N \, / \, L \arrow{\tau}{\rm a} P \, / \, L} \qquad\qquad\qquad
\inference[]{N \arrow{a}{\rm a} P \quad a \notin L}
{N \, / \, L \arrow{a}{\rm a} P \, / \, L} \\[0.4cm]
\emph{Constant} & \inference[]{\mathit{N\!K} \eqdef N \quad N \arrow{a}{\rm a} P}
{\mathit{N\!K} \arrow{a}{\rm a} P} \\[0.4cm]
\hline
\end{array}\]

\caption{Operational semantic rules for nondeterministic processes (action transitions)}
\label{tab:prob_op_sem_nd}

	\end{table}

	\begin{table}[t]

\[\begin{array}{|rc|}
\hline
& \\[-0.3cm]
\emph{ProbChoice} & \inference[]{j \in I}
{\bigoplus_{i \in I} [p_{i}] N_{i} \arrow{p_{j}}{\rm p} N_{j}} \\[0.6cm]
\emph{ProbSynch} & \inference[]{P_{1} \arrow{p_{1}}{\rm p} N_{1} \quad P_{2} \arrow{p_{2}}{\rm p} N_{2}}
{P_{1} \pco{L} P_{2} \arrow{p_{1} \cdot p_{2}}{\rm p} N_{1} \pco{L} N_{2}} \\[0.4cm]
\emph{ProbRestriction} & \inference[]{P \arrow{p}{\rm p} N}
{P \setminus L \arrow{p}{\rm p} N \setminus L} \\[0.4cm]
\emph{ProbHiding} & \inference[]{P \arrow{p}{\rm p} N}
{P \, / \, L \arrow{p}{\rm p} N \, / \, L} \\[0.4cm]
\emph{ProbConstant} & \inference[]{\mathit{P\!K} \eqdef P \quad P \arrow{p}{\rm p} N}
{\mathit{P\!K} \arrow{p}{\rm p} N} \\[0.4cm]
\hline
\end{array}\]

\caption{Operational semantic rules for probabilistic processes (probabilistic transitions)}
\label{tab:prob_op_sem_pr}

	\end{table}

The operational semantic rules for the process language are shown in Tables~\ref{tab:prob_op_sem_nd}
and~\ref{tab:prob_op_sem_pr} for nondeterministic and probabilistic processes, respectively. Together they
produce the PLTS $(\procs, \cala_{\tau}, \! \arrow{}{} \!)$ where $\arrow{}{} \! = \arrow{}{\rm a} \cup
\arrow{}{\rm p} \!$, to which the bisimulation equivalences defined in Section~\ref{sec:prob_bisim} are
applicable. While $\arrow{}{\rm a} \! \subseteq \procs_{\rm n} \times \cala_{\tau} \times \procs_{\rm p}$ is
a relation, $\arrow{}{\rm p} \! \subseteq \procs_{\rm p} \times \realns_{]0, 1]} \times \procs_{\rm n}$ is a
multirelation~\cite{HJ90} \blue{with the multiplicity of every transition being given by the number of
different ways of deriving that transition via the operational semantic
rules}~\blue{\cite{Hil96,Her02,BB03}}; e.g., from $[p_{1}] N \oplus [p_{2}] N$ to $N$ \blue{there are two
transitions when $p_{1} \neq p_{2}$, while there is only one with multiplicity two} when $p_{1} = p_{2}$.
Note that in the \emph{Parallel} rules the nondeterministic subprocess that does not move has to be prefixed
by~$[1]$ to make it probabilistic within the overall target process~\cite{HJ90}; after all, $[1] N \sbis{\rm
pm} N$. \linebreak We let $[1] N \in \reach(E)$ whenever $N \in \reach(E)$.

%
%
\section{Probabilistic Information-Flow Security Properties}
\label{sec:prob_bisim_sec_prop}
%
%

In this section we recast the definitions of noninteference properties of~\cite{FG01,FR06,EABR25} --
\emph{Nondeterministic Non-Interference} (NNI) and \emph{Non-Deducibility on Composition} (NDC) -- by taking
as behavioral equivalence the weak or branching bisimilarity \blue{recalled in}
Section~\ref{sec:prob_bisim}. The intuition behind noninterference in a two-level security system is that,
if a group of agents at the high security level performs some actions, the effect of those actions should
not be \blue{observable} by any agent at the low security level. To formalize this, the restriction and
hiding operators play a central role.

	\begin{defi}\label{def:prob_bisim_sec_prop}

Let $E \in \procs$ and $\wbis{} \: \in \{ \wbis{\rm pw}, \wbis{\rm pb} \}$:

		\begin{itemize} 

\item $E \in \mathrm{BSNNI}_{\wbis{}} \Longleftrightarrow E \setminus \cala_{\calh} \wbis{} E \, / \,
\cala_{\calh}$.

\item $E \in \mathrm{BNDC}_{\wbis{}} \Longleftrightarrow$ for all $F \in \procs$ such that each action
\blue{occurring in $F$} belongs to $\cala_{\calh}$ and for all $L \subseteq \cala_{\calh}$, $E \setminus
\cala_{\calh} \wbis{} ((E \pco{L} F) \, / \, L) \setminus \cala_{\calh}$ when $E, F \in \procs_{\rm n}$ or
$E, F \in \procs_{\rm p}$.

\item $E \in \mathrm{SBSNNI}_{\wbis{}} \Longleftrightarrow$ for all $E' \in \reach(E)$, $E' \in
\mathrm{BSNNI}_{\wbis{}}$.

\item $E \in \mathrm{P\_BNDC}_{\wbis{}} \Longleftrightarrow$ for all $E' \in \reach(E)$, $E' \in
\mathrm{BNDC}_{\wbis{}}$.

\item $E \in \mathrm{SBNDC}_{\wbis{}} \Longleftrightarrow$ for all $E', E'' \in \reach(E)$ such that $E'
\arrow{h}{\rm a} E''$, $E' \setminus \cala_{\calh} \wbis{} E'' \setminus \cala_{\calh}$.
\fullbox

		\end{itemize}

	\end{defi}

\emph{Bisimulation-based Strong Nondeterministic Non-Interference} (BSNNI) has been one of the first and
most intuitive proposals. Basically, it is satisfied by any process $E$ that behaves the same when its
high-level actions are prevented (as modeled by $E \setminus \cala_{\calh}$) or when they are considered as
hidden, unobservable actions (as modeled by $E \, / \, \cala_{\calh}$). The equivalence between these two
low-level views of $E$ states that a low-level agent cannot deduce the high-level behavior of the system.
For instance, in our probabilistic setting, a low-level agent that observes the execution of $l$ in $E{'} =
l \, . \, \nil + l \, . \, ([0.5] h \, . \, [1] l_{1} \, . \, \nil \oplus [0.5] h \, . \, [1] l_{2} \, . \,
\nil) + l \, . \, ([0.5] l_{1} \, . \, \nil \oplus [0.5] l_{2} \, . \, \nil)$ cannot infer anything about
the execution of $h$. Indeed, after the execution of $l$, what the low-level agent observes is either a
terminal state, reached with probability $1$, or the execution of either $l_{1}$ or $l_{2}$, both with
probability $0.5$. Formally, $E{'} \setminus \cala_{\calh} \wbis{} E{'} \, / \, \cala_{\calh}$ because $l \,
. \, \nil + l \, . \, ([0.5] \nil \oplus [0.5] \nil) + l \, . \, ([0.5] l_{1} \, . \, \nil \oplus [0.5]
l_{2} \, . \, \nil) \wbis{} l \, . \, \nil + l \, . \, ([0.5] \tau \, . \, [1] l_{1} \, . \, \nil \oplus
[0.5] \tau \, . \, [1] l_{2} \, . \, \nil) + l \, . \, ([0.5] l_{1} \, . \, \nil \oplus [0.5] l_{2} \, . \,
\nil)$ as can be seen in Figure~\ref{fig:bsnni}.

	\begin{figure}[t]

\begin{center}
\begin{tikzpicture}[modal, node distance = 0.7cm]

\node[point] (root) [label = 90: {$E' \setminus \cala_{\calh}$}] {};
\node[point] (p2)   [below = of root] 			         {};
\node[point] (p1)   [left = 2.4cm of p2]           	         {};
\node[point] (p3)   [right = 2.4cm of p2] 	 	         {};
\node[point] (p4)   [below = of p1]                              {};
\node[point] (p5)   [below left = 1cm of p2, yshift = -2pt]      {};
\node[point] (p6)   [below right = 1cm of p2, yshift = -2pt]     {};
\node[point] (p7)   [below left = 1cm of p3, yshift = -2pt]      {};
\node[point] (p8)   [below right = 1cm of p3, yshift = -2pt]     {};
\node[point] (p11)  [below = of p7]                              {};
\node[point] (p12)  [below = of p8]                              {};
\node[point] (p15)  [below = of p11]                             {};
\node[point] (p16)  [below = of p12]                             {};

\path[->] (root) edge node[above left]  {$l$}     (p1);
\path[->] (root) edge node[right]       {$l$}     (p2);
\path[->] (root) edge node[above right] {$l$}     (p3);
\path[->] (p1)   edge node[left]        {$1$}     (p4);
\path[->] (p2) 	 edge node[above left]  {$0.5$}   (p5);
\path[->] (p2)   edge node[above right] {$0.5$}   (p6);
\path[->] (p3)   edge node[above left]  {$0.5$}   (p7);
\path[->] (p3)   edge node[above right] {$0.5$}   (p8);
\path[->] (p7)   edge node[left]        {$l_{1}$} (p11);
\path[->] (p8)   edge node[right]       {$l_{2}$} (p12);
\path[->] (p11)  edge node[left]        {$1$}     (p15);
\path[->] (p12)  edge node[right]       {$1$}     (p16);

\node[point] (qroot) [label = 90: {$E' \, / \, \cala_{\calh}$}, right = 8cm of root] {};
\node[point] (q2)    [below = of qroot] 				             {};
\node[point] (q1)    [left = 2.4cm of q2]           			    	     {};
\node[point] (q3)    [right = 2.4cm of q2] 	 			  	     {};
\node[point] (q4)    [below = of q1]                     		             {};
\node[point] (q5)    [below left = 1cm of q2, yshift = -2pt]                         {};
\node[point] (q6)    [below right = 1cm of q2, yshift = -2pt]                        {};
\node[point] (q7)    [below left = 1cm of q3, yshift = -2pt]                         {};
\node[point] (q8)    [below right = 1cm of q3, yshift = -2pt]                        {};
\node[point] (q9)    [below = of q5]                          		  	     {};
\node[point] (q10)   [below = of q6]                         		  	     {};
\node[point] (q11)   [below = of q7]                         			     {};
\node[point] (q12)   [below = of q8]                         			     {};
\node[point] (q13)   [below = of q9]                          			     {};
\node[point] (q14)   [below = of q10]                          			     {};
\node[point] (q15)   [below = of q11]                         			     {};
\node[point] (q16)   [below = of q12]                         			     {};
\node[point] (q17)   [below = of q13]                         			     {};
\node[point] (q18)   [below = of q14]                         			     {};
\node[point] (q19)   [below = of q17]                         			     {};
\node[point] (q20)   [below = of q18]                          			     {};

\path[->] (qroot) edge node[above left]  {$l$}     (q1);
\path[->] (qroot) edge node[right]       {$l$}     (q2);
\path[->] (qroot) edge node[above right] {$l$}     (q3);
\path[->] (q1)    edge node[left]        {$1$}     (q4);
\path[->] (q2) 	  edge node[above left]  {$0.5$}   (q5);
\path[->] (q2)    edge node[above right] {$0.5$}   (q6);
\path[->] (q3)    edge node[above left]  {$0.5$}   (q7);
\path[->] (q3)    edge node[above right] {$0.5$}   (q8);
\path[->] (q5)    edge node[left]        {$\tau$}  (q9);
\path[->] (q6)    edge node[right]       {$\tau$}  (q10);
\path[->] (q7)    edge node[left]        {$l_{1}$} (q11);
\path[->] (q8)    edge node[right]       {$l_{2}$} (q12);
\path[->] (q9)    edge node[left]        {$1$}     (q13);
\path[->] (q10)   edge node[right]       {$1$}     (q14);
\path[->] (q11)   edge node[left]        {$1$}     (q15);
\path[->] (q12)   edge node[right]       {$1$}     (q16);
\path[->] (q13)   edge node[left]        {$l_{1}$} (q17);
\path[->] (q14)   edge node[right]       {$l_{2}$} (q18);
\path[->] (q17)   edge node[left]        {$1$}     (q19);
\path[->] (q18)   edge node[right]       {$1$}     (q20);

\end{tikzpicture}
\end{center}

\caption{Processes $E' \setminus \cala_{\calh}$ and $E' \, / \, \cala_{\calh}$ are identified by $\wbis{\rm
pw}$ and $\wbis{\rm pb}$}
\label{fig:bsnni}

	\end{figure}

BSNNI is not powerful enough to capture covert channels that derive from the behavior of a high-level agent
interacting with the system. For instance, $E'' = \, l \, . \, \nil + l \, . \, ([0.5] h_{1} \, . \, [1]
l_{1} \, . \, \nil \oplus [0.5] h_{2} \, . \, [1] l_{2} \, . \, \nil) + l \, . \, ([0.5] l_{1} \, . \, \nil
\oplus [0.5] l_{2} \, . \, \nil)$ is BSNNI$_{\wbis{}}$ \blue{for} the same reason \blue{as} discussed for
$E'$. However, a high-level agent could decide to enable only $h_{1}$, thus turning the low-level view of
the system into $l \, . \, \nil + l \, . \, ([0.5] \tau \, . \, [1] l_{1} \, . \, \nil \oplus [0.5] \nil) +
l \, . \, ([0.5] l_{1} \, . \, \nil \oplus [0.5] l_{2} \, . \, \nil)$, which is clearly distinguishable from
$l \, . \, \nil + l \, . \, ([0.5] \nil \oplus [0.5] \nil) + l \, . \, ([0.5] l_{1} \, . \, \nil \oplus
[0.5] l_{2} \, . \, \nil$), as in the former there is a case in which the low-level agent can observe
$l_{1}$ but not $l_{2}$ after the execution of~$l$. To avoid such a limitation, the most obvious solution
consists of checking explicitly the interaction on any action set $L \subseteq \cala_{\calh}$ between the
system and every possible high-level agent $F$. The resulting property is \emph{Bisimulation-based
Non-Deducibility on Composition} (BNDC), which features a universal quantification over $F$ containing only
high-level actions.

To overcome the verification problems related to such a \blue{quantification}, several properties have been
proposed that are stronger than BNDC. They all express some persistency conditions, stating that the
security checks have to be extended to all the processes reachable from a secure one. Three of the most
representative ones among such properties are the variant of BSNNI that requires every reachable process to
satisfy BSNNI itself, called \emph{Strong} BSNNI (SBSNNI), the variant of BNDC that requires every reachable
process to satisfy BNDC itself, called \emph{Persistent} BNDC (P\_BNDC), and \emph{Strong} BNDC (SBNDC),
which requires the low-level view of every reachable process to be the same before and after the execution
of any high-level action, meaning that the execution of high-level actions must be completely transparent to
low-level agents. In the nondeterministic setting, P\_BNDC and SBSNNI have been proven to coincide
in~\cite{FR06} for their weak bisimilarity variants and in~\cite{EABR25} for their branching bisimilarity
variants.

%
%
\section{Characteristics of Probabilistic Security Properties}
\label{sec:prob_bisim_sec_prop_char} 
%
%

In this section we investigate preservation and compositionality characteristics of the noninterference
properties introduced in the previous section (Section~\ref{sec:prob_pres_comp}) as well as their taxonomy,
i.e., the inclusion relationships among the ones based on $\wbis{\rm pw}$ and the ones based on $\wbis{\rm
pb}$ (Section~\ref{sec:prob_taxonomy}). Then we relate the resulting probabilistic taxonomy with the
nondeterministic one of~\cite{EABR25} (Section~\ref{sec:prob_taxonomy_rel}).

%
\subsection{Preservation and Compositionality}
\label{sec:prob_pres_comp}
%

\mbox{Almost} all the probabilistic noninterference properties of
Definition~\ref{def:prob_bisim_sec_prop} turn out to be preserved by the bisimilarity employed in their
definition. This means that if a process $E_{1}$ is secure under any of such properties for which
preservation holds, then every other bisimilar process $E_{2}$ is secure too according to the
same property. This is very useful for automated property verification, as it allows us to work with the
process with the smallest state space \blue{modulo bisimilarity}.

The preservation result of the forthcoming Theorem~\ref{thm:prob_preservation} immediately follows from
Lemma~\ref{lem:prob_bisim_congr} below, which establishes when $\wbis{\rm pw}$ and $\wbis{\rm pb}$ are
congruences with respect to the operators occurring in the aforementioned noninterference properties.
Congruence with respect to action prefix is also addressed as it will be exploited in the proof of the
compositionality result of Theorem~\ref{thm:prob_compositionality}. Similar congruence results in the
non-strictly alternating model have been proven in~\cite{AGT12} for $\wbis{\rm pb}$\blue{,} whereas they
have not been investigated in~\cite{PLS00} for $\wbis{\rm pw}$.

The congruence lemma is preceded by the following lemma about the relationship between parallel composition
of processes and product of probabilities.

	\begin{lem}\label{lem:prob_prod_par_comp}

Let $E_{1} \pco{L} E_{2}, E'_{1} \pco{L} E'_{2} \in \procs$. Then $\proba(E_{1} \pco{L} E_{2}, E'_{1}
\pco{L} E'_{2}) = \proba(E_{1}, E'_{1}) \cdot \proba(E_{2}, E'_{2})$.

		\begin{proof}

There are two cases:

			\begin{itemize}

\item If $E_{1}$ and $E_{2}$ are both nondeterministic, then $\proba(E_{1}, E'_{1}) \cdot \proba(E_{2},
E'_{2}) = 1$ if $E_{1} = E'_{1}$ and $E_{2} = E'_{2}$, while $\proba(E_{1}, E'_{1}) \cdot \proba(E_{2},
E'_{2}) = 0$ otherwise. From this fact it follows that $\proba(E_{1} \pco{L} E_{2}, E'_{1} \pco{L} E'_{2}) =
1$ if $E_{1} \pco{L} E_{2} = E'_{1} \pco{L} E'_{2}$, i.e., $E_{1} = E'_{1}$ and $E_{2} = E'_{2}$, while
$\proba(E_{1} \pco{L} E_{2}, E'_{1} \pco{L} E'_{2}) = 0$ otherwise.

\item If $E_{1}$ and $E_{2}$ are both probabilistic, then $\proba(E_{1}, E'_{1}) = \sum_{E_{1} \arrow{p}{\rm
p} E'_{1}} p$ and $\proba(E_{2}, E'_{2}) \linebreak = \sum_{E_{2} \arrow{q}{\rm p} E'_{2}} q$ and hence
$\proba(E_{1}, E'_{1}) \cdot \proba(E_{2}, E'_{2}) = \sum_{E_{1} \arrow{p}{\rm p} E'_{1}} p \cdot
\sum_{E_{2} \arrow{q}{\rm p} E'_{2}} q = \sum_{E_{1} \arrow{p}{\rm p} E'_{1}} \sum_{E_{2} \arrow{q}{\rm p}
E'_{2}} p \cdot q$ by distributivity, which is equal to $\proba(E_{1} \pco{L} E_{2}, E'_{1} \pco{L} E'_{2})$
according to the operational semantic rule \blue{\textit{ProbSynch}} of Table~\ref{tab:prob_op_sem_pr}.
\qedhere

			\end{itemize}

		\end{proof}

	\end{lem}

	\begin{lem}\label{lem:prob_bisim_congr}

Let $E_{1}, E_{2} \in \procs$ and $\wbis{} \: \in \{ \wbis{\rm pw}, \wbis{\rm pb} \}$. If $E_{1} \wbis{}
E_{2}$ then:

		\begin{enumerate}

\item $a \, . \, E_{1} \wbis{} a \, . \, E_{2}$ for all $a \in \cala_{\tau}$, when $E_{1}, E_{2} \in
\procs_{\rm p}$.

\item $E_{1} \pco{L} E \wbis{} E_{2} \pco{L} E$ and $E \pco{L} E_{1} \wbis{} E \pco{L} E_{2}$ for all $L \!
\subseteq \! \cala$ and $E \! \in \! \procs$, when $E_{1}, E_{2}, E \in \procs_{\rm n}$ \linebreak or
$E_{1}, E_{2}, E \in \procs_{\rm p}$, provided that $\wbis{} \; = \; \wbis{\rm pb}$.

\item $E_{1} \setminus L \wbis{} E_{2} \setminus L$ for all $L \subseteq \cala$.

\item $E_{1} \, / \, L \wbis{} E_{2} \, / \, L$ for all $L \subseteq \cala$.

		\end{enumerate}

		\begin{proof}

\blue{First we} prove the \blue{three} results for the $\wbis{\rm pw}$-based properties. Let
\blue{$\cali_{\procs}$ be the identity relation over $\procs$ and} $\calb$ be a weak probabilistic
bisimulation witnessing $E_{1} \wbis{\rm pw} E_{2}$:

			\begin{enumerate}

\item The equivalence relation $\calb' = (\calb \cup \{ (a \, . \, F_{1}, a \, . \, F_{2}) \mid (F_{1},
F_{2}) \in \calb \land F_{1}, F_{2} \in \procs_{\rm p} \})^{+}$ is a weak probabilistic bisimulation too,
\blue{where $^{+}$ denotes transitive closure}. The result immediately follows from the fact that, given $(a
\, . \, F_{1}, a \, . \, F_{2}) \in \calb'$ with $(F_{1}, F_{2}) \in \calb$, $a \, . \, F_{1} \arrow{a}{\rm
a} F_{1}$ is matched by $a \, . \, F_{2}$ $\warrow{}{} a \, . \, F_{2} \arrow{a}{\rm a} F_{2} \warrow{}{}
F_{2}$ with $(F_{1}, F_{2}) \in \calb'$ -- which can be scheduled with probability $1$ -- as well as
$\proba(a \, . \, F_{1}, \bar{C}) = \proba(a \, . \, F_{2}, \bar{C}) = 1$ for $\bar{C} = [a \, . \,
F_{1}]_{\calb'}$ while $\proba(a \, . \, F_{1}, C') = \proba(a \, . \, F_{2}, C') = 0$ for any other $C' \in
\procs / \calb'$.

\item $\wbis{\rm pw}$ is not a congruence with respect to parallel composition. As a counterexample,
consider $H_{1} = a \, . \, [1] (\tau \, . \, [1] \, . \, \tau \, . \, [1] \nil + \tau \, . \, [1] \, . \, b
\, . \, [1] \nil)$ and $H_{2} = H_{1} + a \, . \, [1] \nil$. It holds that $H_{1} \wbis{\rm pw} H_{2}$
because the additional $a$-transition of $H_{2}$ to $[1] \nil$ is matched by $H_{1} \! \arrow{a}{\rm a} \!
\arrow{1}{\rm p} \! \arrow{\tau}{\rm a} \! \arrow{1}{\rm p} \! \arrow{\tau}{\rm a} [1] \nil$, which can be
scheduled with probability $1$; note that $H_{1} \not\wbis{\rm pb} H_{2}$. However, for $H = a \, . \,
([0.5] c \, . \, [1] \nil \oplus [0.5] d \, . \, [1] \nil)$ it holds that $H_{1} \pco{\{ a \}} H
\not\wbis{\rm pw} H_{2} \pco{\{ a \}} H$ because the additional $a$-transition of $H_{2} \pco{\{ a \}} H$ to
$[1] \nil \pco{\{ a \}} ([0.5] c \, . \, [1] \nil \oplus [0.5] d \, . \, [1] \nil)$ cannot be matched by
$H_{1} \pco{\{ a \}} H$ because the process the latter evolves into after performing $a$ can subsequently
perform $b$ before or after $c$ or $d$ whereas this is not possible from $[1] \nil \pco{\{ a \}} ([0.5] c \,
. \, [1] \nil \oplus [0.5] d \, . \, [1] \nil)$.

\item The equivalence relation $\calb' = \cali_{\procs} \cup \{ (F_{1} \setminus L, F_{2} \setminus L) \mid
(F_{1}, F_{2}) \in \calb \}$ is a weak probabilistic bisimulation too. Given $(F_{1} \setminus L, F_{2}
\setminus L) \in \calb'$ with $(F_{1}, F_{2}) \in \calb$, there are two cases for action transitions based
on the operational semantic rules in Table~\ref{tab:prob_op_sem_nd}:

				\begin{itemize}

\item If $F_{1} \setminus L \arrow{\tau}{\rm a} F'_{1} \setminus L$ with $F_{1} \arrow{\tau}{\rm a} F'_{1}$,
then there exists $\sigma \in \sched(F_{2})$ such that $\proba_{\sigma}(F_{2}, \tau^{*} \, \hat{\tau} \,
\tau^{*}, [F'_{1}]_{\calb}, \varepsilon_{F_{2}}) = 1$. Since the restriction operator does not affect $\tau$
and probabilistic transitions, we have that there exists $\sigma' \in \sched(F_{2} \setminus L)$ such that
$\proba_{\sigma'}(F_{2} \setminus L, \tau^{*} \, \hat{\tau} \, \tau^{*}, [F'_{1} \setminus L]_{\calb'},
\varepsilon_{F_{2} \, \setminus \, L}) = 1$.

\item If $F_{1} \setminus L \arrow{a}{\rm a} F'_{1} \setminus L$ with $F_{1} \arrow{a}{\rm a} F'_{1}$ and $a
\notin L \cup \{ \tau \}$, then there exists $\sigma \in \sched(F_{2})$ such that $\proba_{\sigma}(F_{2},
\tau^{*} \, \hat{a} \, \tau^{*}, [F'_{1}]_{\calb}, \varepsilon_{F_{2}}) = 1$.  Since the restriction
operator does not affect $\tau$ and probabilistic transitions and $a \notin L$, we have that there exists
$\sigma' \in \sched(F_{2} \setminus L)$ such that $\proba_{\sigma'}(F_{2} \setminus L, \tau^{*} \,
\blue{\hat{a}} \, \tau^{*}, [F'_{1} \setminus L]_{\calb'}, \varepsilon_{F_{2} \, \setminus \, L}) = 1$.

				\end{itemize}

\noindent
As for probabilities, to avoid trivial cases consider an equivalence class $C' = C \setminus L = \{ F
\setminus L \mid F \in C \}$ for some $C \in \procs / \calb$. From $(F_{1}, F_{2}) \in \calb$ it follows
that $\proba(F_{1}, C) = \proba(F_{2}, C)$. Since the restriction operator does not affect probabilistic
transitions, \linebreak we have that $\proba(F_{1} \setminus L, C') = \proba(F_{1}, C) = \proba(F_{2}, C) =
\proba(F_{2} \setminus L, C')$.

\item The equivalence relation $\calb' = \cali_{\procs} \cup \{ (F_{1} \, / \, L, F_{2} \, / \, L) \mid
(F_{1}, F_{2}) \in \calb \}$ is a weak probabilistic bisimulation too. Given $(F_{1} \, / \, L, F_{2} \, /
\, L) \in \calb'$ with $(F_{1}, F_{2}) \in \calb$, there are two cases for action transitions based on the
operational semantic rules in Table~\ref{tab:prob_op_sem_nd}:

				\begin{itemize}

\item If $F_{1} \, / \, L \arrow{\tau}{\rm a} F'_{1} \, / \, L$ with $F_{1} \arrow{\tau}{\rm a} F'_{1}$,
then there exists $\sigma \in \sched(F_{2})$ such that $\proba_{\sigma}(F_{2}, \tau^{*} \, \hat{\tau} \,
\tau^{*}, [F'_{1}]_{\calb}, \varepsilon_{F_{2}}) = 1$. Since the hiding operator does not affect $\tau$ and
probabilistic transitions, we have that there exists $\sigma' \in \sched(F_{2} \, / \, L)$ such that
$\proba_{\sigma'}(F_{2} \, / \, L, \tau^{*} \, \hat{\tau} \, \tau^{*}, [F'_{1} \, / \, L]_{\calb'},
\varepsilon_{F_{2} \, / \, L}) = 1$.

\item If $F_{1} \, / \, L \arrow{a}{\rm a} F'_{1} \, / \, L$ with $F_{1} \arrow{b}{\rm a} F'_{1}$ and $b \in
L \land a = \tau$ or $b \notin L \cup \{ \tau \} \land a = b$, then there exists $\sigma \in \sched(F_{2})$
such that $\proba_{\sigma}(F_{2}, \tau^{*} \, \hat{b} \, \tau^{*}, [F'_{1}]_{\calb}, \varepsilon_{F_{2}}) =
1$. Since the hiding operator does not affect $\tau$ and probabilistic transitions, we have that there
exists $\sigma' \in \sched(F_{2} \, / \, L)$ such that $\proba_{\sigma'}(F_{2} \, / \, L, \tau^{*} \,
\hat{a} \, \tau^{*}, [F'_{1} \, / \, L]_{\calb'}, \varepsilon_{F_{2} \, / \, L}) = 1$.

				\end{itemize}

\noindent
As for probabilities, to avoid trivial cases consider an equivalence class $C' = C \, / \, L = \{ F \, / \,
L \mid F \in C \}$ for some $C \in \procs / \calb$. From $(F_{1}, F_{2}) \in \calb$ it follows that
$\proba(F_{1}, C) = \proba(F_{2}, C)$. Since the hiding operator does not affect probabilistic transitions,
we have that $\proba(F_{1} \, / \, L, C') = \proba(F_{1}, C) = \proba(F_{2}, C) = \proba(F_{2} \, / \, L,
C')$.

			\end{enumerate}

\noindent
\blue{Next we} prove the four results for the $\wbis{\rm pb}$-based properties. Let $\calb$ be a
probabilistic branching bisimulation witnessing $E_{1} \wbis{\rm pb} E_{2}$. We show that the equivalence
relations $\calb'$ considered for the $\wbis{\rm pw}$-based properties are probabilistic branching
bisimulations too:

			\begin{enumerate}

\item The result immediately follows from the fact that, given $(a \, . \, F_{1}, a \, . \, F_{2}) \in
\calb'$ with $(F_{1}, F_{2}) \in \calb$, $a \, . \, F_{1} \arrow{a}{\rm a} F_{1}$ is matched by $a \, . \,
F_{2} \warrow{}{} a \, . \, F_{2} \arrow{a}{\rm a} F_{2}$ with $(a \, . \, F_{1}, a \, . \, F_{2}) \in
\calb'$ and $(F_{1}, F_{2}) \in \calb'$, as well as $\proba(a \, . \, F_{1}, \bar{C}) = \proba(a \, . \,
F_{2}, \bar{C}) = 1$ for $\bar{C} = [a \, . \, F_{1}]_{\calb'}$ while $\proba(a \, . \, F_{1}, C') =
\proba(a \, . \, F_{2}, C') = 0$ for any other $C' \in \procs / \calb'$.

\item The equivalence relation $\calb' = \cali_{\procs} \cup \{ (F_{1} \! \pco{L} \! F, F_{2} \! \pco{L} \!
F) \mid (F_{1}, F_{2}) \! \in \! \calb \land (F_{1} \! \pco{L} \! F, F_{2} \! \pco{L} \! F \!  \in \!
\procs_{\rm n} \linebreak \lor F_{1} \pco{L} F, F_{2} \pco{L} F \in \procs_{\rm p}) \} \cup \{ (F_{1}
\pco{L} [1] F, F_{2} \pco{L} F) \mid (F_{1}, F_{2}) \in \calb \land F_{1} \in \procs_{\rm p} \land F_{2}
\pco{L} F \in \procs_{\rm n} \} \linebreak \cup \{ (F_{1} \pco{L} F, F_{2} \pco{L} [1] F) \mid (F_{1},
F_{2}) \in \calb \land F_{2} \in \procs_{\rm p} \land F_{1} \pco{L} F \in \procs_{\rm n} \}$ and its variant
$\calb''$ in which $F$ occurs to the left of parallel composition in each pair are probabilistic branching
bisimulations too. Let us focus on $\calb'$. Given $(F_{1} \pco{L} F, F_{2} \pco{L} F) \in \calb'$ with
$(F_{1}, F_{2}) \in \calb$, there are three cases for action transitions based on the operational semantic
rules in Table~\ref{tab:prob_op_sem_nd}:

				\begin{itemize}

\item If $F_{1} \pco{L} F \arrow{a}{\rm a} F'_{1} \pco{L} [1] F$ with $F_{1} \arrow{a}{\rm a} F'_{1}$ and $a
\notin \! L$, then either $a = \tau$ and $(F'_{1}, F_{2}) \! \in \calb$, \linebreak or there exists $F_{2}
\warrow{}{} \bar{F}_{2} \arrow{a}{\rm a} F'_{2}$ such that $(F_{1}, \bar{F}_{2}) \in \calb$ and $(F'_{1},
F'_{2}) \in \calb$. Note that the action transition from $F_{1} \pco{L} F$ implies that $F_{1} \pco{L} F \in
\procs_{\rm n}$, i.e., $F_{1}, F \in \procs_{\rm n}$, hence $F_{2} \pco{L} F \in \procs_{\rm n}$ too. Since
synchronization does not affect $\tau$ and $a \notin L$, in the former subcase $F_{2} \pco{L} F$ is allowed
to stay idle with $(F'_{1} \pco{L} [1] F, F_{2} \pco{L} F) \in \calb'$, while in the latter subcase $F_{2}
\pco{L} F \warrow{}{} \bar{F}_{2} \pco{L} F \arrow{a}{\rm a} F'_{2} \pco{L} [1] F$ with $(F_{1} \pco{L} F,
\bar{F}_{2} \pco{L} F) \in \calb'$ and $(F'_{1} \pco{L} [1] F, F'_{2} \pco{L} [1] F) \in \calb'$, in which
subcase the right subprocess alternates between $F$ and $[1] F$ before $a$ is performed thus allowing the
probabilistic transitions along $F_{2} \warrow{}{} \bar{F}_{2}$ to synchronize with the only one of $[1] F$.

\item The case $F_{1} \pco{L} F \arrow{a}{\rm a} [1] F_{1} \pco{L} F'$ with $F \arrow{a}{\rm a} F'$ and $a
\notin L$ is trivial.

\item If $F_{1} \pco{L} F \arrow{a}{\rm a} F'_{1} \pco{L} F'$ with $F_{1} \arrow{a}{\rm a} F'_{1}$, $F
\arrow{a}{\rm a} F'$, and $a \in L$, then there exists $F_{2} \warrow{}{} \bar{F}_{2} \arrow{a}{\rm a}
F'_{2}$ such that $(F_{1}, \bar{F}_{2}) \in \calb$ and $(F'_{1}, F'_{2}) \in \calb$. Since synchronization
does not affect $\tau$ and $a \in L$, we have that $F_{2} \pco{L} F \warrow{}{} \bar{F}_{2} \pco{L} F
\arrow{a}{\rm a} F'_{2} \pco{L} F'$ with $(F_{1} \pco{L} F, \bar{F}_{2} \pco{L} F) \in \calb'$ and $(F'_{1}
\pco{L} F', F'_{2} \pco{L} F') \in \calb'$, where the right subprocess alternates between $F$ and $[1] F$
before performing $a$ thus allowing the probabilistic transitions along $F_{2} \warrow{}{} \bar{F}_{2}$ to
synchronize with the only one of $[1] F$.

				\end{itemize}

\noindent
As for probabilities, to avoid trivial cases let $F_{1}, F_{2}, F \in \procs_{\rm p}$ and consider an
equivalence class $C' = C \pco{L} F' = \{ H \pco{L} F' \mid H \in C \} \in \procs / \calb'$ for some $C \in
\procs / \calb$ with $F' \in \procs_{\rm n}$. \linebreak By virtue of Lemma~\ref{lem:prob_prod_par_comp} we
obtain $\proba(F_{k} \pco{L} F, C') = \sum_{H \pco{L} F' \in C'} \proba(F_{k} \pco{L} F, H \pco{L} F')
\linebreak = \sum_{H \pco{L} F' \in C'} \proba(F_{k}, H) \cdot \proba(F, F') = \sum_{H \in C} \proba(F_{k},
H) \cdot \proba(F, F') = \linebreak (\sum_{H \in C} \proba(F_{k}, H)) \cdot \proba(F, F') = \proba(F_{k}, C)
\cdot \proba(F, F')$ for $k \in \{ 1, 2 \}$. From $(F_{1}, F_{2}) \in \calb$ it follows that $\proba(F_{1},
C) = \proba(F_{2}, C)$, hence $\proba(F_{1} \pco{L} F, C') = \proba(F_{2} \pco{L} F, C')$.

\item Given $(F_{1} \setminus L, F_{2} \setminus L) \in \calb'$ with $(F_{1}, F_{2}) \in \calb$, there are
two cases for action transitions based on the operational semantic rules in Table~\ref{tab:prob_op_sem_nd}:

				\begin{itemize}

\item If $F_{1} \setminus L \arrow{\tau}{\rm a} F'_{1} \setminus L$ with $F_{1} \arrow{\tau}{\rm a} F'_{1}$,
then either $(F'_{1}, F_{2}) \in \calb$, or there exists $F_{2} \warrow{}{} \bar{F}_{2} \arrow{\tau}{\rm a}
F'_{2}$ such that $(F_{1}, \bar{F}_{2}) \in \calb$ and $(F'_{1}, F'_{2}) \in \calb$. Since the restriction
operator does not affect $\tau$ and probabilistic transitions, in the former subcase $F_{2} \setminus L$ is
allowed to stay idle with $(F'_{1} \setminus L, F_{2} \setminus L) \in \calb'$, while in the latter subcase
$F_{2} \setminus L \warrow{}{} \bar{F}_{2} \setminus L \arrow{\tau}{\rm a} F'_{2} \setminus L$ with $(F_{1}
\setminus L, \bar{F}_{2} \setminus L) \in \calb'$ and $(F'_{1} \setminus L, F'_{2} \setminus L) \in \calb'$.

\item If $F_{1} \setminus L \arrow{a}{\rm a} F'_{1} \setminus L$ with $F_{1} \arrow{a}{\rm a} F'_{1}$ and $a
\notin L \cup \{ \tau \}$, then there exists $F_{2} \warrow{}{} \bar{F}_{2} \arrow{a}{\rm a} F'_{2}$ such
that $(F_{1}, \bar{F}_{2}) \in \calb$ and $(F'_{1}, F'_{2}) \in \calb$. Since the restriction operator does
not affect~$\tau$ and probabilistic transitions and $a \notin L$, we have that $F_{2} \setminus L
\warrow{}{} \bar{F}_{2} \setminus L \arrow{a}{\rm a} F'_{2} \setminus L$ with $(F_{1} \setminus L,
\bar{F}_{2} \setminus L) \in \calb'$ and $(F'_{1} \setminus L, F'_{2} \setminus L) \in \calb'$.

				\end{itemize}

\noindent
As for probabilities, we reason like in the proof of the corresponding result for $\wbis{\rm pw}$.

\item Given $(F_{1} \, / \, L, F_{2} \, / \, L) \in \calb'$ with $(F_{1}, F_{2}) \in \calb$, there are two
cases for action transitions based on the operational semantic rules in Table~\ref{tab:prob_op_sem_nd}:

				\begin{itemize}

\item If $F_{1} \, / \, L \arrow{\tau}{\rm a} F'_{1} \, / \, L$ with $F_{1} \arrow{\tau}{\rm a} F'_{1}$,
then either $(F'_{1}, F_{2}) \in \calb$, or there exists $F_{2} \warrow{}{} \bar{F}_{2} \linebreak
\arrow{\tau}{\rm a} F'_{2}$ such that $(F_{1}, \bar{F}_{2}) \in \calb$ and $(F'_{1}, F'_{2}) \in \calb$.
Since the hiding operator does not affect $\tau$ and probabilistic transitions, in the former subcase $F_{2}
\, / \, L$ is allowed to stay idle with $(F'_{1} \, / \, L, F_{2} \, / \, L) \in \calb'$, while in the
latter subcase $F_{2} \, / \, L \warrow{}{} \bar{F}_{2} \, / \, L \arrow{\tau}{\rm a} F'_{2} \, / \, L$ with
$(F_{1} \, / \, L, \bar{F}_{2} \, / \, L) \in \calb'$ and $(F'_{1} \, / \, L, F'_{2} \, / \, L) \in \calb'$.

\item If $F_{1} \, / \, L \arrow{a}{\rm a} F'_{1} \, / \, L$ with $F_{1} \arrow{b}{\rm a} F'_{1}$ and $b \in
L \land a = \tau$ or $b \notin L \cup \{ \tau \} \land a = b$, then there exists $F_{2} \warrow{}{}
\bar{F}_{2} \arrow{b}{\rm a} F'_{2}$ such that $(F_{1}, \bar{F}_{2}) \in \calb$ and $(F'_{1}, F'_{2}) \in
\calb$. Since the hiding operator does not affect $\tau$ and probabilistic transitions, we have that $F_{2}
\, / \, L \warrow{}{} \bar{F}_{2} \, / \, L \arrow{a}{\rm a} F'_{2} \, / \, L$ with $(F_{1} \, / \, L ,
\bar{F}_{2} \, / \, L) \in \calb'$ and $(F'_{1} \, / \, L, F'_{2} \, / \, L) \in \calb'$.

				\end{itemize}

\noindent
As for probabilities, we reason like in the proof of the corresponding result for $\wbis{\rm pw}$.
\qedhere

			\end{enumerate}
		
		\end{proof}			
	
	\end{lem}

	\begin{thm}\label{thm:prob_preservation}

Let $E_{1}, E_{2} \in \procs$, $\wbis{} \: \in \{ \wbis{\rm pw}, \wbis{\rm pb} \}$, and $\calp \in \{
\mathrm{BSNNI}_{\wbis{}}, \mathrm{BNDC}_{\wbis{}}, \mathrm{SBSNNI}_{\wbis{}}, \linebreak
\mathrm{P\_BNDC}_{\wbis{}}, \mathrm{SBNDC}_{\wbis{}} \}$. If $E_{1} \wbis{} E_{2}$ then $E_{1} \in \calp
\Longleftrightarrow E_{2} \in \calp$ provided that $\calp \neq \mathrm{BNDC}_{\wbis{\rm pw}}$.

		\begin{proof}

A straightforward consequence of the definition of the various properties, i.e.,
Definition~\ref{def:prob_bisim_sec_prop}, and Lemma~\ref{lem:prob_bisim_congr}. Note that the preservation
of $\mathrm{P\_BNDC}_{\wbis{}}$ holds when $\wbis{} \; = \; \wbis{\rm pw}$ because
Theorem~\ref{thm:prob_taxonomy_1} will establish that $\mathrm{P\_BNDC}_{\wbis{}} =
\mathrm{SBSNNI}_{\wbis{}}$.
\qedhere

		\end{proof}

	\end{thm}

As far as modular verification is concerned, like in the nondeterministic setting~\cite{FG01,EABR25} the
forthcoming Theorem~\ref{thm:prob_compositionality} establishes that only the local properties
SBSNNI$_{\wbis{}}$, P\_BNDC$_{\wbis{}}$, and SBNDC$_{\wbis{}}$ are compositional, i.e., are preserved by
some operators of the calculus in certain circumstances. \blue{A counterexample for BSNNI is $l \, . \,
\nil + h_{1} \, . \, h_{2} \, . \, l \, . \, \nil$, while for BNDC the reader is referred to Proposition~$1$
of~\cite{Mar98}.} Moreover, similar to~\cite{EABR25} compositionality with respect to parallel composition
is limited, for SBSNNI$_{\wbis{\rm pb}}$ and P\_BNDC$_{\wbis{\rm pb}}$, to the case in which
synchronizations can take place only among low-level actions, i.e., $L \subseteq \cala_{\call}$. A
limitation to low-level actions applies to action prefix and hiding as well, whilst this is not the case for
restriction. Another analogy with the nondeterministic setting~\cite{FG01,EABR25} is that none of the
considered noninterference properties is compositional with respect to alternative composition. As an
example, let us examine processes $E_{1} = l \, . \, \nil$ and $E_{2} = h \, . \, \nil$.  For $\wbis{} \:
\in \{ \wbis{\rm pw}, \wbis{\rm pb} \}$ both processes are BSNNI$_{\wbis{}}$, as $(l \, . \, \nil) \setminus
\cala_{\calh} \wbis{} (l \, . \, \nil) \, / \, \cala_{\calh}$ and $(h \, . \, \nil) \setminus \cala_{\calh}
\wbis{} (h \, . \, \nil) \, / \, \cala_{\calh}$, but $E_{1} + E_{2} \notin \mathrm{BSNNI}_{\wbis{}}$,
because $(l \, . \, \nil + h \, . \, \nil) \setminus \cala_{\calh} \wbis{} l \, . \, \nil \not\wbis{} l \, .
\, \nil + \tau \, . \, \nil \wbis{} (l \, . \, \nil + h \, . \, \nil) \, / \, \cala_{\calh}$. \linebreak It
is easy to check that $E_{1} + E_{2} \notin \calp$ also for $\calp \in \{ \mathrm{BNDC}_{\wbis{}},
\mathrm{SBSNNI}_{\wbis{}}, \mathrm{P\_BNDC}_{\wbis{}}, \mathrm{SBNDC}_{\wbis{}} \}$. Moreover,
compositionality fails for probabilistic composition as well, for which it is sufficient to consider $[p]
E_{1} \oplus [1 - p] E_{2}$.

As already mentioned, the compositionality of SBSNNI$_{\wbis{\rm pb}}$ for the parallel operator holds only
for all $L \subseteq \cala_{\call}$. For example, both $E_{1} = \tau \, . \, \nil + l_{1} \, . \, \nil + h
\, . \, \nil$ and $E_{2} = \tau \, . \, \nil + l_{2} \, . \, \nil + h \, . \, \nil$ are SBSNNI$_{\wbis{\rm
pb}}$, but $E_{1} \pco{\{ h \}} E_{2}$ is not because the transition $(E_{1} \pco{\{ h \}} E_{2}) \, / \,
\cala_{\calh} \arrow{\tau}{\rm a} ([1] \nil \pco{\{ h \}} [1] \nil) \linebreak \, / \, \cala_{\calh}$
arising from the synchronization between the two $h$-actions cannot be matched by $(E_{1} \pco{\{ h \}}
E_{2}) \setminus \cala_{\calh}$. \blue{Due to the CSP-like parallel composition in our process language},
the only two $\tau$-possibilities are $(E_{1} \pco{\{ h \}} E_{2}) \setminus \cala_{\calh} \warrow{}{}
(E_{1} \pco{\{ h \}} E_{2}) \setminus \cala_{\calh} \arrow{\tau}{\rm a} ([1] \nil \pco{\{ h \}} [1] E_{2})
\setminus \cala_{\calh} \linebreak \arrow{1}{\rm p} (\nil \pco{\{ h \}} E_{2}) \setminus \cala_{\calh}
\arrow{\tau}{\rm a} ([1] \nil \pco{\{ h \}} [1] \nil) \setminus \cala_{\calh}$ and $(E_{1} \pco{\{ h \}}
E_{2}) \setminus \cala_{\calh} \warrow{}{} (E_{1} \pco{\{ h \}} E_{2}) \setminus \cala_{\calh} \linebreak
\arrow{\tau}{\rm a} ([1] E_{1} \pco{\{ h \}} [1] \nil) \setminus \cala_{\calh} {\arrow{1}{\rm p}} (E_{1}
\pco{\{ h \}} \nil) \setminus \cala_{\calh} \arrow{\tau}{\rm a} ([1] \nil \pco{\{ h \}} [1] \nil) \setminus
\cala_{\calh}$, but neither \linebreak $([1] \nil \pco{\{ h \}} [1] E_{2}) \setminus \cala_{\calh}$ nor
$([1] E_{1} \pco{\{ h \}} [1] \nil) \setminus \cala_{\calh}$ is $\wbis{\rm pb}$-equivalent to $(E_{1}
\pco{\{ h \}} E_{2}) \setminus \cala_{\calh}$ \linebreak when $l_{1} \neq l_{2}$ as can be seen in
Figure~\ref{fig:sbsnni_par}. Note that $(E_{1} \pco{\{ h \}} E_{2}) \, / \, \cala_{\calh} \wbis{\rm pw}
(E_{1} \pco{\{ h \}} E_{2}) \setminus \cala_{\calh}$ because $(E_{1} \pco{\{ h \}} E_{2}) \, / \,
\cala_{\calh} \arrow{\tau}{\rm a} ([1] \nil \pco{\{ h \}} [1] \nil) \, / \, \cala_{\calh}$ is matched by
$(E_{1} \pco{\{ h \}} E_{2}) \setminus \cala_{\calh} \warrow{}{}$ \linebreak $([1] \nil \pco{\{ h \}} [1]
\nil) \setminus \cala_{\calh}$. As observed in~\cite{EABR25}, it is not only a matter of the higher
discriminating power of $\wbis{\rm pb}$ with respect to $\wbis{\rm pw}$. If we \blue{had} used the CCS
parallel composition operator~\cite{Mil89a}, which turns the synchronization of two actions into $\tau$ thus
combining communication with hiding, then the parallel composition of $E_{1}$ and $E_{2}$ with restriction
on $\cala_{\calh}$ would be able to respond, in the probabilistic branching bisimulation game, with a single
$\tau$-transition reaching the parallel composition of $[1] \nil$ and $[1] \nil$ with restriction on
$\cala_{\calh}$.

	\begin{figure}[t]

\begin{center}
\begin{tikzpicture}[modal]

\node[point]        (root) [label = 90: {$(E_{1} \pco{\{ h \}} E_{2}) \setminus \cala_{\calh}$}] {};
\node[point, white] (p2)   [below = of root]                                         	         {};
\node[point]        (p1)   [left = of p2, xshift = -25pt]      				         {};
\node[point]        (p3)   [right = of p2, xshift = 25pt]      				         {};
\node[point]        (p4)   [below = of p1]                                                       {};
\node[point]        (p6)   [below = of p3]                                         	         {};
\node[point]        (p7)   [below left = of p4]                                        	         {};
\node[point]        (p8)   [below right = of p4]                                                 {};
\node[point]        (p9)   [below left = of p6]                                                  {};
\node[point]        (p10)  [below right = of p6]                                                 {};
\node[point]        (p11)  [below = of p7]                                         	         {};
\node[point]        (p12)  [below = of p8]                                          	         {};
\node[point]        (p13)  [below = of p9]                                          	         {};
\node[point]        (p14)  [below = of p10]                                          	         {};

\path[->] (root)  edge[bend right = 25] node[above left]                   {$\tau$}  (p1);
\path[->] (root)  edge                  node[below right]                  {$l_{1}$} (p1);
\path[->] (root)  edge[bend left = 25]  node[above right]                  {$\tau$}  (p3);
\path[->] (root)  edge                  node[below left]                   {$l_{2}$} (p3);
\path[->] (p1)    edge                  node[left]  	                   {$1$}     (p4);
\path[->] (p3)    edge                  node[right] 	                   {$1$}     (p6);
\path[->] (p4)    edge                  node[above left]                   {$\tau$}  (p7);
\path[->] (p4)    edge                  node[above right, yshift = -1.6pt] {$l_{2}$} (p8);
\path[->] (p6)    edge                  node[above left]                   {$\tau$}  (p9);
\path[->] (p6)    edge                  node[above right, yshift = -1.6pt] {$l_{1}$} (p10);
\path[->] (p7)    edge                  node[left]                         {$1$}     (p11);
\path[->] (p8)    edge                  node[right]                        {$1$}     (p12);
\path[->] (p9)    edge                  node[left]                         {$1$}     (p13);
\path[->] (p10)   edge                  node[right]                        {$1$}     (p14);

\node[point] (qroot) [right = 7.5cm of root, label = 90: {$(E_1 \pco{\{ h \}} E_2) \, / \, \cala_\calh$}] {};
\node[point] (q2)    [below = of qroot]                                         			  {};
\node[point] (q1)    [left = of q2, xshift = -25pt]      						  {};
\node[point] (q3)    [right = of q2, xshift = 25pt]      						  {};
\node[point] (q4)    [below = of q1]                                                			  {};
\node[point] (q5)    [below = of q2]                                              			  {};
\node[point] (q6)    [below = of q3]                                         				  {};
\node[point] (q7)    [below left = of q4]                                        			  {};
\node[point] (q8)    [below right = of q4]                                         			  {};
\node[point] (q9)    [below left = of q6]                                          			  {};
\node[point] (q10)   [below right = of q6]                                        			  {};
\node[point] (q11)   [below = of q7]                                         				  {};
\node[point] (q12)   [below = of q8]                                          				  {};
\node[point] (q13)   [below = of q9]                                          				  {};
\node[point] (q14)   [below = of q10]                                          				  {};

\path[->] (qroot) edge[bend right = 25] node[above left]                   {$\tau$}  (q1);
\path[->] (qroot) edge                  node[below right]                  {$l_{1}$} (q1);
\path[->] (qroot) edge                  node[right]                        {$\tau$}  (q2);
\path[->] (qroot) edge[bend left = 25]  node[above right]                  {$\tau$}  (q3);
\path[->] (qroot) edge                  node[below left]                   {$l_{2}$} (q3);
\path[->] (q1)    edge                  node[left]  	                   {$1$}     (q4);
\path[->] (q2)    edge                  node[right]                        {$1$}     (q5);
\path[->] (q3)    edge                  node[right] 	                   {$1$}     (q6);
\path[->] (q4)    edge                  node[above left]                   {$\tau$}  (q7);
\path[->] (q4)    edge                  node[above right, yshift = -1.6pt] {$l_{2}$} (q8);
\path[->] (q6)    edge                  node[above left]                   {$\tau$}  (q9);
\path[->] (q6)    edge                  node[above right, yshift = -1.6pt] {$l_{1}$} (q10);
\path[->] (q7)    edge                  node[left]                         {$1$}     (q11);
\path[->] (q8)    edge                  node[right]                        {$1$}     (q12);
\path[->] (q9)    edge                  node[left]                         {$1$}     (q13);
\path[->] (q10)   edge                  node[right]                        {$1$}     (q14);

\end{tikzpicture}
\end{center}

\caption{Processes $(E_{1} \pco{\{ h \}} E_{2}) \setminus \cala_{\calh}$ and $(E_{1} \pco{\{ h \}} E_{2}) \,
/ \, \cala_{\calh}$ are distinguished by $\wbis{\rm pb}$ but related by $\wbis{\rm pw}$}
\label{fig:sbsnni_par}

	\end{figure}

To establish compositionality, we first prove some ancillary results about parallel composition,
restriction, and hiding under SBSNNI and SBNDC similar to those in~\cite{EABR25}.

	\begin{lem}\label{lem:prob_compositionality}

Let $E_{1}, E_{2} \in \procs_{\rm n}$ or $E_{1}, E_{2} \in \procs_{\rm p}$, $E \in \procs$, and $\wbis{} \:
\in \{ \wbis{\rm pw}, \wbis{\rm pb} \}$. Then:

		\begin{enumerate}

\item If $E_{1}, E_{2} \in \mathrm{SBSNNI}_{\blue{\wbis{}}}$ and $L \subseteq \cala_{\call}$, then $(F_{1}
\pco{L} F_{2}) \setminus \cala_{\calh} \blue{\, \wbis{} \,} (G_{1} \pco{L} G_{2}) \, / \, \cala_{\calh}$ for
all $F_{1}, G_{1} \! \in \! \reach(E_{1})$ and $F_{2}, G_{2} \! \in \! \reach(E_{2})$ such that $F_{1}
\pco{L} F_{2}, G_{1} \pco{L} G_{2} \! \in \! \reach(E_{1} \pco{L} E_{2})$, $F_{1} \setminus \cala_{\calh}
\blue{\, \wbis{} \,} G_{1} \, / \, \cala_{\calh}$, and $F_{2} \setminus \cala_{\calh} \blue{\, \wbis{} \,}
G_{2} \, / \, \cala_{\calh}$\blue{, provided that $\wbis{} \; = \; \wbis{\rm pb}$}.

\item If $E \in \mathrm{SBSNNI_{\wbis{}}}$ and $L \subseteq \cala$, then $(F \, / \, \cala_{\calh})
\setminus L \wbis{} (G \setminus L) \, / \, \cala_{\calh}$ for all $F, G \in \reach(E)$ such that $F \, / \,
\cala_{\calh} \wbis{} G \setminus \cala_{\calh}$.

\item If $E_{1}, E_{2} \in \mathrm{SBNDC}_{\wbis{}}$ and $L \subseteq \cala$, then $(F_{1} \pco{L} F_{2})
\setminus \cala_{\calh} \wbis{} (G_{1} \pco{L} G_{2}) \setminus \cala_{\calh}$ for all $F_{1}, G_{1} \in
\reach(E_{1})$ and $F_{2}, G_{2} \in \reach(E_{2})$ such that $F_{1} \pco{L} F_{2}, G_{1} \pco{L} G_{2} \in
\reach(E_{1} \pco{L} E_{2})$, \linebreak $F_{1} \setminus \cala_{\calh} \wbis{} G_{1} \setminus
\cala_{\calh}$\blue{,} and $F_{2} \setminus \cala_{\calh} \wbis{} G_{2} \setminus \cala_{\calh}$.

		\end{enumerate}

		\begin{proof}


\blue{First we} prove the \blue{two} results for the $\wbis{\rm pw}$-based properties. Let $\calb$ be an
equivalence relation containing all the pairs of processes mentioned at beginning of the considered result
that have to be shown to be $\wbis{\rm pw}$-equivalent under the constraints mentioned at the end of the
result itself:

			\begin{enumerate}


\item The result does not hold for $\wbis{\rm pw}$ as can be seen by taking $E_{1}$ equal to $\tau \, . \,
[1] H_{1} + h \, . \, [1] H_{2}$ and $E_{2}$ equal to $H$, with processes $H_{1}$, $H_{2}$, and $H$ being
defined in the proof of Lemma~\ref{lem:prob_bisim_congr}(2) and all of their observable actions belonging to
$\cala_{\call}$, as well as $F_{1}$ and $G_{1}$ both equal to $E_{1}$ and $F_{2}$ and $G_{2}$ both equal to
$E_{2}$. It holds that $F_{1} \setminus \cala_{\calh} \wbis{\rm pw} G_{1} \, / \, \cala_{\calh}$ and $F_{2}
\setminus \cala_{\calh} \wbis{\rm pw} G_{2} \, / \, \cala_{\calh}$ -- as $E_{1}, E_{2} \in
\mathrm{SBSNNI}_{\wbis{\rm pw}}$ -- but $(F_{1} \pco{\{ a \}} F_{2}) \setminus \cala_{\calh} \not\wbis{\rm
pw} (G_{1} \pco{\{ a \}} G_{2}) \, / \, \cala_{\calh}$ because $(F_{1} \pco{\{ a \}} F_{2}) \setminus
\cala_{\calh}$ is isomorphic to $\tau \, . \, [1] H_{1} \pco{\{ a \}} H$, $(G_{1} \pco{\{ a \}} G_{2}) \, /
\, \cala_{\calh}$ is isomorphic to $(\tau \, . \, [1] H_{1} + \tau \, . \, [1] H_{2}) \pco{\{ a \}} H$, and
whenever $(\tau \, . \, [1] H_{1} + \tau \, . \, [1] H_{2}) \pco{\{ a \}} H \arrow{\tau}{\rm a} [1] H_{2}
\pco{\{ a \}} [1] H$ then $\tau \, . \, [1] H_{1} \pco{\{ a \}} H$ can either stay idle or perform $\tau \,
. \, [1] H_{1} \pco{\{ a \}} H \arrow{\tau}{\rm a} \! \arrow{1}{\rm p} H_{1} \pco{\{ a \}} H$ -- both
schedulable with probability $1$ -- but $[1] H_{2} \pco{\{ a \}} [1] H \not\wbis{\rm pw} \tau \, . \, [1]
H_{1} \pco{\{ a \}} H \wbis{\rm pw} H_{1} \pco{\{ a \}} H$ as observed in the proof of
Lemma~\ref{lem:prob_bisim_congr}(2).


\item Starting from $(F \, / \, \cala_{\calh}) \setminus L$ and $(G \setminus L ) \, / \, \cala_{\calh}$
related by $\calb$, so that $F \, / \, \cala_{\calh} \wbis{\rm pw} G \setminus \cala_{\calh}$, there are six
cases for action transitions based on the operational semantic rules in Table~\ref{tab:prob_op_sem_nd}. In
the first three cases, it is $(F \, / \, \cala_{\calh}) \setminus L$ to move first:

				\begin{itemize}

\item If $(F \, / \, \cala_{\calh}) \setminus L \arrow{l}{\rm a} (F' \, / \, \cala_{\calh}) \setminus L$
with $F \arrow{l}{\rm a} F'$ and $l \notin L$, then $F \, / \, \cala_{\calh} \arrow{l}{\rm a} F' \, / \,
\cala_{\calh}$ as $l \notin \cala_{\calh}$. From $F \, / \, \cala_{\calh} \wbis{\rm pw} G \setminus
\cala_{\calh}$ it follows that there exists $\sigma \in \sched(G \setminus \cala_{\calh})$ such that
$\proba_{\sigma}(G \setminus \cala_{\calh}, \tau^{*} \, \hat{l} \, \tau^{*}, [F' \, / \,
\cala_{\calh}]_{\wbis{\rm pw}}, \varepsilon_{G \, \setminus \, \cala_{\calh}}) = 1$. Since the restriction
and hiding operators do not affect $\tau$, $l$, and probabilistic transitions, we have that there exists
$\sigma' \in \sched((G \setminus L) \, / \, \cala_{\calh})$ such that $\proba_{\sigma'}((G \setminus L) \, /
\, \cala_{\calh}, \tau^{*} \, \hat{l} \, \tau^{*}, [(F' \, / \, \cala_{\calh}) \setminus L]_{\calb},
\linebreak \varepsilon_{(G \, \setminus \, L) \, / \, \cala_{\calh}}) = 1$.

\item If $(F \, / \, \cala_{\calh}) \setminus L \arrow{\tau}{\rm a} (F' \, / \, \cala_{\calh}) \setminus L$
with $F \arrow{\tau}{\rm a} F'$, then $F \, / \, \cala_{\calh} \arrow{\tau}{\rm a} F' \, / \, \cala_{\calh}$
as $\tau \notin \cala_{\calh}$. \linebreak From $F \, / \, \cala_{\calh} \wbis{\rm pw} G \setminus
\cala_{\calh}$ it follows that there exists $\sigma \in \sched(G \setminus \cala_{\calh})$ such that
$\proba_{\sigma}(G \setminus \cala_{\calh}, \tau^{*} \, \hat{\tau} \, \tau^{*}, [F' \, / \,
\cala_{\calh}]_{\wbis{\rm pw}}, \varepsilon_{G \, \setminus \, \cala_{\calh}}) = 1$. Since the restriction
and hiding operators do not affect $\tau$ and probabilistic transitions, we have that there exists $\sigma'
\in \sched((G \setminus L) \, / \, \cala_{\calh})$ such that $\proba_{\sigma'}((G \setminus L) \, / \,
\cala_{\calh}, \tau^{*} \, \hat{\tau} \, \tau^{*}, [(F' \, / \, \cala_{\calh}) \setminus L]_{\calb},
\linebreak \varepsilon_{(G \, \setminus \, L) \, / \, \cala_{\calh}}) = 1$.

\item If $(F \, / \, \cala_{\calh}) \setminus L \arrow{\tau}{\rm a} (F' \, / \, \cala_{\calh}) \setminus L$
with $F \arrow{h}{\rm a} F'$, then $F \, / \, \cala_{\calh} \arrow{\tau}{\rm a} F' \, / \, \cala_{\calh}$ as
$h \in \cala_{\calh}$. The rest of the proof is like the one of the previous case.

				\end{itemize}

\noindent
In the other three cases, instead, it is $(G \setminus L) \, / \, \cala_{\calh}$ to move first:

				\begin{itemize}

\item If $(G \setminus L) \, / \, \cala_{\calh} \arrow{l}{\rm a} (G' \setminus L) \, / \, \cala_{\calh}$
with $G \arrow{l}{\rm a} G'$ and $l \notin L$, then $G \setminus \cala_{\calh} \arrow{l}{\rm a} G' \setminus
\cala_{\calh}$ as $l \! \notin \! \cala_{\calh}$. From $G \setminus \cala_{\calh} \wbis{\rm pw} F /
\cala_{\calh}$ it follows that there exists $\sigma \in \sched(F \, / \, \cala_{\calh})$ such that
$\proba_{\sigma}(F \, / \, \cala_{\calh}, \tau^{*} \, \hat{l} \, \tau^{*}, [G' \setminus
\cala_{\calh}]_{\wbis{\rm pw}}, \varepsilon_{F \, / \, \cala_{\calh}}) = 1$. Since the restriction operator
does not affect $\tau$, $l$, and probabilistic transitions, we have that there exists $\sigma' \in \sched((F
\, / \, \cala_{\calh}) \setminus L)$ such that $\proba_{\sigma'}((F \, / \, \cala_{\calh}) \setminus L,
\tau^{*} \, \hat{l} \, \tau^{*}, [(G' \setminus L) \, / \, \cala_{\calh}]_{\calb}, \varepsilon_{(F \, / \,
\cala_{\calh}) \, \setminus \, L}) \linebreak = 1$.

\item If $(G \setminus L) \, / \, \cala_{\calh} \arrow{\tau}{\rm a}(G' \setminus L) \, / \, \cala_{\calh}$
with $G \arrow{\tau}{\rm a} G'$, then $G \setminus \cala_{\calh} \arrow{\tau}{\rm a} G' \setminus
\cala_{\calh}$ as $\tau \notin \cala_{\calh}$. \linebreak From $G \setminus \cala_{\calh} \wbis{\rm pw} F \,
/ \, \cala_{\calh}$ it follows that there exists $\sigma \in \sched(F \, / \, \cala_{\calh})$ such that
$\proba_{\sigma}(F \, / \, \cala_{\calh}, \tau^{*} \, \hat{\tau} \, \tau^{*}, [G' \setminus
\cala_{\calh}]_{\wbis{\rm pw}}, \varepsilon_{F \, / \, \cala_{\calh}}) = 1$. Since the restriction operator
does not affect $\tau$ and probabilistic transitions, we have that there exists $\sigma' \in \sched((F \, /
\, \cala_{\calh}) \setminus L)$ such that $\proba_{\sigma'}((F \, / \, \cala_{\calh}) \setminus L, \tau^{*}
\, \hat{\tau} \, \tau^{*}, [(G' \setminus L) \, / \, \cala_{\calh}]_{\calb}, \varepsilon_{(F \, / \,
\cala_{\calh}) \, \setminus \, L}) \linebreak = 1$.

\item If $(G \setminus L) \, / \, \cala_{\calh} \arrow{\tau}{\rm a} (G' \setminus L) \, / \, \cala_{\calh}$
with $G \arrow{h}{\rm a} G'$ and $h \notin L$, then $G \, / \, \cala_{\calh} \arrow{\tau}{\rm a} G' \, / \,
\cala_{\calh}$ \linebreak as $h \in \cala_{\calh}$ (note that $G \setminus \cala_{\calh}$ cannot perform
$h$). From $G \, / \, \cala_{\calh} \wbis{\rm pw} G \setminus \cala_{\calh}$ -- as $E \in
\mathrm{SBSNNI}_{\wbis{\rm pw}}$ and $G \in \reach(E)$ -- and $G \setminus \cala_{\calh} \wbis{\rm pw} F \,
/ \, \cala_{\calh}$ -- hence $G \, / \, \cala_{\calh} \wbis{\rm pw} F \, / \, \cala_{\calh}$ as $\wbis{\rm
pw}$ is transitive -- it follows that there exists $\sigma \in \sched(F \, / \, \cala_{\calh})$ such that
$\proba_{\sigma}(F \, / \, \cala_{\calh}, \tau^{*} \, \hat{\tau} \, \tau^{*}, [G' \, / \,
\cala_{\calh}]_{\wbis{\rm pw}}, \varepsilon_{F \, / \, \cala_{\calh}}) = 1$ where $[G' \, / \,
\cala_{\calh}]_{\wbis{\rm pw}} = [G' \setminus \cala_{\calh}]_{\wbis{\rm pw}}$ because $G' \, / \,
\cala_{\calh} \wbis{\rm pw} G' \setminus \cala_{\calh}$ due to $E \in \mathrm{SBSNNI}_{\wbis{\rm pw}}$ and
$G' \in \reach(E)$. Since the restriction operator does not affect $\tau$ and probabilistic transitions, we
have that there exists $\sigma' \in \sched((F \, / \, \cala_{\calh}) \setminus L)$ such that
$\proba_{\sigma'}((F \, / \, \cala_{\calh}) \setminus L, \tau^{*} \, \hat{\tau} \, \tau^{*},$ \linebreak
$[(G' \setminus L) \, / \, \cala_{\calh}]_{\calb}, \varepsilon_{(F \, / \, \cala_{\calh}) \, \setminus \,
L}) = 1$.

				\end{itemize}

\noindent
As for probabilities, to avoid trivial cases let $F, G \in \procs_{\rm p}$ and consider an equivalence class
$C \in \procs / \calb$ that involves nondeterministic processes reachable from $E$, specifically $C = \{
(H_{i} \, / \, \cala_{\calh}) \setminus L, (H_{j} \setminus L) \, / \, \cala_{\calh} \mid H_{\blue{k}} \in
\reach(E) \land H_{i} \setminus \cala_{\calh} \wbis{\rm pw} H_{j} \, / \, \cala_{\calh} \}$. Since the
restriction and hiding operators do not affect probabilistic transitions, we have that:
\cws{0}{\hspace*{-0.6cm}\begin{array}{rcl}
\proba((F \, / \, \cala_{\calh}) \setminus L, C) & \!\! = \!\! & \proba(F \setminus \cala_{\calh}, \bar{C})
\\
\proba((G \setminus L) \, / \, \cala_{\calh}, C) & \!\! = \!\! & \proba(G \, / \, \cala_{\calh}, \bar{C}) \\
\end{array}}
where:
\cws{0}{\hspace*{-0.6cm}\begin{array}{rcl}
\bar{C} & \!\! = \!\! & \{ H_{i} \setminus \cala_{\calh} \mid (H_{i} \, / \, \cala_{\calh}) \setminus L \in
C \} \cup \{ H_{j} \, / \, \cala_{\calh} \mid (H_{j} \setminus L) \, / \, \cala_{\calh} \in C \} \\
\end{array}}
Since $F \setminus \cala_{\calh} \wbis{\rm pw} G \, / \, \cala_{\calh}$ and $\bar{C}$ is the union of some
$\wbis{\rm pw}$-equivalence classes, we have that:
\cws{10}{\hspace*{-0.6cm}\begin{array}{rcl}
\proba(F \setminus \cala_{\calh}, \bar{C}) & \!\! = \!\! & \proba(G \, / \, \cala_{\calh}, \bar{C}) \\
\end{array}}


\item Starting from $(F_{1} \pco{L} F_{2}) \setminus \cala_{\calh}$ and $(G_{1} \pco{L} G_{2}) \setminus
\cala_{\calh}$ related by $\calb$, so that $F_{1} \setminus \cala_{\calh} \wbis{\rm pw} G_{1} \setminus
\cala_{\calh}$ and $F_{2} \setminus \cala_{\calh} \wbis{\rm pw} G_{2} \setminus \cala_{\calh}$, there are
five cases for action transitions based on the operational semantic rules in Table~\ref{tab:prob_op_sem_nd}:

				\begin{itemize}

\item If $(F_{1} \pco{L} F_{2}) \setminus \cala_{\calh} \arrow{l}{\rm a} (F'_{1} \pco{L} [1] F_{2})
\setminus \cala_{\calh}$ with $F_{1} \arrow{l}{\rm a} F'_{1}$ and $l \notin L$, then $F_{1} \setminus
\cala_{\calh} \arrow{l}{\rm a}$ \linebreak $F'_{1} \setminus \cala_{\calh}$ as $l \notin \cala_{\calh}$.
From $F_{1} \setminus \cala_{\calh} \wbis{\rm pw} G_{1} \setminus \cala_{\calh}$ it follows that there
exists $\sigma \in \sched(G_{1} \setminus \cala_{\calh})$ such that $\proba_{\sigma}(G_{1} \setminus
\cala_{\calh}, \tau^{*} \, \hat{l} \, \tau^{*}, [F'_{1} \setminus \cala_{\calh}]_{\wbis{\rm pw}},
\varepsilon_{G_{1} \, \setminus \, \cala_{\calh}}) = 1$. Since synchronization does not affect $\tau$ and $l
\notin L$, we have that there exists $\sigma' \in \sched((G_{1} \pco{L} G_{2}) \setminus \cala_{\calh})$
such that $\proba_{\sigma'}((G_{1} \pco{L} G_{2}) \setminus \cala_{\calh}, \tau^{*} \, \hat{l} \, \tau^{*},
[(F'_{1} \pco{L} [1] F_{2}) \setminus \cala_{\calh}]_{\calb}, \varepsilon_{(G_{1} \pco{L} G_{2}) \,
\setminus \, \cala_{\calh}}) = 1$, where the right subprocess alternates between $G_{2}$ and $[1] G_{2}$
thus allowing the probabilistic transitions along $G_{1} \setminus \cala_{\calh} \warrow{}{} \!
\arrow{l}{\rm a} \! \warrow{}{} G'_{1} \setminus \cala_{\calh}$ to synchronize with the only one of~$[1]
G_{2}$.

\item If $(F_{1} \pco{L} F_{2}) \setminus \cala_{\calh} \arrow{l}{\rm a} ([1] F_{1} \pco{L} F'_{2})
\setminus \cala_{\calh}$ with $F_{2} \arrow{l}{\rm a} F'_{2}$ and $l \notin L$, then the proof is similar to
the one of the previous case.

\item If $(F_{1} \pco{L} F_{2}) \setminus \cala_{\calh} \arrow{l}{\rm a} (F'_{1} \pco{L} F'_{2}) \setminus
\cala_{\calh}$ with $F_{i} \arrow{l}{\rm a} F'_{i}$ for $i \in \{ 1, 2 \}$ and $l \in L$, then \linebreak
$F_{i} \setminus \cala_{\calh} \arrow{l}{\rm a} F'_{i} \setminus \cala_{\calh}$ as $l \notin \cala_{\calh}$.
From $F_{i} \setminus \cala_{\calh} \wbis{\rm pw} G_{i} \setminus \cala_{\calh}$ it follows that there
exists $\sigma_{i} \in \sched(G_{i} \setminus \cala_{\calh})$ such that $\proba_{\sigma_{i}}(G_{i} \setminus
\cala_{\calh}, \tau^{*} \, \hat{l} \, \tau^{*}, [F'_{i} \setminus \cala_{\calh}]_{\wbis{\rm pw}},
\varepsilon_{G_{i} \, \setminus \, \cala_{\calh}}) = 1$. \linebreak Since synchronization does not affect
$\tau$ and $l \in L$, we have that there exists $\sigma' \in \sched((G_{1} \pco{L} G_{2}) \setminus
\cala_{\calh})$ such that $\proba_{\sigma'}((G_{1} \pco{L} G_{2}) \setminus \cala_{\calh}, \tau^{*} \,
\hat{l} \, \tau^{*}, [(F'_{1} \pco{L} F'_{2}) \setminus \cala_{\calh}]_{\calb}, \varepsilon_{(G_{1} \pco{L}
G_{2}) \, \setminus \, \cala_{\calh}}) = 1$, where subprocess $i$ alternates between $G_{i}$ and $[1] G_{i}$
before performing $l$ or between $G'_{i}$ and $[1] G'_{i}$ after performing $l$ thus allowing the
probabilistic transitions along $G_{j} \setminus \cala_{\calh} \warrow{}{} \! \arrow{l}{\rm a} \!
\warrow{}{} G'_{j} \setminus \cala_{\calh}$ for $j \neq i$ to synchronize with the only one of $[1] G_{i}$
before performing $a$ or the only one of $[1] G'_{i}$ after performing~$a$.

\item If $(F_{1} \pco{L} F_{2}) \setminus \cala_{\calh} \arrow{\tau}{\rm a} (F'_{1} \pco{L} [1] F_{2})
\setminus \cala_{\calh}$ with $F_{1} \arrow{\tau}{\rm a} F'_{1}$, then $F_{1} \setminus \cala_{\calh}
\arrow{\tau}{\rm a} F'_{1} \setminus \cala_{\calh}$ as $\tau \notin \cala_{\calh}$. From $F_{1} \setminus
\cala_{\calh} \wbis{\rm pw} G_{1} \setminus \cala_{\calh}$ it follows that there exists $\sigma \in
\sched(G_{1} \setminus \cala_{\calh})$ such that $\proba_{\sigma}(G_{1} \setminus \cala_{\calh}, \tau^{*} \,
\hat{\tau} \, \tau^{*}, [F'_{1} \setminus \cala_{\calh}]_{\wbis{\rm pw}}, \varepsilon_{G_{1} \, \setminus \,
\cala_{\calh}}) = 1$. Since synchronization does not affect $\tau$, we have that there exists $\sigma' \in
\sched((G_{1} \pco{L} G_{2}) \setminus \cala_{\calh})$ such that $\proba_{\sigma'}((G_{1} \pco{L} G_{2})
\setminus \cala_{\calh}, \tau^{*} \, \hat{\tau} \, \tau^{*}, [(F'_{1} \pco{L} [1] F_{2}) \setminus
\cala_{\calh}]_{\calb}, \varepsilon_{(G_{1} \pco{L} G_{2}) \, \setminus \, \cala_{\calh}}) = 1$, where the
right subprocess alternates between $G_{2}$ and $[1] G_{2}$ thus allowing the probabilistic transitions
along $G_{1} \setminus \cala_{\calh} \warrow{}{} G'_{1} \setminus \cala_{\calh}$ to synchronize with the
only one of $[1] G_{2}$.

\item If $(F_{1} \pco{L} F_{2}) \setminus \cala_{\calh} \arrow{\tau}{\rm a} ([1] F_{1} \pco{L} F'_{2})
\setminus \cala_{\calh}$ with $F_{2} \arrow{\tau}{\rm a} F'_{2}$, then the proof is similar to the one of
the previous case.

				\end{itemize}

\noindent
As for probabilities, to avoid trivial cases let $F_{1}, F_{2}, G_{1}, G_{2} \in \procs_{\rm p}$ and
consider an equivalence class $C \in \procs / \calb$ that involves nondeterministic processes reachable from
$E_{1} \pco{L} E_{2}$, specifically $C = \{ (H_{1, i} \pco{L} H_{2, i}) \setminus \cala_{\calh} \mid H_{k,
\blue{\iota}} \in \reach(E_{k}) \land H_{1, \blue{\iota}} \pco{L} H_{2, \blue{\iota}} \in \reach(E_{1}
\pco{L} E_{2}) \land H_{k, i} \setminus \cala_{\calh} \wbis{\rm pw} H_{k, j} \setminus \cala_{\calh} \}$.
Since the restriction operator does not affect probabilistic transitions, we have that:
\cws{0}{\hspace*{-0.6cm}\begin{array}{rcl}
\proba((F_{1} \pco{L} F_{2}) \setminus \cala_{\calh}, C) & \!\! = \!\! & \proba((F_{1} \setminus
\cala_{\calh}) \pco{L} (F_{2} \setminus \cala_{\calh}), C) \\
\proba((G_{1} \pco{L} G_{2}) \setminus \cala_{\calh}, C) & \!\! = \!\! & \proba((G_{1} \setminus
\cala_{\calh}) \pco{L} (G_{2} \setminus \cala_{\calh}), C) \\
\end{array}}
and hence by virtue of Lemma~\ref{lem:prob_prod_par_comp} we have that:
\cws{0}{\hspace*{-0.6cm}\begin{array}{rcl}
\proba((F_{1} \setminus \cala_{\calh}) \pco{L} (F_{2} \setminus \cala_{\calh}), C) & \!\! = \!\! &
\proba(F_{1} \setminus \cala_{\calh}, C_{1}) \cdot \proba(F_{2} \setminus \cala_{\calh}, C_{2}) \\
\proba((G_{1} \setminus \cala_{\calh}) \pco{L} (G_{2} \setminus \cala_{\calh}), C) & \!\! = \!\! &
\proba(G_{1} \setminus \cala_{\calh}, C_{1}) \cdot \proba(G_{2} \setminus \cala_{\calh}, C_{2}) \\
\end{array}}
where:
\cws{0}{\hspace*{-0.6cm}\begin{array}{rcl}
C_{1} & \!\! = \!\! & \{ H_{1, \blue{\iota}} \setminus \cala_{\calh} \mid (H_{1, \blue{\iota}} \pco{L} H_{2,
\blue{\iota}}) \setminus \cala_{\calh} \in C \} \\
C_{2} & \!\! = \!\! & \{ H_{2, \blue{\iota}} \setminus \cala_{\calh} \mid (H_{1, \blue{\iota}} \pco{L} H_{2,
\blue{\iota}}) \setminus \cala_{\calh} \in C \} \\
\end{array}}
Since $F_{k} \setminus \cala_{\calh} \wbis{\rm pw} G_{k} \setminus \cala_{\calh}$ and $C_{k}$ is the union
of some $\wbis{\rm pw}$-equivalence classes for $k \in \{ 1, 2 \}$, we have that:
\cws{6}{\hspace*{-0.6cm}\begin{array}{rcl}
\proba(F_{1} \setminus \cala_{\calh}, C_{1}) & \!\! = \!\! & \proba(G_{1} \setminus \cala_{\calh}, C_{1}) \\
\proba(F_{2} \setminus \cala_{\calh}, C_{2}) & \!\! = \!\! & \proba(G_{2} \setminus \cala_{\calh}, C_{2}) \\
\end{array}}

		\end{enumerate}


\noindent
\blue{Next we} prove the three results for the $\wbis{\rm pb}$-based properties. Let $\calb$ be an
equivalence relation containing all the pairs of processes mentioned at beginning of the considered result
that have to be shown to be $\wbis{\rm pb}$-equivalent under the constraints mentioned at the end of the
result itself:

		\begin{enumerate}


\item Starting from $(F_{1} \pco{L} F_{2}) \setminus \cala_{\calh}$ and $(G_{1} \pco{L} G_{2}) \, / \,
\cala_{\calh}$ related by $\calb$, so that $F_{1} \setminus \cala_{\calh} \wbis{\rm pb} G_{1} \, / \,
\cala_{\calh}$ and $F_{2} \setminus \cala_{\calh} \wbis{\rm pb} G_{2} \, / \, \cala_{\calh}$, there are
twelve cases for action transitions based on the operational semantic rules in
Table~\ref{tab:prob_op_sem_nd}. In the first five cases, it is $(F_{1} \pco{L} F_{2}) \setminus
\cala_{\calh}$ to move first:

				\begin{itemize}

\item If $(F_{1} \pco{L} F_{2}) \setminus \cala_{\calh} \arrow{l}{\rm a} (F'_{1} \pco{L} [1] F_{2})
\setminus \cala_{\calh}$ with $F_{1} \arrow{l}{\rm a} F'_{1}$ and $l \notin L$, then $F_{1} \setminus
\cala_{\calh} \linebreak \arrow{l}{\rm a} F'_{1} \setminus \cala_{\calh}$ as $l \notin \cala_{\calh}$. From
$F_{1} \setminus \cala_{\calh} \wbis{\rm pb} G_{1} \, / \, \cala_{\calh}$ it follows that there exists
$G_{1} \, / \, \cala_{\calh} \warrow{}{} \bar{G}_{1} \, / \, \cala_{\calh} \arrow{l}{\rm a} G'_{1} \, / \,
\cala_{\calh}$ such that $F_{1} \setminus \cala_{\calh} \wbis{\rm pb} \bar{G}_{1} \, / \, \cala_{\calh}$ and
$F'_{1} \setminus \cala_{\calh} \linebreak \wbis{\rm pb} G'_{1} \, / \, \cala_{\calh}$. Since
synchronization does not affect $\tau$ and $l \notin L$, we have that $(G_{1} \pco{L} G_{2}) \, / \,
\cala_{\calh} \warrow{}{} (\bar{G}_{1} \pco{L} G_{2}) \, / \, \cala_{\calh} {\arrow{l}{\rm a}} (G'_{1}
\pco{L} [1] G_{2}) \, / \, \cala_{\calh}$ with $((F_{1} \pco{L} F_{2}) \setminus \cala_{\calh}, \linebreak
(\bar{G}_{1} \pco{L} G_{2}) \, / \, \cala_{\calh}) \in \calb$ and $((F'_{1} \pco{L} [1] F_{2}) \setminus
\cala_{\calh}, (G'_{1} \pco{L} [1] G_{2}) \, / \, \cala_{\calh}) \in \calb$, where the right subprocess
alternates between $G_{2}$ and $[1] G_{2}$ thus allowing the probabilistic transitions along $G_{1} \, / \,
\cala_{\calh} \warrow{}{} \bar{G}_{1} \, / \, \cala_{\calh}$ to synchronize with the only one of $[1]
G_{2}$.

\item If $(F_{1} \pco{L} F_{2}) \setminus \cala_{\calh} \arrow{l}{\rm a} ([1] F_{1} \pco{L} F'_{2})
\setminus \cala_{\calh}$ with $F_{2} \arrow{l}{\rm a} F'_{2}$ and $l \notin L$, then the proof is similar to
the one of the previous case.

\item If $(F_{1} \pco{L} F_{2}) \setminus \cala_{\calh} \arrow{l}{\rm a} (F'_{1} \pco{L} F'_{2}) \setminus
\cala_{\calh}$ with $F_{i} \arrow{l}{\rm a} F'_{i}$ for $i \in \{ 1, 2 \}$ and $l \in L$, then $F_{i}
\setminus \cala_{\calh} \arrow{l}{\rm a} F'_{i} \setminus \cala_{\calh}$ as $l \notin \cala_{\calh}$. From
$F_{i} \setminus \cala_{\calh} \wbis{\rm pb} G_{i} \, / \, \cala_{\calh}$ it follows that there exists
$G_{i} \, / \, \cala_{\calh} \warrow{}{} \bar{G}_{i} \, / \, \cala_{\calh} \arrow{l}{\rm a} G'_{i} \, / \,
\cala_{\calh}$ such that $F_{i} \setminus \cala_{\calh} \wbis{\rm pb} \bar{G}_{i} \, / \, \cala_{\calh}$ and
$F'_{i} \setminus \cala_{\calh} \wbis{\rm pb} G'_{i} \, / \, \cala_{\calh}$. Since synchronization does not
affect $\tau$ and $l \in L$, we have that $(G_{1} \pco{L} G_{2}) \, / \, \cala_{\calh} \warrow{}{}
(\bar{G}_{1} \pco{L} \bar{G}_{2}) \, / \, \cala_{\calh} \arrow{l}{\rm a} (G'_{1} \pco{L} G'_{2}) \, / \,
\cala_{\calh}$ with $((F_{1} \pco{L} F_{2}) \setminus \cala_{\calh}, \linebreak (\bar{G}_{1} \pco{L}
\bar{G}_{2}) \, / \, \cala_{\calh}) \in \calb$ and $((F'_{1} \pco{L} F'_{2}) \setminus \cala_{\calh},
(G'_{1} \pco{L} G'_{2}) \, / \, \cala_{\calh}) \in \calb$, where subprocess $i$ alternates between $G_{i}$
and $[1] G_{i}$ thus allowing the probabilistic transitions along $G_{j} \, / \, \cala_{\calh} \warrow{}{}
\bar{G}_{j} \, / \, \cala_{\calh}$ for $j \neq i$ to synchronize with the only one of $[1] G_{i}$.

\item If $(F_{1} \pco{L} F_{2}) \setminus \cala_{\calh} \arrow{\tau}{\rm a} (F'_{1} \pco{L} [1] F_{2})
\setminus \cala_{\calh}$ with $F_{1} \arrow{\tau}{\rm a} F'_{1}$, then $F_{1} \setminus \cala_{\calh}
{\arrow{\tau}{\rm a}} F'_{1} \setminus \cala_{\calh}$ as $\tau \notin \cala_{\calh}$. From $F_{1} \setminus
\cala_{\calh} \wbis{\rm pb} G_{1} \, / \, \cala_{\calh}$ it follows that either $F'_{1} \setminus
\cala_{\calh} \wbis{\rm pb} G_{1} \, / \, \cala_{\calh}$, or there exists $G_{1} \, / \, \cala_{\calh}
\warrow{}{} \bar{G}_{1} \, / \, \cala_{\calh} \arrow{\tau}{\rm a} G'_{1} \, / \, \cala_{\calh}$ such that
$F_{1} \setminus \cala_{\calh} \wbis{\rm pb} \bar{G}_{1} \, / \, \cala_{\calh}$ and $F'_{1} \setminus
\cala_{\calh} \linebreak \wbis{\rm pb} G'_{1} \, / \, \cala_{\calh}$. Since synchronization does not affect
$\tau$, in the former subcase $(G_{1} \pco{L} G_{2}) \linebreak \, / \, \cala_{\calh}$ is allowed to stay
idle with $((F'_{1} \pco{L} [1] F_{2}) \setminus \cala_{\calh}, (G_{1} \pco{L} G_{2}) \, / \, \cala_{\calh})
\in \calb$, while in the latter subcase $(G_{1} \pco{L} G_{2}) \, / \, \cala_{\calh} \warrow{}{}
(\bar{G}_{1} \pco{L} G_{2}) \, / \, \cala_{\calh} \arrow{\tau}{\rm a} (G'_{1} \pco{L} [1] G_{2}) \, / \,
\cala_{\calh}$ with $((F_{1} \pco{L} F_{2}) \setminus \cala_{\calh}, (\bar{G}_{1} \pco{L} G_{2}) \, / \,
\cala_{\calh}) \in \calb$ and $((F'_{1} \pco{L} [1] F_{2}) \setminus \cala_{\calh}, (G'_{1} \pco{L} [1]
G_{2}) \, / \, \cala_{\calh}) \linebreak \in \calb$, where the right subprocess alternates between $G_{2}$
and $[1] G_{2}$ thus allowing the probabilistic transitions along $G_{1} \, / \, \cala_{\calh} \warrow{}{}
\bar{G}_{1} \, / \, \cala_{\calh}$ to synchronize with the only one of $[1] G_{2}$.

\item If $(F_{1} \pco{L} F_{2}) \setminus \cala_{\calh} \arrow{\tau}{\rm a} ([1] F_{1} \pco{L} F'_{2})
\setminus \cala_{\calh}$ with $F_{2} \arrow{\tau}{\rm a} F'_{2}$, then the proof is similar to the one of
the previous case. 

				\end{itemize}

\noindent
In the other seven cases, instead, it is $(G_{1} \pco{L} G_{2}) \, / \, \cala_{\calh}$ to move first:

				\begin{itemize}

\item If $(G_{1} \pco{L} G_{2}) \, / \, \cala_{\calh} \arrow{l}{\rm a} (G'_{1} \pco{L} [1] G_{2}) \, / \,
\cala_{\calh}$ with $G_{1} \arrow{l}{\rm a} G'_{1}$ and $l \notin L$, then $G_{1} \, / \, \cala_{\calh}
\linebreak \arrow{l}{\rm a} G'_{1} \, / \, \cala_{\calh}$ as $l \notin \cala_{\calh}$. From $G_{1} \, / \,
\cala_{\calh} \wbis{\rm pb} F_{1} \setminus \cala_{\calh}$ it follows that there exists $F_{1} \setminus
\cala_{\calh}$ \linebreak $\warrow{}{} \bar{F}_{1} \setminus \cala_{\calh} \arrow{l}{\rm a} F'_{1} \setminus
\cala_{\calh}$ such that $G_{1} \, / \, \cala_{\calh} \wbis{\rm pb} \bar{F}_{1} \setminus \cala_{\calh}$ and
$G'_{1} \, / \, \cala_{\calh} \wbis{\rm pb} F'_{1} \setminus \cala_{\calh}$. Since synchronization does not
affect $\tau$ and $l \notin L$, we have that $(F_{1} \pco{L} F_{2}) \setminus \cala_{\calh} \linebreak
\warrow{}{} (\bar{F}_{1} \pco{L} F_{2}) \setminus \cala_{\calh} {\arrow{l}{\rm a}} (F'_{1} \pco{L} [1]
F_{2}) \setminus \cala_{\calh}$ with $((G_{1} \pco{L} G_{2}) / \cala_{\calh}, (\bar{F}_{1} \pco{L} F_{2})
\setminus \cala_{\calh}) \in \calb$ and $((G'_{1} \pco{L} [1] G_{2}) / \cala_{\calh}, (F'_{1} \pco{L} [1]
F_{2}) \setminus \cala_{\calh}) \in \calb$, where the right subprocess alternates between $F_{2}$ and $[1]
F_{2}$ thus allowing the probabilistic transitions along $F_{1} \setminus \cala_{\calh} \warrow{}{}$
\linebreak $\bar{F}_{1} \setminus \cala_{\calh}$ to synchronize with the only one of $[1] F_{2}$.

\item If $(G_{1} \pco{L} G_{2}) \, / \, \cala_{\calh} \arrow{l}{\rm a} ([1] G_{1} \pco{L} G'_{2}) \, / \,
\cala_{\calh}$ with $G_{2} \arrow{l}{\rm a} G'_{2}$ and $l \notin L$, then the proof is similar to the one
of the previous case. 

\item If $(G_{1} \pco{L} G_{2}) \, / \, \cala_{\calh} \arrow{l}{\rm a} (G'_{1} \pco{L} G'_{2}) \, / \,
\cala_{\calh}$ with $G_{i} \arrow{l}{\rm a} G'_{i}$ for $i \in \{ 1, 2 \}$ and $l \in L$, then $G_{i} \, /
\, \cala_{\calh} \arrow{l}{\rm a} G'_{i} \, / \, \cala_{\calh}$ as $l \notin \cala_{\calh}$.  From $G_{i} \,
/ \, \cala_{\calh} \wbis{\rm pb} F_{i} \setminus \cala_{\calh}$ it follows that there exists $F_{i}
\setminus \cala_{\calh} \warrow{}{} \bar{F}_{i} \setminus \cala_{\calh} \arrow{l}{\rm a} F'_{i} \setminus
\cala_{\calh}$ such that $G_{i} / \cala_{\calh} \wbis{\rm pb} \bar{F}_{i} \setminus \cala_{\calh}$ and
$G'_{i} / \cala_{\calh} \wbis{\rm pb} F'_{i} \setminus \cala_{\calh}$. \linebreak Since synchronization does
not affect $\tau$ and $l \in L$, we have that $(F_{1} \pco{L} F_{2}) \setminus \cala_{\calh} \linebreak
\warrow{}{} (\bar{F}_{1} \pco{L} \bar{F}_{2}) \setminus \cala_{\calh} {\arrow{l}{\rm a}} (F'_{1} \pco{L}
F'_{2}) \setminus \cala_{\calh}$ with $((G_{1} \pco{L} G_{2}) / \cala_{\calh}, (\bar{F}_{1} \pco{L}
\bar{F}_{2}) \setminus \cala_{\calh}) \in \calb$ and $((G'_{1} \pco{L} G'_{2}) \, / \, \cala_{\calh},
(F'_{1} \pco{L} F'_{2}) \setminus \cala_{\calh}) \in \calb$, where subprocess~$i$ alternates between $F_{i}$
and $[1] F_{i}$ thus allowing the probabilistic transitions along $F_{j} \setminus \cala_{\calh} \warrow{}{}
\bar{F}_{j} \setminus \cala_{\calh}$ for $j \neq i$ to synchronize with the only one of $[1] F_{i}$.

\item If $(G_{1} \pco{L} G_{2}) \, / \, \cala_{\calh} \arrow{\tau}{\rm a} (G'_{1} \pco{L} [1] G_{2}) \, / \,
\cala_{\calh}$ with $G_{1} \arrow{\tau}{\rm a} G'_{1}$, then $G_{1} \, / \cala_{\calh} \arrow{\tau}{\rm a}
G'_{1} \, / \cala_{\calh}$ as $\tau \notin \cala_{\calh}$. From $G_{1} \, / \, \cala_{\calh} \wbis{\rm pb}
F_{1} \setminus \cala_{\calh}$ it follows that either $G'_{1} \, / \, \cala_{\calh} \wbis{\rm pb} F_{1}
\setminus \cala_{\calh}$, or there exists $F_{1} \setminus \cala_{\calh} \warrow{}{} \bar{F}_{1} \setminus
\cala_{\calh} \arrow{\tau}{\rm a} F'_{1} \setminus \cala_{\calh}$ such that $G_{1} \, / \, \cala_{\calh}
\wbis{\rm pb} \bar{F}_{1} \setminus \cala_{\calh}$ and $G'_{1} \, / \, \cala_{\calh} \wbis{\rm pb} F'_{1}
\setminus \cala_{\calh}$. Since synchronization does not affect $\tau$, in the former subcase $(F_{1}
\pco{L} F_{2}) \setminus \cala_{\calh}$ is allowed to stay idle with $((G'_{1} \pco{L} [1] G_{2}) /
\cala_{\calh}, (F_{1} \pco{L} F_{2}) \setminus \cala_{\calh}) \in \calb$, while in the latter subcase
$(F_{1} \pco{L} F_{2}) \setminus \cala_{\calh} \warrow{}{} (\bar{F}_{1} \pco{L} F_{2}) \setminus
\cala_{\calh} \arrow{\tau}{\rm a} (F'_{1} \pco{L} [1] F_{2}) \setminus \cala_{\calh}$ with $((G_{1} \pco{L}
G_{2}) / \cala_{\calh}, (\bar{F}_{1} \pco{L} F_{2}) \setminus \cala_{\calh}) \in \calb$ and $((G'_{1}
\pco{L} [1] G_{2}) / \cala_{\calh}, (F'_{1} \pco{L} [1] F_{2}) \setminus \cala_{\calh}) \linebreak \in
\calb$, where the right subprocess alternates between $F_{2}$ and $[1] F_{2}$ thus allowing the
probabilistic transitions along $F_{1} \setminus \cala_{\calh} \warrow{}{} \bar{F}_{1} \setminus
\cala_{\calh}$ to synchronize with the only one of $[1] F_{2}$.

\item If $(G_{1} \pco{L} G_{2}) \, / \, \cala_{\calh} \arrow{\tau}{\rm a} ([1] G_{1} \pco{L} G'_{2}) \, / \,
\cala_{\calh}$ with $G_{2} \arrow{\tau}{\rm a} G'_{2}$, then the proof is similar to the one of the previous
case.

\item If $(G_{1} \pco{L} G_{2}) \, / \, \cala_{\calh} \arrow{\tau}{\rm a} (G'_{1} \pco{L} [1] G_{2}) \, / \,
\cala_{\calh}$ with $G_{1} \arrow{h}{\rm a} G'_{1}$ and $h \notin L$, then $G_{1} \, / \, \cala_{\calh}
\linebreak \arrow{\tau}{\rm a} G'_{1} \, / \, \cala_{\calh}$ as $h \in \cala_{\calh}$. The rest of the proof
is like the one of the fourth case.

\item If $(G_{1} \pco{L} G_{2}) \, / \, \cala_{\calh} \arrow{\tau}{\rm a} ([1] G_{1} \pco{L} G'_{2}) \, / \,
\cala_{\calh}$ with $G_{2} \arrow{h}{\rm a} G'_{2}$ and $h \notin L$, then the proof is similar to the one
of the previous case.

				\end{itemize}

\noindent
As for probabilities, to avoid trivial cases let $F_{1}, F_{2}, G_{1}, G_{2} \in \procs_{\rm p}$ and
consider an equivalence class $C \in \procs / \calb$ that involves nondeterministic processes reachable from
$E_{1} \pco{L} E_{2}$, specifically $C = \{ (H_{1, i} \pco{L} H_{2, i}) \setminus \cala_{\calh}, (H_{1, j}
\pco{L} H_{2, j}) \, / \, \cala_{\calh} \mid H_{k, \blue{\iota}} \in \reach(E_{k}) \land H_{1, \blue{\iota}}
\pco{L} H_{2, \blue{\iota}} \in \reach(E_{1} \pco{L} E_{2}) \land H_{k, i} \setminus \cala_{\calh} \wbis{\rm
pw} H_{k, j} \, / \, \cala_{\calh} \}$. Since the restriction and hiding operators do not affect
probabilistic transitions, we have that:
\cws{0}{\hspace*{-0.6cm}\begin{array}{rcl}
\proba((F_{1} \pco{L} F_{2}) \setminus \cala_{\calh}, C) & \!\! = \!\! & \proba((F_{1} \setminus
\cala_{\calh}) \pco{L} (F_{2} \setminus \cala_{\calh}), C) \\
\proba((G_{1} \pco{L} G_{2}) \, / \, \cala_{\calh}, C) & \!\! = \!\! & \proba((G_{1} \, / \, \cala_{\calh})
\pco{L} (G_{2} \, / \, \cala_{\calh}), C) \\
\end{array}}
and hence by virtue of Lemma~\ref{lem:prob_prod_par_comp}:
\cws{0}{\hspace*{-0.6cm}\begin{array}{rcl}
\proba((F_{1} \setminus \cala_{\calh}) \pco{L} (F_{2} \setminus \cala_{\calh}), C) & \!\! = \!\! &
\proba(F_{1} \setminus \cala_{\calh}, C_{1}) \cdot \proba(F_{2} \setminus \cala_{\calh}, C_{2}) \\
\proba((G_{1} \, / \, \cala_{\calh}) \pco{L} (G_{2} \, / \, \cala_{\calh}), C) & \!\! = \!\! & \proba(G_{1}
\, / \, \cala_{\calh}, C_{1}) \cdot \proba(G_{2} \, / \, \cala_{\calh}, C_{2}) \\
\end{array}}
where:
\cws{0}{\hspace*{-0.6cm}\begin{array}{rcl}
C_{1} & \!\! = \!\! & \{ H_{1, \blue{\iota}} \setminus \cala_{\calh} \mid (H_{1, \blue{\iota}} \pco{L} H_{2,
\blue{\iota}}) \setminus \cala_{\calh} \in C \} \cup \{ H_{1, \blue{\iota}} \, / \, \cala_{\calh} \mid
(H_{1, \blue{\iota}} \pco{L} H_{2, \blue{\iota}}) \, / \, \cala_{\calh} \in C \} \\
C_{2} & \!\! = \!\! & \{ H_{2, \blue{\iota}} \setminus \cala_{\calh} \mid (H_{1, \blue{\iota}} \pco{L} H_{2,
\blue{\iota}}) \setminus \cala_{\calh} \in C \} \cup \{ H_{2, \blue{\iota}} \, / \, \cala_{\calh} \mid
(H_{1, \blue{\iota}} \pco{L} H_{2, \blue{\iota}}) \, / \, \cala_{\calh} \in C \} \\
\end{array}}
Since $F_{k} \setminus \cala_{\calh} \wbis{\rm pb} G_{k} \, / \, \cala_{\calh}$ and $C_{k}$ is the union of
some $\wbis{\rm pb}$-equivalence classes for $k \in \{ 1, 2 \}$, we have that:
\cws{6}{\hspace*{-0.6cm}\begin{array}{rcl}
\proba(F_{1} \setminus \cala_{\calh}, C_{1}) & \!\! = \!\! & \proba(G_{1} \, / \, \cala_{\calh}, C_{1}) \\
\proba(F_{2} \setminus \cala_{\calh}, C_{2}) & \!\! = \!\! & \proba(G_{2} \, / \, \cala_{\calh}, C_{2}) \\
\end{array}}


\item Starting from $(F \, / \, \cala_{\calh}) \setminus L$ and $(G \setminus L) \, / \, \cala_{\calh}$
related by $\calb$, so that $F \, / \, \cala_{\calh} \wbis{\rm pb} G \setminus \cala_{\calh}$, there are six
cases for action transitions based on the operational semantic rules in Table~\ref{tab:prob_op_sem_nd}. In
the first three cases, it is $(F \, / \, \cala_{\calh}) \setminus L$ to move first:

				\begin{itemize}

\item If $(F \, / \, \cala_{\calh}) \setminus L \arrow{l}{\rm a} (F' \, / \, \cala_{\calh}) \setminus L$
with $F \arrow{l}{\rm a} F'$ and $l \notin L$, then $F \, / \, \cala_{\calh} \arrow{l}{\rm a} F' \, / \,
\cala_{\calh}$ as $l \notin \cala_{\calh}$. From $F \, / \, \cala_{\calh} \wbis{\rm pb} G \setminus
\cala_{\calh}$ it follows that there exists $G \setminus \cala_{\calh} \warrow{}{} \bar{G} \setminus
\cala_{\calh} \linebreak \arrow{l}{\rm a} G' \setminus \cala_{\calh}$ such that $F \, / \, \cala_{\calh}
\wbis{\rm pb} \bar{G} \setminus \cala_{\calh}$ and $F' \, / \, \cala_{\calh} \wbis{\rm pb} G' \setminus
\cala_{\calh}$. Since the restriction and hiding operators do not affect $\tau$, $l$, and probabilistic
transitions, we have that $(G \setminus L) \, / \, \cala_{\calh} \warrow{}{} (\bar{G} \setminus L) \, / \,
\cala_{\calh} \arrow{l}{\rm a} (G' \setminus L) \, / \, \cala_{\calh}$ with $((F \, / \, \cala_{\calh})
\setminus L, \linebreak (\bar{G} \setminus L) \, / \, \cala_{\calh}) \in \calb$ and $((F' \, / \,
\cala_{\calh}) \setminus L, (G' \setminus L) \, / \, \cala_{\calh}) \in \calb$.

\item If $(F \, / \, \cala_{\calh}) \setminus L \arrow{\tau}{\rm a} (F' \, / \, \cala_{\calh}) \setminus L$
with $F {\arrow{\tau}{\rm a}} F'$, then $F \, / \, \cala_{\calh} \arrow{\tau}{\rm a} F' \, / \,
\cala_{\calh}$ as $\tau \notin \cala_{\calh}$. \linebreak From $F \, / \, \cala_{\calh} \wbis{\rm pb} G
\setminus \cala_{\calh}$ it follows that either $F' \, / \, \cala_{\calh} \wbis{\rm pb} G \setminus
\cala_{\calh}$, or there exists $G \setminus \cala_{\calh} \warrow{}{} \bar{G} \setminus \cala_{\calh}
\arrow{\tau}{\rm a} G' \setminus \cala_{\calh}$ such that $F \, / \, \cala_{\calh} \wbis{\rm pb} \bar{G}
\setminus \cala_{\calh}$ and $F' \, / \, \cala_{\calh} \wbis{\rm pb} G' \setminus \cala_{\calh}$. Since the
restriction and hiding operators do not affect $\tau$ and probabilistic transitions, in the former subcase
$(G \setminus L) \, / \, \cala_{\calh}$ is allowed to stay idle with $((F' \, / \, \cala_{\calh}) \setminus
L, \linebreak (G \setminus L) \, / \, \cala_{\calh}) \in \calb$, while in the latter subcase $(G \setminus
L) \, / \, \cala_{\calh} \warrow{}{} (\bar{G} \setminus L) \, / \, \cala_{\calh} \arrow{\tau}{\rm a}$
\linebreak $(G' \setminus L) \, / \, \cala_{\calh}$ with $((F \, / \, \cala_{\calh}) \setminus L, (\bar{G}
\setminus L) \, / \, \cala_{\calh}) \in \calb$ and $((F' \, / \, \cala_{\calh}) \setminus L, (G' \setminus
L) \, / \, \cala_{\calh}) \in \calb$.

\item If $(F \, / \, \cala_{\calh}) \setminus L \arrow{\tau}{\rm a} (F' \, / \, \cala_{\calh}) \setminus L$
with $F \arrow{h}{\rm a} F'$, then $F \, / \, \cala_{\calh} {\arrow{\tau}{\rm a}} F' \, / \, \cala_{\calh}$
as $h \in \cala_{\calh}$. The rest of the proof is like the one of the previous case.

				\end{itemize}

\noindent
In the other three cases, instead, it is $(G \setminus L) \, / \, \cala_{\calh}$ to move first:

				\begin{itemize}

\item If $(G \setminus L) \, / \, \cala_{\calh} \arrow{l}{\rm a} (G' \setminus L) \, / \, \cala_{\calh}$
with $G \arrow{l}{\rm a} G'$ and $l \notin L$, then $G \setminus \cala_{\calh} \arrow{l}{\rm a} G' \setminus
\cala_{\calh}$ as $l \notin \cala_{\calh}$. From $G \setminus \cala_{\calh} \wbis{\rm pb} F \, / \,
\cala_{\calh}$ it follows that there exists $F \, / \, \cala_{\calh} \warrow{}{} \bar{F} \, / \,
\cala_{\calh} \arrow{l}{\rm a}$ \linebreak $F' \, / \, \cala_{\calh}$ such that $G \setminus \cala_{\calh}
\wbis{\rm pb} \bar{F} \, / \, \cala_{\calh}$ and $G' \setminus \cala_{\calh} \wbis{\rm pb} F' \, / \,
\cala_{\calh}$. Since the restriction operator does not affect $\tau$, $l$, and probabilistic transitions,
we have that $(F \, / \, \cala_{\calh}) \setminus L \linebreak \warrow{}{} (\bar{F} \, / \, \cala_{\calh})
\setminus L \arrow{l}{\rm a} (F' \, / \, \cala_{\calh}) \setminus L$ with $((G \setminus L) \, / \,
\cala_{\calh}, (\bar{F} \, / \, \cala_{\calh}) \setminus L) \in \calb$ and \linebreak $((G' \setminus L) \,
/ \, \cala_{\calh}, (F' \, / \, \cala_{\calh}) \setminus L) \in \calb$.

\item If $(G \setminus L) \, / \, \cala_{\calh} \arrow{\tau}{\rm a} (G' \setminus L) \, / \, \cala_{\calh}$
with $G \arrow{\tau}{\rm a} G'$, then $G \setminus \cala_{\calh} \arrow{\tau}{\rm a} G' \setminus
\cala_{\calh}$ as $\tau \notin \cala_{\calh}$. From $G \setminus \cala_{\calh} \wbis{\rm pb} F \, / \,
\cala_{\calh}$ it follows that either $G' \setminus \cala_{\calh} \wbis{\rm pb} F \, / \, \cala_{\calh}$, or
there exists $F \, / \, \cala_{\calh} {\warrow{}{}} \bar{F} \, / \, \cala_{\calh} \arrow{\tau}{\rm a} F' \,
/ \, \cala_{\calh}$ such that $G \setminus \cala_{\calh} \wbis{\rm pb} \bar{F} \, / \, \cala_{\calh}$ and
$G' \setminus \cala_{\calh} \linebreak \wbis{\rm pb} F' \, / \, \cala_{\calh}$. Since the restriction
operator does not affect $\tau$ and probabilistic transitions, in the former subcase $(F \, / \,
\cala_{\calh}) \setminus L$ is allowed to stay idle with $((G' \setminus L) \, / \, \cala_{\calh},
\linebreak (F \, / \, \cala_{\calh}) \setminus L) \in \calb$, while in the latter subcase $(F \, / \,
\cala_{\calh}) \setminus L \warrow{}{} (\bar{F} \, / \, \cala_{\calh}) \setminus L \arrow{\tau}{\rm a}$
\linebreak $(F' \, / \, \cala_{\calh}) \setminus L$ with $((G \setminus L) \, / \, \cala_{\calh}, (\bar{F}
\, / \, \cala_{\calh}) \setminus L) \in \calb$ and $((G' \setminus L) \, / \, \cala_{\calh}, (F' \, / \,
\cala_{\calh}) \setminus L) \in \calb$.

\item If $(G \setminus L) \, / \, \cala_{\calh} \arrow{\tau}{\rm a} (G' \setminus L) \, / \, \cala_{\calh}$
with $G \arrow{h}{\rm a} G'$ and $h \notin L$, then $G \, / \, \cala_{\calh} \arrow{\tau}{\rm a} G' \, / \,
\cala_{\calh}$ as $h \in \cala_{\calh}$ (note that $G \setminus \cala_{\calh}$ cannot perform $h$). From $G
\, / \, \cala_{\calh} \wbis{\rm pb} G \setminus \cala_{\calh}$ -- as $E \in \mathrm{SBSNNI}_{\wbis{\rm pb}}$
and $G \in \reach(E)$ -- and $G \setminus \cala_{\calh} \wbis{\rm pb} F \, / \, \cala_{\calh}$ -- hence $G
\, / \, \cala_{\calh} \wbis{\rm pb} F \, / \, \cala_{\calh}$ as $\wbis{\rm pb}$ is transitive -- it follows
that either $G' \, / \, \cala_{\calh} \wbis{\rm pb} F \, / \, \cala_{\calh}$ and hence $G' \setminus
\cala_{\calh} \wbis{\rm pb} F \, / \, \cala_{\calh}$ -- as $G' \, / \, \cala_{\calh} \wbis{\rm pb} G'
\setminus \cala_{\calh}$ due to $E \in \mathrm{SBSNNI}_{\wbis{\rm pb}}$ and $G' \in \reach(E)$ -- or there
exists $F \, / \, \cala_{\calh} \warrow{}{} \bar{F} \, / \, \cala_{\calh} \arrow{\tau}{\rm a} F' \, / \,
\cala_{\calh}$ such that $G \, / \, \cala_{\calh} \wbis{\rm pb} \bar{F} \, / \, \cala_{\calh}$ and $G' \, /
\, \cala_{\calh} \wbis{\rm pb} F' \, / \, \cala_{\calh}$ and hence $G \setminus \cala_{\calh} \wbis{\rm pb}
\bar{F} \, / \, \cala_{\calh}$ and $G' \setminus \cala_{\calh} \wbis{\rm pb} F' \, / \, \cala_{\calh}$.
Since the restriction operator does not affect $\tau$ and probabilistic transitions, in the former subcase
$(F \, / \, \cala_{\calh}) \setminus L$ is allowed to stay idle with $((G' \setminus L) \, / \,
\cala_{\calh}, (F \, / \, \cala_{\calh}) \setminus L) \in \calb$, while in the latter subcase $(F \, / \,
\cala_{\calh}) \setminus L \warrow{}{} (\bar{F} \, / \, \cala_{\calh}) \setminus L \arrow{\tau}{\rm a} (F'
\, / \, \cala_{\calh}) \setminus L$ with $((G \setminus L) \, / \, \cala_{\calh}, (\bar{F} \, / \,
\cala_{\calh}) \setminus L) \in \calb$ and $((G' \setminus L) \, / \, \cala_{\calh}, (F' \, / \,
\cala_{\calh}) \setminus L) \in \calb$.

				\end{itemize}

\noindent
As for probabilities, we reason like in the proof of the corresponding result for $\wbis{\rm pw}$.


\item Starting from $(F_{1} \pco{L} F_{2}) \setminus \cala_{\calh}$ and $(G_{1} \pco{L} G_{2}) \setminus
\cala_{\calh}$ related by $\calb$, so that $F_{1} \setminus \cala_{\calh} \wbis{\rm pb} G_{1} \setminus
\cala_{\calh}$ and $F_{2} \setminus \cala_{\calh} \wbis{\rm pb} G_{2} \setminus \cala_{\calh}$, there are
five cases for action transitions based on the operational semantic rules in Table~\ref{tab:prob_op_sem_nd}:

				\begin{itemize}

\item If $(F_{1} \pco{L} F_{2}) \setminus \cala_{\calh} \arrow{l}{\rm a} (F'_{1} \pco{L} [1] F_{2})
\setminus \cala_{\calh}$ with $F_{1} \arrow{l}{\rm a} F'_{1}$ and $l \notin L$, then $F_{1} \setminus
\cala_{\calh} \linebreak \arrow{l}{\rm a} F'_{1} \setminus \cala_{\calh}$ as $l \notin \cala_{\calh}$. From
$F_{1} \setminus \cala_{\calh} \wbis{\rm pb} G_{1} \setminus \cala_{\calh}$ it follows that there exists
$G_{1} \setminus \cala_{\calh}$ \linebreak $\warrow{}{} \bar{G}_{1} \setminus \cala_{\calh} \arrow{l}{\rm a}
G'_{1} \setminus \cala_{\calh}$ such that $F_{1} \setminus \cala_{\calh} \wbis{\rm pb} \bar{G}_{1} \setminus
\cala_{\calh}$ and $F'_{1} \setminus \cala_{\calh} \wbis{\rm pb} G'_{1} \setminus \cala_{\calh}$. Since
synchronization does not affect $\tau$ and $l \notin L$, we have that $(G_{1} \pco{L} G_{2}) \setminus
\cala_{\calh} \linebreak \warrow{}{} (\bar{G}_{1} \pco{L} G_{2}) \setminus \cala_{\calh} \arrow{l}{\rm a}
(G'_{1} \pco{L} [1] G_{2}) \setminus \cala_{\calh}$ with $((F_{1} \pco{L} F_{2}) \setminus \cala_{\calh},
(\bar{G}_{1} \pco{L} G_{2}) \setminus \cala_{\calh}) \in \calb$ and $((F'_{1} \pco{L} [1] F_{2}) \setminus
\cala_{\calh}, (G'_{1} \pco{L} [1] G_{2}) \setminus \cala_{\calh}) \in \calb$, where the right subprocess
alternates between $G_{2}$ and $[1] G_{2}$ thus allowing the probabilistic transitions along $G_{1}
\setminus \cala_{\calh} \warrow{}{}$ \linebreak $\bar{G}_{1} \setminus \cala_{\calh}$ to synchronize with
the only one of $[1] G_{2}$.

\item If $(F_{1} \pco{L} F_{2}) \setminus \cala_{\calh} \arrow{l}{\rm a} ([1] F_{1} \pco{L} F'_{2})
\setminus \cala_{\calh}$ with $F_{2} \arrow{l}{\rm a} F'_{2}$ and $l \notin L$, then the proof is similar to
the one of the previous case.

\item If $(F_{1} \pco{L} F_{2}) \setminus \cala_{\calh} \arrow{l}{\rm a} (F'_{1} \pco{L} F'_{2}) \setminus
\cala_{\calh}$ with $F_{i} \arrow{l}{\rm a} F'_{i}$ for $i \in \{ 1, 2 \}$ and $l \in L$, then $F_{i}
\setminus \cala_{\calh} \arrow{l}{\rm a} F'_{i} \setminus \cala_{\calh}$ as $l \notin \cala_{\calh}$. From
$F_{i} \setminus \cala_{\calh} \wbis{\rm pb} G_{i} \setminus \cala_{\calh}$ it follows that there exists
$G_{i} \setminus \cala_{\calh} \warrow{}{} \bar{G}_{i} \setminus \cala_{\calh} \arrow{l}{\rm a} G'_{i}
\setminus \cala_{\calh}$ such that $F_{i} \setminus \cala_{\calh} \wbis{\rm pb} \bar{G}_{i} \setminus
\cala_{\calh}$ and $F'_{i} \setminus \cala_{\calh} \wbis{\rm pb} G'_{i} \setminus \cala_{\calh}$. Since
synchronization does not affect $\tau$ and $l \in L$, we have that $(G_{1} \pco{L} G_{2}) \setminus
\cala_{\calh} \warrow{}{} (\bar{G}_{1} \pco{L} \bar{G}_{2}) \setminus \cala_{\calh} \arrow{l}{\rm a} (G'_{1}
\pco{L} G'_{2}) \setminus \cala_{\calh}$ with $((F_{1} \pco{L} F_{2}) \setminus \cala_{\calh}, \linebreak
(\bar{G}_{1} \pco{L} \bar{G}_{2}) \setminus \cala_{\calh}) \in \calb$ and $((F'_{1} \pco{L} F'_{2})
\setminus \cala_{\calh}, (G'_{1} \pco{L} G'_{2}) \setminus \cala_{\calh}) \in \calb$, where subprocess~$i$
alternates between $G_{i}$ and $[1] G_{i}$ thus allowing the probabilistic transitions along $G_{j}
\setminus \cala_{\calh} \warrow{}{} \bar{G}_{i} \setminus \cala_{\calh}$ for $j \neq i$ to synchronize with
the only one of $[1] G_{i}$.

\item If $(F_{1} \pco{L} F_{2}) \setminus \cala_{\calh} \arrow{\tau}{\rm a} (F'_{1} \pco{L} [1] F_{2})
\setminus \cala_{\calh}$ with $F_{1} \arrow{\tau}{\rm a} F'_{1}$, then $F_{1} \setminus \cala_{\calh}
\arrow{\tau}{\rm a} F'_{1} \setminus \cala_{\calh}$ \linebreak as $\tau \notin \cala_{\calh}$. From $F_{1}
\setminus \cala_{\calh} \wbis{\rm pb} G_{1} \setminus \cala_{\calh}$ it follows that either $F'_{1}
\setminus \cala_{\calh} \wbis{\rm pb} G_{1} \setminus \cala_{\calh}$, \linebreak or there exists $G_{1}
\setminus \cala_{\calh} \warrow{}{} \bar{G}_{1} \setminus \cala_{\calh} \arrow{\tau}{\rm a} G'_{1} \setminus
\cala_{\calh}$ such that $F_{1} \setminus \cala_{\calh} \wbis{\rm pb} \bar{G}_{1} \setminus \cala_{\calh}$
and $F'_{1} \setminus \cala_{\calh} \wbis{\rm pb} G'_{1} \setminus \cala_{\calh}$. Since synchronization
does not affect $\tau$, in the former subcase $(G_{1} \pco{L} G_{2}) \setminus \cala_{\calh}$ is allowed to
stay idle with $((F'_{1} \pco{L} [1] F_{2}) \setminus \cala_{\calh}, (G_{1} \pco{L} G_{2}) \setminus
\cala_{\calh}) \in \calb$, while in the latter subcase $(G_{1} \pco{L} G_{2}) \setminus \cala_{\calh}
\warrow{}{} (\bar{G}_{1} \pco{L} G_{2}) \setminus \cala_{\calh} \arrow{\tau}{\rm a} (G'_{1} \pco{L} [1]
G_{2}) \setminus \cala_{\calh}$ with $((F_{1} \pco{L} F_{2}) \setminus \cala_{\calh}, (\bar{G}_{1} \pco{L}
G_{2}) \setminus \cala_{\calh}) \in \calb$ and $((F'_{1} \pco{L} [1] F_{2}) \setminus \cala_{\calh}, (G'_{1}
\pco{L} [1] G_{2}) \setminus \cala_{\calh}) \linebreak \in \calb$, where the right subprocess alternates
between $G_{2}$ and $[1] G_{2}$ thus allowing the probabilistic transitions along $G_{1} \setminus
\cala_{\calh} \warrow{}{} \bar{G}_{1} \setminus \cala_{\calh}$ to synchronize with the only one of $[1]
G_{2}$.

\item If $(F_{1} \pco{L} F_{2}) \setminus \cala_{\calh} \arrow{\tau}{\rm a} ([1] F_{1} \pco{L} F'_{2})
\setminus \cala_{\calh}$ with $F_{2} \arrow{\tau}{\rm a} F'_{2}$, then the proof is similar to the one of
the previous case.

				\end{itemize}

\noindent		
As for probabilities, we reason like in the proof of the corresponding result for $\wbis{\rm pw}$.
\qedhere

			\end{enumerate}

		\end{proof}

	\end{lem}

	\begin{thm}\label{thm:prob_compositionality} 

Let $E_{1}, E_{2} \in \procs_{\rm n}$ or $E_{1}, E_{2} \in \procs_{\rm p}$, $E \in \procs$, $\wbis{} \: \in
\{ \wbis{\rm pw}, \wbis{\rm pb} \}$, and $\calp \in \{ \mathrm{SBSNNI}_{\wbis{}},
\mathrm{P\_BNDC}_{\wbis{}}, \mathrm{SBNDC}_{\wbis{}} \}$. Then:

		\begin{enumerate}

\item $E \in \calp \Longrightarrow a \, . \, E \in \calp$ for all $a \in \cala_{\call} \cup \{ \tau \}$,
when $E \in \procs_{\rm p}$.

\item $E_{1}, E_{2} \in \calp \Longrightarrow E_{1} \pco{L} E_{2} \in \calp$ for all $L \subseteq
\cala_{\call}$ if $\calp \in \{ \mathrm{SBSNNI}_{\blue{\wbis{}}}, \mathrm{P\_BNDC}_{\blue{\wbis{}}} \}$ or
\linebreak for all $L \subseteq \cala$ if $\calp = \mathrm{SBNDC}_{\blue{\wbis{}}}$\blue{, provided that
$\wbis{} \; = \; \wbis{\rm pb}$}.

\item $E \in \calp \Longrightarrow E \setminus L \in \calp$ for all $L \subseteq \cala$.

\item $E \in \calp \Longrightarrow E \, / \, L \in \calp$ for all $L \subseteq \cala_{\call}$.

		\end{enumerate}

		\begin{proof}

\blue{First we} prove the four results for SBSNNI$_{\wbis{}}$, from which it will follow that they hold for
P\_BNDC$_{\wbis{}}$ too due to P\_BNDC$_{\wbis{}}$ = SBSNNI$_{\wbis{}}$ as will be established by
Theorem~\ref{thm:prob_taxonomy_1}:

			\begin{enumerate}

\item Given an arbitrary $E \in \procs_{\rm p} \, \cap \, \mathrm{SBSNNI}_{\wbis{}}$ and an arbitrary $a \in
\cala_{\call} \cup \{ \tau \}$, from $E \setminus \cala_{\calh} \linebreak \wbis{} E \, / \, \cala_{\calh}$
we derive that $a \, . \, (E \setminus \cala_{\calh}) \wbis{} a \, . \, (E \, / \, \cala_{\calh})$ because
$\wbis{}$ is a congruence with respect to action prefix by virtue of Lemma~\ref{lem:prob_bisim_congr}(1),
from which it follows that $(a \, . \, E) \setminus \cala_{\calh} \linebreak \wbis{} (a \, . \, E) \, / \,
\cala_{\calh}$, i.e., $a \, . \, E \in \mathrm{BSNNI}_{\wbis{}}$, because $a \notin \cala_{\calh}$. To
conclude the proof, it suffices to observe that all the processes reachable from $a \, . \, E$ after
performing $a$ are processes reachable from $E$, which are known to be BSNNI$_{\wbis{}}$.

\item Given two arbitrary $E_{1}, E_{2} \in \procs_{\rm n}$ or $E_{1}, E_{2} \in \procs_{\rm p}$ such that
$E_{1}, E_{2} \in \mathrm{SBSNNI}_{\wbis{\rm pb}}$ and an arbitrary $L \subseteq \cala_{\call}$, the result
follows from Lemma~\ref{lem:prob_compositionality}(1) by taking $F_{1}$ and $G_{1}$ identical to $E_{1}$ and
$F_{2}$ and $G_{2}$ identical to $E_{2}$. \\
The result does not hold for $\mathrm{SBSNNI}_{\wbis{\rm pw}}$ as can be seen by considering $H' \pco{\{ a
\}} H$ where $H' = \tau \, . \, [1] H_{1} + h \, . \, [1] H_{2}$, with processes $H_{1}$, $H_{2}$, and $H$
being defined in the proof of Lemma~\ref{lem:prob_bisim_congr}(2) and all of their observable actions
belonging to $\cala_{\call}$. The reason is that $(H' \pco{\{ a \}} H) \setminus \cala_{\calh} \not\wbis{\rm
pw} (H' \pco{\{ a \}} H) \, / \, \cala_{\calh}$ because $(H' \pco{\{ a \}} H) \setminus \cala_{\calh}$ is
isomorphic to $\tau \, . \, [1] H_{1} \pco{\{ a \}} H$, $H' \, / \, \cala_{\calh}$ is isomorphic to $(\tau
\, . \, [1] H_{1} + \tau \, . \, [1] H_{2}) \pco{\{ a \}} H$, and whenever $(\tau \, .  \, [1] H_{1} + \tau
\, . \, [1] H_{2}) \pco{\{ a \}} H \arrow{\tau}{\rm a} [1] H_{2} \pco{\{ a \}} [1] H$ then $\tau \, . \, [1]
H_{1} \pco{\{ a \}} H$ can either stay idle or perform $\tau \, . \, [1] H_{1} \pco{\{ a \}} H
\arrow{\tau}{\rm a} \! \arrow{1}{\rm p} H_{1} \pco{\{ a \}} H$ -- both schedulable with probability $1$ --
but $[1] H_{2} \pco{\{ a \}} [1] H \not\wbis{\rm pw} \tau \, . \, [1] H_{1} \pco{\{ a \}} H \wbis{\rm pw}
H_{1} \pco{\{ a \}} H$ due to what has been observed in the proof of Lemma~\ref{lem:prob_bisim_congr}(2).

\item Given an arbitrary $E \in \mathrm{SBSNNI}_{\wbis{}}$ and an arbitrary $L \subseteq \cala$, the result
follows from Lemma~\ref{lem:prob_compositionality}(2) by taking $F$ identical to $G$ -- which will be
denoted by $E'$ -- because:

				\begin{itemize}

\item $(E' \setminus L) \setminus \cala_{\calh} \wbis{} (E' \setminus \cala_{\calh}) \setminus L$ as the
order in which restriction sets are considered \linebreak is unimportant.

\item $(E' \setminus \cala_{\calh}) \setminus L \wbis{} (E' \, / \, \cala_{\calh}) \setminus L$ because $E'
\setminus \cala_{\calh} \wbis{} E' \, / \, \cala_{\calh}$ -- as $E \in \mathrm{SBSNNI}_{\wbis{}}$ and $E'
\in \reach(E)$ -- and $\wbis{}$ is a congruence with respect to the restriction operator due to
Lemma~\ref{lem:prob_bisim_congr}(3).

\item $(E' \, / \, \cala_{\calh}) \setminus L \wbis{} (E' \setminus L) \, / \, \cala_{\calh}$ as a
consequence of Lemma~\ref{lem:prob_compositionality}(2).

\item From the transitivity of $\wbis{}$ we obtain that $(E' \setminus L) \setminus \cala_{\calh} \wbis{}
(E' \setminus L) \, / \, \cala_{\calh}$.

				\end{itemize}

\item Given an arbitrary $E \in \mathrm{SBSNNI}_{\wbis{}}$ and an arbitrary $L \subseteq \cala_{\call}$, for
every $E' \in \reach(E)$ \linebreak it holds that $E' \setminus \cala_{\calh} \wbis{} E' \, / \,
\cala_{\calh}$, from which we derive that $(E' \setminus \cala_{\calh}) \, / \, L \wbis{} (E' /
\cala_{\calh}) \, / \, L$ because $\wbis{}$ is a congruence with respect to the hiding operator due to
Lemma~\ref{lem:prob_bisim_congr}(4).  Since $L \cap \cala_{\calh} = \emptyset$, we have that $(E' \setminus
\cala_{\calh}) \, / \, L$ is isomorphic to $(E' \, / \, L) \setminus \cala_{\calh}$ and $(E' \, / \,
\cala_{\calh}) \, / \, L$ is isomorphic to $(E' \, / \, L) \, / \, \cala_{\calh}$ -- where \blue{two
processes are isomorphic iff their} underlying PLTSs \blue{coincide} up to \blue{the} processes associated
with \blue{their} states \blue{by the operational semantics} -- hence $(E' \, / \, L) \setminus
\cala_{\calh} \wbis{} (E' \, / \, L) \, / \, \cala_{\calh}$, i.e., $E' \, / \, L$ is BSNNI$_{\wbis{}}$.

			\end{enumerate}

\noindent
\blue{Next we} prove the four results for SBNDC$_{\wbis{}}$:

			\begin{enumerate}

\item Given an arbitrary $E \in \procs_{\rm p} \, \cap \, \mathrm{SBNDC}_{\wbis{}}$ and an arbitrary $a \in
\cala_{\call} \cup \{ \tau \}$, it trivially holds that $a \, . \, E \in \mathrm{SBNDC}_{\wbis{}}$ because
$a \blue{\, \notin \cala_{\calh}}$ and all the processes reachable from $a \, . \, E$ after performing $a$
are processes reachable from $E$, which is known to be SBNDC$_{\wbis{}}$.

\item Given two arbitrary $E_{1}, E_{2} \in \procs_{\rm n}$ or $E_{1}, E_{2} \in \procs_{\rm p}$ such that
$E_{1}, E_{2} \in \mathrm{SBNDC}_{\wbis{\rm pb}}$ and an arbitrary $L \subseteq \cala$, the result follows
from Lemma~\ref{lem:prob_compositionality}(3) as can be seen by observing that whenever $E'_{1} \pco{L}
E'_{2} \arrow{h}{\rm a} E''_{1} \pco{L} E''_{2} $ for $E'_{1} \pco{L} E'_{2} \in \reach(E_{1} \pco{L}
E_{2})$:

				\begin{itemize}

\item If $E'_{1} \arrow{h}{\rm a} E''_{1}$, $E''_{2} = E'_{2}$ (hence $E'_{2} \setminus \cala_{\calh}
\wbis{} E''_{2} \setminus \cala_{\calh}$), and $h \notin L$, then from \linebreak $E_{1} \in
\mathrm{SBNDC}_{\wbis{\rm pb}}$ it follows that $E'_{1} \setminus \cala_{\calh} \wbis{\rm pb} E''_{1}
\setminus \cala_{\calh}$, which in turn entails that $(E'_{1} \pco{L} E'_2) \setminus \cala_{\calh}
\wbis{\rm pb} (E''_{1} \pco{L} E''_{2}) \setminus \cala_{\calh}$ because $\wbis{\rm pb}$ is a congruence
with respect to the parallel composition operator due to Lemma~\ref{lem:prob_bisim_congr}(2) and restriction
distributes over parallel composition.

\item If $E'_{2} \arrow{h}{\rm a} E''_{2}$, $E''_{1} = E'_{1}$, and $h \notin L$, then we reason like in the
previous case.

\item If $E'_{1} \arrow{h}{\rm a} E''_{1}$, $E'_{2} \arrow{h}{\rm a} E''_{2}$, and $h \in L$, then from
$E_{1}, E_{2} \in \mathrm{SBNDC}_{\wbis{\rm pb}}$ it follows that $E'_{1} \setminus \cala_{\calh} \wbis{\rm
pb} E''_{1} \setminus \cala_{\calh}$ and $E'_{2} \setminus \cala_{\calh} \wbis{\rm pb} E''_{2} \setminus
\cala_{\calh}$, which in turn entail that $(E'_{1} \pco{L} E'_{2}) \setminus \cala_{\calh} \wbis{\rm pb}
(E''_{1} \pco{L} E''_{2}) \setminus \cala_{\calh}$ because $\wbis{\rm pb}$ is a congruence with respect to
the parallel composition operator due to Lemma~\ref{lem:prob_bisim_congr}(2) and restriction distributes
over parallel composition.

				\end{itemize} 

\noindent
The result does not hold for $\mathrm{SBNDC}_{\wbis{\rm pw}}$ as can be seen by considering $H' \pco{\{ a
\}} H$ where $H' = \tau \, . \, [1] H_{1} + h \, . \, [1] H_{2}$ with processes $H_{1}$, $H_{2}$, and $H$
being defined in the proof of Lemma~\ref{lem:prob_bisim_congr}(2) and all of their observable actions
belonging to $\cala_{\call}$. The reason is that $H' \pco{\{ a \}} H \arrow{h}{\rm a} [1] H_{2} \pco{\{ a
\}} [1] H$ with $(H' \pco{\{ a \}} H) \setminus \cala_{\calh} \not\wbis{\rm pw} ([1] H_{2} \pco{\{ a \}} [1]
H) \setminus \cala_{\calh}$ because $(H' \pco{\{ a \}} H) \setminus \cala_{\calh}$ is isomorphic to $\tau \,
. \, [1] H_{1} \pco{\{ a \}} H$, $([1] H_{2} \pco{\{ a \}} [1] H) \setminus \cala_{\calh}$ is isomorphic to
$[1] H_{2} \pco{\{ a \}} [1] H$, and whenever $\tau \, . \, [1] H_{1} \pco{\{ a \}} H \arrow{\tau}{\rm a}
[1] H_{1} \pco{\{ a \}} [1] H$ then the only matching possibility is that $[1] H_{2} \pco{\{ a \}} [1] H$
stays idle -- which can be scheduled with probability $1$ -- but $[1] H_{1} \pco{\{ a \}} [1] H
\not\wbis{\rm pw} [1] H_{2} \pco{\{ a \}} [1] H$ due to what has been observed in the proof of
Lemma~\ref{lem:prob_bisim_congr}(2).

\item Given an arbitrary $E \in \mathrm{SBNDC}_{{\wbis{}}}$ and an arbitrary $L \subseteq \cala$, for every
$E' \in \reach(E)$ and for every $E''$ such that $E' \arrow{h}{\rm a} E''$ it holds that $E' \setminus
\cala_{\calh} \wbis{} E'' \setminus \cala_{\calh}$, from which we derive that $(E' \setminus \cala_{\calh})
\setminus L \wbis{} (E'' \setminus \cala_{\calh}) \setminus L$ because $\wbis{}$ is a congruence with
respect to the restriction operator due to Lemma~\ref{lem:prob_bisim_congr}(3). Since $(E' \setminus
\cala_{\calh}) \setminus L$ is isomorphic to $(E' \setminus L) \setminus \cala_{\calh}$ and $(E'' \setminus
\cala_{\calh}) \setminus L$ is isomorphic to $(E'' \setminus L) \setminus \cala_{\calh}$, we have that $(E'
\setminus L) \setminus \cala_{\calh} \wbis{} (E'' \setminus L) \setminus \cala_{\calh}$.

\item Given an arbitrary $E \in \mathrm{SBNDC}_{\wbis{}}$ and an arbitrary $L \subseteq \cala_{\call}$, for
every $E' \in \reach(E)$ and for every $E''$ such that $E' \arrow{h}{\rm a} E''$ it holds that $E' \setminus
\cala_{\calh} \wbis{} E'' \setminus \cala_{\calh}$, from which we derive that $(E' \setminus \cala_{\calh})
\, / \, L \wbis{} (E'' \setminus \cala_{\calh}) \, / \, L$ because $\wbis{}$ is a congruence with respect to
the hiding operator due to Lemma~\ref{lem:prob_bisim_congr}(4). Since $L \cap \cala_{\calh} = \emptyset$, we
have that $(E' \setminus \cala_{\calh}) \, / \, L$ is isomorphic to $(E' \, / \, L) \setminus \cala_{\calh}$
and $(E'' \setminus \cala_{\calh}) \, / \, L$ is isomorphic to $(E'' \, / \, L) \setminus \cala_{\calh}$,
hence $(E' \, / \, L) \setminus \cala_{\calh} \wbis{} (E'' \, / \, L) \setminus \cala_{\calh}$.
\qedhere

			\end{enumerate}

		\end{proof}

	\end{thm}

%
\subsection{Taxonomy of Security Properties}
\label{sec:prob_taxonomy}
%

Similar to the nondeterministic setting~\cite{FG01,EABR25}, the noninterference properties in
Definition~\ref{def:prob_bisim_sec_prop} turn out to be increasingly finer. This holds both for those based
on $\wbis{\rm pw}$ and for those based on $\wbis{\rm pb}$.

In~\cite{EAB24} part of the proof of the forthcoming taxonomy Theorem~\ref{thm:prob_taxonomy_1} proceeded by
induction on the depth of the \blue{PLTS} underlying the process under examination. Now that the language
includes recursion, which may introduce cycles, we have to follow a different proof strategy. This relies on
the bisimulation-up-to technique~\cite{SM92} and requires introducing probabilistic variants of up-to
weak~\cite{Mil89a} and branching~\cite{Gla93} bisimulations. In doing so in our quantitative setting, we
have to take into account some technicalities mentioned in~\cite{BBG98b,HL97,GSS95}. In particular, given
$\wbis{} \: \in \{ \wbis{\rm pw}, \wbis{\rm pb} \}$ and a related bisimulation $\calb$, we cannot consider
the relation composition $\wbis{} \! \calb \! \wbis{}$ like in the fully nondeterministic case as it may not
be transitive and this would make it \blue{im}possible to work with equivalence classes for the
probabilistic part. Rather we have to consider $(\calb \cup \calb^{-1} \, \cup \wbis{})^{+} = \, \bigcup_{n
= 1}^{\infty} (\calb \cup \calb^{-1} \, \cup \wbis{})^{n}$ to ensure transitivity in addition to reflexivity
and symmetry, where $\calb^{-1}$ is the inverse of $\calb$ and $\calb$ is no longer required to be an
equivalence relation thus avoiding redundant information in it. We remind that $(\calb \cup \calb^{-1} \,
\cup \wbis{})^{n}$ for $n > 1$ is the composition of relations $(\calb \cup \calb^{-1} \, \cup \wbis{})^{n -
1}$ and $\calb \cup \calb^{-1} \, \cup \wbis{}$.

	\begin{defi}\label{def:prob_weak_up-to}

A relation $\calb$ over $\procs$ is a \emph{weak probabilistic bisimulation up to $\wbis{\rm pw}$} iff,
whenever $(E_{1}, E_{2}) \in \calb$, then: 

		\begin{itemize}

\item For each $E_{1} \warrow{a}{} E'_{1}$ \blue{due to some $\sigma_{1} \in \sched(E_{1})$} there exists
$\sigma_{2} \in \sched(E_{2})$ such that $\proba_{\sigma_{1}}(E_{1}, \tau^{*} \, a \, \tau^{*}, \blue{\{
E'_{1} \}}, \varepsilon_{E_{1}}) = \proba_{\sigma_{2}}(E_{2}, \tau^{*} \, \hat{a} \, \tau^{*},
[E'_{1}]_{(\calb \cup \calb^{-1} \cup \wbis{\rm pw})^{+}}, \varepsilon_{E_{2}})$, and vice versa.

\item $\proba(E_{1}, C) = \proba(E_{2}, C)$ for all equivalence classes $C \in \procs / (\calb \cup
\calb^{-1} \, \cup \wbis{\rm pw})^{+}$. 
\fullbox

		\end{itemize}

	\end{defi}

	\begin{defi}\label{def:prob_branching_up-to}

A relation $\calb$ over $\procs$ is a \emph{probabilistic branching bisimulation up to $\wbis{\rm pb}$} iff,
whenever $(E_{1}, E_{2}) \in \calb$, then: 

		\begin{itemize}

\item For each $E_{1} \warrow{}{} \bar{E}_{1} \arrow{a}{\rm a} E'_{1}$ with $E_{1} \wbis{\rm pb}
\bar{E}_{1}$:

			\begin{itemize}

\item either $a = \tau$ and $\bar{E}_{1} \wbis{\rm pb} E'_{1}$;

\item or there exists $E_{2} \warrow{}{} \bar{E}_{2} \arrow{a}{\rm a} E'_{2}$ such that $(\bar{E}_{1},
\bar{E}_{2}) \in (\calb \cup \calb^{-1} \cup \wbis{\rm pb})^{+}$ and $(E'_{1}, E'_{2}) \in (\calb \cup
\calb^{-1} \, \cup \wbis{\rm pb})^{+}$;

			\end{itemize}

\noindent
and vice versa.

\item $\proba(E_{1}, C) = \proba(E_{2}, C)$ for all equivalence classes $C \in \procs / (\calb \cup
\calb^{-1} \, \cup \wbis{\rm pb})^{+}$.
\fullbox

		\end{itemize}

	\end{defi}

As for Definition~\ref{def:prob_branching_up-to}, in the case that $a = \tau$ and $\bar{E}_{1} \wbis{\rm pb}
E'_{1}$ it holds that $E'_{1} \wbis{\rm pb} \bar{E}_{1} \wbis{\rm pb} E_{1} \; \calb \; E_{2}$, i.e.,
$(E'_{1}, E_{2}) \in (\calb \cup \calb^{-1} \, \cup \wbis{\rm pb})^{+}$, because $\wbis{\rm pb}$ is
symmetric. We now prove that the two notions above are correct, i.e., they imply the respective
bisimilarities.

	\begin{prop}\label{prop:prob_weak_up-to}

Let $E_{1}, E_{2} \in \procs$ and $\calb$ be a weak probabilistic bisimulation up to $\wbis{\rm pw}$.
\linebreak If $(E_{1}, E_{2}) \in \calb$ then $E_{1} \wbis{\rm pw} E_{2}$.

		\begin{proof}

It suffices to prove that the equivalence relation $(\calb' \, \cup \wbis{\rm pw})^{+}$ is a weak
probabilistic bisimulation, where $\calb' = \calb \cup \calb^{-1}$. Given $(E_{1}, E_{2}) \in (\calb' \,
\cup \wbis{\rm pw})^{+}$ and considering the smallest $n \in \natns_{> 0}$ for which $(E_{1}, E_{2}) \in
(\calb' \, \cup \wbis{\rm pw})^{n}$, we proceed by induction on $n$:

			\begin{itemize}

\item If $n = 1$ then there are two cases:
 
				\begin{itemize}

\item Let $(E_{1}, E_{2}) \in \calb'$. If $E_{1} \arrow{a}{\rm a} E'_{1}$, hence $E_{1} \warrow{a}{} E'_{1}$
can be scheduled with probability~$1$ \blue{by some $\sigma_{1} \in \sched(E_{1})$}, then from the fact that
$\calb'$ is a weak probabilistic bisimulation up to $\wbis{\rm pw}$ it follows that there exist\blue{s}
$\sigma_{2} \in \sched(E_{2})$ such that $\proba_{\sigma_{1}}(E_{1}, \tau^{*} \, a \, \tau^{*}, \blue{\{
E'_{1} \}}, \varepsilon_{E_{1}}) = \proba_{\sigma}(E_{2}, \tau^{*} \, \hat{a} \, \tau^{*}, [E'_{1}]_{(\calb'
\cup \wbis{\rm pw})^{+}}, \varepsilon_{E_{2}})$, hence $\proba_{\sigma}(E_{2}, \tau^{*} \, \hat{a} \,
\tau^{*}, [E'_{1}]_{(\calb' \cup \wbis{\rm pw})^{+}}, \varepsilon_{E_{2}}) = 1$. Moreover, since $\calb'$ is
a weak probabilistic bisimulation up to $\wbis{\rm pw}$, we have that $\proba(E_{1}, C) = \proba(E_{2}, C)$
for all $C \in \procs / (\calb' \, \cup \wbis{\rm pw})^{+}$.

\item Let $E_{1} \wbis{\rm pw} E_{2}$. If $E_{1} \arrow{a}{\rm a} E'_{1}$ then there exists $\sigma \! \in
\! \sched(E_{2})$ such that $\proba_{\sigma}(E_{2}, \tau^{*} \, \hat{a} \, \tau^{*},$ \linebreak
$[E'_{1}]_{\wbis{\rm pw}}, \varepsilon_{E_{2}}) \! = \! 1$, hence $\proba_{\sigma}(E_{2}, \tau^{*} \,
\hat{a} \, \tau^{*}, [E'_{1}]_{(\calb' \cup \wbis{\rm pw})^{+}}, \varepsilon_{E_{2}}) \! = \! 1$ as
$\wbis{\rm pw} \, \subseteq (\calb' \, \cup \wbis{\rm pw})^{+}$ \blue{and the probability of reaching from
$E_{2}$ a process in $[E'_{1}]_{(\calb' \cup \wbis{\rm pw})^{+}}$ cannot exceed $1$}. Moreover, since
$\wbis{\rm pw} \, \subseteq (\calb' \, \cup \wbis{\rm pw})^{+}$ implies that every equivalence class of
$(\calb' \, \cup \wbis{\rm pw})^{+}$ is the union of some equivalence classes of $\wbis{\rm pw}$, we have
that $\proba(E_{1}, C) = \proba(E_{2}, C)$ for all $C \in \procs / (\calb' \, \cup \wbis{\rm pw})^{+}$.

				\end{itemize}

\item If $n > 1$ then from $(E_{1}, E_{2}) \in (\calb' \, \cup \wbis{\rm pw})^{n}$ and the minimality of $n$
it follows that there exists $\tilde{E} \in \procs$ such that $(E_{1}, \tilde{E}) \in (\calb' \, \cup
\wbis{\rm pw})^{n - 1}$ and $(\tilde{E}, E_{2}) \in (\calb' \, \cup \wbis{\rm pw})$. \linebreak If $E_{1}
\arrow{a}{\rm a} E'_{1}$ then by the induction hypothesis applied to $(E_{1}, \tilde{E}) \in (\calb' \, \cup
\wbis{\rm pw})^{n - 1}$ there exists $\tilde{\sigma} \in \sched(\tilde{E})$ such that
$\proba_{\tilde{\sigma}}(\tilde{E}, \tau^{*} \, \hat{a} \, \tau^{*}, [E'_{1}]_{(\calb' \cup \wbis{\rm
pw})^{+}}, \varepsilon_{\tilde{E}}) = 1$. Therefore, by the induction hypothesis applied to $(\tilde{E},
E_{2}) \in (\calb' \, \cup \wbis{\rm pw})$, for each $\tilde{E} \warrow{\hat{a}}{} \tilde{E}'$ such that
$(E'_{1}, \tilde{E}') \in (\calb' \, \cup \wbis{\rm pw})^{+}$ there exists $\sigma \in \sched(E_{2})$ such
that $\proba_{\sigma}(E_{2}, \tau^{*} \, \hat{a} \, \tau^{*}, [E'_{1}]_{(\calb' \cup \wbis{\rm pw})^{+}},
\varepsilon_{E_{2}}) = 1$. \linebreak Moreover, from the induction hypothesis applied to $(E_{1}, \tilde{E})
\in (\calb' \, \cup \wbis{\rm pw})^{n - 1}$ and \linebreak $(\tilde{E}, E_{2}) \in (\calb' \, \cup \wbis{\rm
pw})$ it follows that $\proba(E_{1}, C) = \proba(\tilde{E}, C) = \proba(E_{2}, C)$ for all $C \in \procs /
(\calb' \, \cup \wbis{\rm pw})^{+}$.
\qedhere

			\end{itemize}

		\end{proof}
		
	\end{prop}

	\begin{prop}\label{prop:prob_branching_up-to}

Let $E_{1}, E_{2} \in \procs$ and $\calb$ be a probabilistic branching bisimulation up to $\wbis{\rm pb}$.
If $(E_{1}, E_{2}) \in \calb$ then $E_{1} \wbis{\rm pb} E_{2}$.

		\begin{proof}

It suffices to prove that the equivalence relation $(\calb' \, \cup \wbis{\rm pb})^{+}$ is a probabilistic
branching bisimulation, where $\calb' = \calb \cup \calb^{-1}$. Given $(E_{1}, E_{2}) \in (\calb' \, \cup
\wbis{\rm pb})^{+}$ and considering the smallest $n \in \natns_{> 0}$ for which $(E_{1}, E_{2}) \in (\calb'
\, \cup \wbis{\rm pb})^{n}$, we proceed by induction on $n$:

			\begin{itemize}

\item If $n = 1$ then there are two cases:
 
				\begin{itemize}

\item Let $(E_{1}, E_{2}) \in \calb'$. If $E_{1} \arrow{a}{\rm a} E'_{1}$, hence $E_{1} \warrow{}{} E_{1}
\arrow{a}{\rm a} E'_{1}$, then from the fact that $\calb'$ is a probabilistic branching bisimulation up to
$\wbis{\rm pb}$ it follows that there are two subcases:

					\begin{itemize}

\item If $a = \tau$ and $E_{1} \wbis{\rm pb} E'_{1}$, hence $(E'_{1}, E_{1}) \in (\calb' \, \cup \wbis{\rm
pb})^{+}$ by symmetry, from $(E_{1}, E_{2}) \in (\calb' \, \cup \wbis{\rm pb})^{+}$ it follows that
$(E'_{1}, E_{2}) \in (\calb' \, \cup \wbis{\rm pb})^{+}$ by transitivity.

\item If there exists $E_{2} \warrow{}{} \bar{E}_{2} \arrow{a}{\rm a} E'_{2}$ such that $(E_{1},
\bar{E}_{2}) \in (\calb' \, \cup \wbis{\rm pb})^{+}$ and $(E'_{1}, E'_{2}) \in (\calb' \, \cup \wbis{\rm
pb})^{+}$, then we are done.

					\end{itemize}

Moreover, since $\calb'$ is a probabilistic branching bisimulation up to $\wbis{\rm pb}$, we have that
$\proba(E_{1}, C) = \proba(E_{2}, C)$ for all $C \in \procs / (\calb' \, \cup \wbis{\rm pb})^{+}$.

\item Let $E_{1} \wbis{\rm pb} E_{2}$. If $E_{1} \arrow{a}{\rm a} E'_{1}$ then there are two subcases:

					\begin{itemize}

\item If $a = \tau$ and $E'_{1} \wbis{\rm pb} E_{2}$, then $(E'_{1}, E_{2}) \in (\calb' \, \cup \wbis{\rm
pb})^{+}$ because $\wbis{\rm pb} \, \subseteq (\calb' \, \cup \wbis{\rm pb})^{+}$.

\item If there exists $E_{2} \warrow{}{} \bar{E}_{2} \arrow{a}{\rm a} E'_{2}$ such that $E_{1} \wbis{\rm pb}
\bar{E}_{2}$ and $E'_{1} \wbis{\rm pb} E'_{2}$, then $(E_{1}, \bar{E}_{2}) \in (\calb' \, \cup \wbis{\rm
pb})^{+}$ and $(E'_{1}, E'_{2}) \in (\calb' \, \cup \wbis{\rm pb})^{+}$ because $\wbis{\rm pb} \, \subseteq
(\calb' \, \cup \wbis{\rm pb})^{+}$.

					\end{itemize}

Moreover, since $\wbis{\rm pb} \, \subseteq (\calb' \, \cup \wbis{\rm pb})^{+}$ implies that every
equivalence class of $(\calb' \, \cup \wbis{\rm pb})^{+}$ is the union of some equivalence classes of
$\wbis{\rm pb}$, we have that $\proba(E_{1}, C) = \proba(E_{2}, C)$ for all $C \in \procs / (\calb' \, \cup
\wbis{\rm pb})^{+}$.

				\end{itemize}

\item If $n > 1$ then from $(E_{1}, E_{2}) \in (\calb' \, \cup \wbis{\rm pb})^{n}$ and the minimality of $n$
it follows that there exists $\tilde{E} \in \procs$ such that $(E_{1}, \tilde{E}) \in (\calb' \, \cup
\wbis{\rm pb})^{n - 1}$ and $(\tilde{E}, E_{2}) \in (\calb' \, \cup \wbis{\rm pb})$. If $E_{1} \arrow{a}{\rm
a} E'_{1}$ then by the induction hypothesis applied to $(E_{1}, \tilde{E}) \in (\calb' \, \cup \wbis{\rm
pb})^{n - 1}$ there are two cases:

				\begin{itemize}

\item If $a = \tau$ and $(E'_{1}, \tilde{E}) \in (\calb' \, \cup \wbis{\rm pb})^{+}$, then from $(\tilde{E},
E_{2}) \in (\calb' \, \cup \wbis{\rm pb})$ it follows that $(E'_{1}, E_{2}) \in (\calb' \, \cup \wbis{\rm
pb})^{+}$ by transitivity.

\item If there exists $\tilde{E} \warrow{}{} \bar{E} \arrow{a}{\rm a} \tilde{E}'$ such that $(E_{1},
\bar{E}) \in (\calb' \, \cup \! \wbis{\rm pb})^{+}$ and $(E'_{1}, \tilde{E}') \in \linebreak (\calb' \, \cup
\wbis{\rm pb})^{+}$, then by the induction hypothesis applied to $(\tilde{E}, E_{2}) \in (\calb' \, \cup
\wbis{\rm pb})$ there are two subcases:

					\begin{itemize}

\item If $a = \tau$ and $(\tilde{E}', E_{2}) \in (\calb' \, \cup \wbis{\rm pb})^{+}$, then from $(E'_{1},
\tilde{E}') \in (\calb' \, \cup \wbis{\rm pb})^{+}$ it follows that $(E'_{1}, E_{2}) \in (\calb' \, \cup
\wbis{\rm pb})^{+}$ by transitivity.

\item If there exists $E_{2} \warrow{}{} \bar{E}_{2} \arrow{a}{\rm a} E'_{2}$ such that $(\bar{E},
\bar{E}_{2}) \in (\calb' \, \cup \wbis{\rm pb})^{+}$ and $(\tilde{E}', E'_{2}) \in (\calb' \, \cup \wbis{\rm
pb})^{+}$, then from $(E_{1}, \bar{E}) \in (\calb' \, \cup \wbis{\rm pb})^{+}$ and $(E'_{1}, \tilde{E}') \in
(\calb' \, \cup \wbis{\rm pb})^{+}$ it follows that $(E_{1}, \bar{E}_{2}) \in (\calb' \, \cup \wbis{\rm
pb})^{+}$ and $(E'_{1}, E'_{2}) \in (\calb' \, \cup \wbis{\rm pb})^{+}$ by transitivity.

					\end{itemize}

				\end{itemize}

\noindent
Moreover, from the induction hypothesis applied to $(E_{1}, \tilde{E}) \in (\calb' \, \cup \wbis{\rm pb})^{n
- 1}$ and $(\tilde{E}, E_{2}) \linebreak \in (\calb' \, \cup \! \wbis{\rm pb})$ it follows that
$\proba(E_{1}, C) = \proba(\tilde{E}, C) = \proba(E_{2}, C)$ for all $C \in \procs \, / \linebreak (\calb'
\, \cup \wbis{\rm pb})^{+}$.
\qedhere

			\end{itemize}
			
		\end{proof}

	\end{prop}

Before presenting the taxonomy, we prove some further ancillary results about parallel composition,
restriction, and hiding under SBSNNI$_{\wbis{}}$ and SBNDC$_{\wbis{}}$.

	\begin{lem}\label{lem:prob_taxonomy}

Let $E, E_{1}, E_{2} \in \procs$ and $\wbis{} \: \in \{ \wbis{\rm pw}, \wbis{\rm pb} \}$. Then:

		\begin{enumerate}

\item If $E \in \mathrm{SBNDC}_{\wbis{}}$, $E' \in \reach(E)$, and $E' \, / \, \cala_{\calh} \warrow{}{} E''
\, / \, \cala_{\calh}$, then $E' \setminus \cala_{\calh} \warrow{}{} \hat{E}'' \setminus \cala_{\calh}$ with
$E'' \setminus \cala_{\calh} \wbis{} \hat{E}'' \setminus \cala_{\calh}$.

\item If $E_{1}, E_{2} \in \mathrm{SBNDC}_{\wbis{}}$ and $E_{1} \setminus \cala_{\calh} \wbis{} E_{2}
\setminus \cala_{\calh}$, then $E_{1} \, / \, \cala_{\calh} \wbis{} E_{2} \, / \, \cala_{\calh}$.

\item If $E_{2} \in \mathrm{SBSNNI_{\wbis{}}}$ and $L \subseteq \cala_{\calh}$, then $E'_{1} \setminus
\cala_{\calh} \wbis{} ((E'_{2} \pco{L} F) \, / \, L) \setminus \cala_{\calh}$ for all $F \in \procs$ having
only actions in $\cala_{\calh}$ and for all $E'_{1} \in \reach(E_{1})$ and $E'_{2} \in \reach(E_{2})$ such
that $E'_{1} \setminus \cala_{\calh} \wbis{} E'_{2} \, / \, \cala_{\calh}$, when $E'_{2}, F \in \procs_{\rm
n}$ or $E'_{2}, F \in \procs_{\rm p}$.

		\end{enumerate}

		\begin{proof}

\blue{First we} prove the three results for the $\wbis{\rm pw}$-based properties:

			\begin{enumerate}

\item We proceed by induction on the number $n \in \natns$ of $\tau$- and probabilistic transitions along
$E' \, / \, \cala_{\calh} \warrow{}{} E'' \, / \, \cala_{\calh}$:

				\begin{itemize}

\item If $n = 0$ then $E' \, / \, \cala_{\calh}$ stays idle and $E'' \, / \, \cala_{\calh}$ is $E' \, / \,
\cala_{\calh}$. Likewise, $E' \setminus \cala_{\calh}$ can stay idle, i.e., $E' \setminus \cala_{\calh}
\warrow{}{} E' \setminus \cala_{\calh}$, with $E' \setminus \cala_{\calh} \wbis{\rm pw} E' \setminus
\cala_{\calh}$ as $\wbis{\rm pw}$ is reflexive.

\item Let $n > 0$ and $E'_{0} \, / \, \cala_{\calh} \warrow{}{} E'_{n - 1} \, / \, \cala_{\calh}
\arrow{\tau}{\rm a} E'_{n} \, / \, \cala_{\calh}$ or $E'_{0} \, / \, \cala_{\calh} \warrow{}{} E'_{n - 1} \,
/ \, \cala_{\calh} \arrow{p}{\rm p}$ \linebreak $E'_{n} \, / \, \cala_{\calh}$ where $E'_{0}$ is $E'$ and
$E'_{n}$ is $E''$. From the induction hypothesis it follows that $E' \setminus \cala_{\calh} \warrow{}{}
\hat{E}'_{n - 1} \setminus \cala_{\calh}$ with $E'_{n - 1} \setminus \cala_{\calh} \wbis{\rm pw} \hat{E}'_{n
- 1} \setminus \cala_{\calh}$. As far as the $n$-th transition is concerned, which is $E'_{n - 1} \, / \,
  \cala_{\calh} \arrow{\tau}{\rm a} E'_{n} \, / \, \cala_{\calh}$ or $E'_{n - 1} \, / \, \cala_{\calh}
\arrow{p}{\rm p} E'_{n} \, / \, \cala_{\calh}$, there are three cases depending on whether it is originated
from $E'_{n - 1} \arrow{\tau}{\rm a} E'_{n}$, $E'_{n - 1} \arrow{h}{\rm a} E'_{n}$, or $E'_{n - 1}
\arrow{p}{\rm p} E'_{n}$:

					\begin{itemize}

\item If $E'_{n - 1} \arrow{\tau}{\rm a} E'_{n}$ then $E'_{n - 1} \setminus \cala_{\calh} \arrow{\tau}{\rm
a} E'_{n} \setminus \cala_{\calh}$. Since $E'_{n - 1} \setminus \cala_{\calh} \wbis{\rm pw} \hat{E}'_{n - 1}
\setminus \cala_{\calh}$, \linebreak there exists $\sigma \in \sched(\hat{E}'_{n - 1} \setminus
\cala_{\calh})$ such that $\proba_{\sigma}(\hat{E}'_{n - 1} \setminus \cala_{\calh}, \tau^{*} \, \hat{\tau}
\, \tau^{*}, [E'_{n} \setminus \cala_{\calh}]_{\wbis{\rm pw}}, \linebreak \varepsilon_{\hat{E}'_{n - 1} \,
\setminus \, \cala_{\calh}}) = 1$. Therefore $E' \setminus \cala_{\calh} \warrow{}{} \hat{E}'_{n} \setminus
\cala_{\calh}$ with $E'' \setminus \cala_{\calh} \wbis{\rm pw} \hat{E}'_{n} \setminus \cala_{\calh}$.

\item If $E'_{n - 1} \arrow{h}{\rm a} E'_{n}$ then from $E \in \mathrm{SBNDC}_{\wbis{\rm pw}}$ and $E'_{n -
1}, E'_{n} \in \reach(E)$ it follows that $E'_{n - 1} \setminus \cala_{\calh} \wbis{\rm pw} E'_{n} \setminus
\cala_{\calh}$. Since $E'_{n - 1} \setminus \cala_{\calh} \wbis{\rm pw} \hat{E}'_{n - 1} \setminus
\cala_{\calh}$ and $\wbis{\rm pw}$ is symmetric and transitive, we obtain $E'_{n} \setminus \cala_{\calh}
\wbis{\rm pw} \hat{E}'_{n - 1} \setminus \cala_{\calh}$. Therefore $E' \setminus \cala_{\calh} \warrow{}{}
\hat{E}'_{n - 1} \setminus \cala_{\calh}$ with $E'' \setminus \cala_{\calh} \wbis{\rm pw} \hat{E}'_{n
- 1} \setminus \cala_{\calh}$.

\item If $E'_{n - 1} \arrow{p}{\rm p} E'_{n}$ then from the fact that $E'_{n - 1} \setminus \cala_{\calh}
\wbis{\rm pw} \hat{E}'_{n - 1} \setminus \cala_{\calh}$ it follows that $\proba(E'_{n - 1} \setminus
\cala_{\calh}, C) = \proba(\hat{E}'_{n - 1} \setminus \cala_{\calh}, C)$ for all $C \in \procs / \!
\wbis{\rm pw}$ and hence there exists $\hat{E}'_{n - 1} \setminus \cala_{\calh} \arrow{q}{\rm p}
\hat{E}'_{n} \setminus \cala_{\calh}$ for some $q \in \realns_{]0, 1]}$ such that $\hat{E}'_{n} \setminus
\cala_{\calh} \in [E'_{n} \setminus \cala_{\calh}]_{\wbis{\rm pw}}$. Therefore $E' \setminus \cala_{\calh}
\warrow{}{} \hat{E}'_{n} \setminus \cala_{\calh}$ with $E'' \setminus \cala_{\calh} \wbis{\rm pw}
\hat{E}'_{n} \setminus \cala_{\calh}$.

					\end{itemize}

				\end{itemize}

\item Let $\calb$ be an equivalence relation containing all the pairs of processes mentioned at end of the
considered result that have to be shown to be $\wbis{\rm pw}$-equivalent under the constraints mentioned at
the beginning of the result itself. Starting from $(E_{1} \, / \, \cala_{\calh}, E_{2} \, / \,
\cala_{\calh}) \in \calb$, \linebreak so that $E_{1} \setminus \cala_{\calh} \wbis{\rm pw} E_{2} \setminus
\cala_{\calh}$, there are three cases for action transitions based on the operational semantic rules in
Table~\ref{tab:prob_op_sem_nd}:

				\begin{itemize}

\item If $E_{1} \, / \, \cala_{\calh} \arrow{\tau}{\rm a} E'_{1} \, / \, \cala_{\calh}$ with $E_{1}
\arrow{h}{\rm a} E'_{1}$, then $E_{1} \setminus \cala_{\calh} \wbis{\rm pw} E'_{1} \setminus \cala_{\calh}$
as $h \in \cala_{\calh}$ and $E_{1} \in \mathrm{SBNDC}_{\wbis{\rm pw}}$. Since $E'_{1} \setminus
\cala_{\calh} \wbis{\rm pw} E_{2} \setminus \cala_{\calh}$, as $E_{1} \setminus \cala_{\calh} \wbis{\rm pw}
E_{2} \setminus \cala_{\calh}$ and $\wbis{\rm pw}$ is symmetric and transitive, with $E'_{1}, E_{2} \in
\mathrm{SBNDC}_{\wbis{\rm pw}}$, we have that $(E'_{1} \, / \, \cala_{\calh}, E_{2} \, / \, \cala_{\calh})
\in \calb$ with $\proba_{\sigma}(E_{2} \, / \, \cala_{\calh}, \tau^{*} \, \hat{\tau} \, \tau^{*}, [E'_{1} \,
/ \, \cala_{\calh}]_{\calb}, \varepsilon_{E_{2} \, / \, \cala_{\calh}}) = 1$ for $\sigma \in \sched(E_{2} \,
/ \, \cala_{\calh})$ such that $\sigma(\varepsilon_{E_{2} \, / \, \cala_{\calh}}) = \bot$.

\item If $E_{1} \, / \, \cala_{\calh} \arrow{l}{\rm a} E'_{1} \, / \, \cala_{\calh}$ with $E_{1}
\arrow{l}{\rm a} E'_{1}$, then $E_{1} \setminus \cala_{\calh} \arrow{l}{\rm a} E'_{1} \setminus
\cala_{\calh}$ as $l \notin \cala_{\calh}$. From $E_{1} \setminus \cala_{\calh} \wbis{\rm pw} E_{2}
\setminus \cala_{\calh}$ it follows that there exists $\sigma \in \sched(E_{2} \setminus \cala_{\calh})$
such that $\proba_{\sigma}(E_{2} \setminus \cala_{\calh}, \tau^{*} \, \hat{l} \, \tau^{*}, [E'_{1} \setminus
\cala_{\calh}]_{\wbis{\rm pw}}, \varepsilon_{E_{2} \, \setminus \, \cala_{\calh}}) = 1$.  Thus there exists
$\sigma' \in \sched(E_{2} \, / \, \cala_{\calh})$ such that $\proba_{\sigma'}(E_{2} \, / \, \cala_{\calh},
\tau^{*} \, \hat{l} \, \tau^{*}, [E'_{1} \, / \, \cala_{\calh}]_{\calb}, \varepsilon_{E_{2} \, / \,
\cala_{\calh}}) = 1$ as $l, \tau \notin \cala_{\calh}$. Since $E'_{1} \setminus \cala_{\calh} \wbis{\rm pw}
E'_{2} \setminus \cala_{\calh}$ with $E'_{1}, E'_{2} \in \mathrm{SBNDC}_{\wbis{\rm pw}}$, we have that
$(E'_{1} \, / \, \cala_{\calh}, E'_{2} \, / \, \cala_{\calh}) \in \calb$.

\item If $E_{1} \, / \, \cala_{\calh} \arrow{\tau}{\rm a} E'_{1} \, / \, \cala_{\calh}$ with $E_{1}
\arrow{\tau}{\rm a} E'_{1}$, then the proof is like the one of the previous case.

				\end{itemize}

\noindent
As for probabilities, to avoid trivial cases let $E_{1}, E_{2} \in \procs_{\rm p}$ and consider an
equivalence class $C \in \procs / \calb$ that involves nondeterministic processes reachable from $E_{1}$ or
$E_{2}$, specifically $C = \{ H_{i} \, / \, \cala_{\calh}, H_{j} \, / \, \cala_{\calh} \mid H_{\blue{k}} \in
\reach(E_{\blue{k}}) \land H_{i} \setminus \cala_{\calh} \wbis{\rm pw} H_{j} \setminus \cala_{\calh} \}$.
Since the hiding and restriction operators do not affect probabilistic transitions, we have that:
\cws{0}{\hspace*{-0.6cm}\begin{array}{rcl}
\proba(E_{\blue{k}} \, / \, \cala_{\calh}, C) & \!\! = \!\! & \proba(E_{\blue{k}} \setminus \cala_{\calh},
\bar{C}) \\
\end{array}}
where:
\cws{0}{\hspace*{-0.6cm}\begin{array}{rcl}
\bar{C} & \!\! = \!\! & \{ H_{\blue{k}} \setminus \cala_{\calh} \mid H_{\blue{k}} \, / \, \cala_{\calh} \in
C \} \\
\end{array}}
Since $E_{1} \setminus \cala_{\calh} \wbis{\rm pw} E_{2} \setminus \cala_{\calh}$ and $\bar{C}$ is the union
of some $\wbis{\rm pw}$-equivalence classes, \linebreak we have that:
\cws{10}{\hspace*{-0.6cm}\begin{array}{rcl}
\proba(E_{1} \setminus \cala_{\calh}, \bar{C}) & \!\! = \!\! & \proba(E_{2} \setminus \cala_{\calh},
\bar{C}) \\
\end{array}}

\item Let $\calb$ be an equivalence relation containing all the pairs of processes mentioned at beginning of
the considered result that have to be shown to be $\wbis{\rm pw}$-equivalent under the constraints mentioned
at the end of the result itself. Starting from $E'_{1} \setminus \cala_{\calh}$ and $((E'_{2} \pco{L} F) \,
/ \, L) \setminus \cala_{\calh}$ related by $\calb$, so that $E'_{1} \setminus \cala_{\calh} \wbis{\rm pw}
E'_{2} \, / \, \cala_{\calh}$, there are six cases for action transitions based on the operational semantic
rules in Table~\ref{tab:prob_op_sem_nd}. In the first two cases, it is $E'_{1} \setminus \cala_{\calh}$ to
move first:

				\begin{itemize}

\item Let $E'_{1} \setminus \cala_{\calh} \arrow{l}{\rm a} E''_{1} \setminus \cala_{\calh}$. We observe that
from $E'_{2} \in \reach(E_{2})$ and $E_{2} \in \mathrm{SBSNNI}_{\wbis{\rm pw}}$ it follows that $E'_{2}
\setminus \cala_{\calh} \wbis{\rm pw} E'_{2} \, / \, \cala_{\calh}$, so that $E'_{1} \setminus \cala_{\calh}
\wbis{\rm pw} E'_{2} \, / \, \cala_{\calh} \wbis{\rm pw} E'_{2} \setminus \cala_{\calh}$, i.e., $E'_{1}
\setminus \cala_{\calh} \wbis{\rm pw} E'_{2} \setminus \cala_{\calh}$, as $\wbis{\rm pw}$ is symmetric and
transitive. As a consequence there exists $\sigma \in \sched(E'_{2} \setminus \cala_{\calh})$ such that
$\proba_{\sigma}(E'_{2} \setminus \cala_{\calh}, \tau^{*} \, \hat{l} \, \tau^{*}, [E'_{1} \setminus
\cala_{\calh}]_{\wbis{\rm pw}}, \varepsilon_{E'_{2} \, \setminus \, \cala_{\calh}}) = 1$. Thus there exists
$\sigma' \in \sched(((E'_{2} \pco{L} F) \, / \, L) \setminus \cala_{\calh})$ such that
$\proba_{\sigma'}(((E'_{2} \pco{L} F) \, / \, L) \setminus \cala_{\calh}, \tau^{*} \, \hat{l} \, \tau^{*},
[E''_{1} \setminus \cala_{\calh}]_{\calb}, \varepsilon_{((E'_{2} \pco{L} F) \, / \, L) \, \setminus \,
\cala_{\calh}}) = 1$ as $l, \tau \notin \cala_{\calh} \cup L$ with $E''_{1} \in \reach(E_{1})$ and $E''_{1}
\setminus \cala_{\calh} \wbis{\rm pw} E''_{2} \, / \, \cala_{\calh}$ as $E_{2} \in
\mathrm{SBSNNI}_{\wbis{\rm pw}}$ for all $E''_{2} \in \reach(E_{2})$ such that $E''_{1} \setminus
\cala_{\calh} \wbis{\rm pw} E''_{2} \setminus \cala_{\calh}$, where the right subprocess alternates between
$F$ and $[1] F$ thus allowing the probabilistic transitions along $E'_{2} \setminus \cala_{\calh}
\warrow{l}{} E''_{2} \setminus \cala_{\calh}$ to synchronize with the only one of $[1] F$.

\item Let $E'_{1} \setminus \cala_{\calh} \arrow{\tau}{\rm a} E''_1 \setminus \cala_{\calh}$. The proof is
like the one of the previous case with $\warrow{}{}$ used in place of $\warrow{l}{} \!$.

				\end{itemize}

\noindent
In the other four cases, instead, it is $((E'_{2} \pco{L} F) \, / \, L) \setminus \cala_{\calh}$ to move
first:

				\begin{itemize}

\item Let $((E'_{2} \pco{L} F) \, / \, L) \setminus \cala_{\calh} \arrow{l}{\rm a} ((E''_{2} \pco{L} [1] F)
\, / \, L) \setminus \cala_{\calh}$ with $E'_{2} \arrow{l}{\rm a} E''_{2}$ so that $E'_{2} \setminus
\cala_{\calh} \linebreak \arrow{l}{\rm a} E''_{2} \setminus \cala_{\calh}$ as $l \notin \cala_{\calh}$. We
observe that from $E'_{2} \in \reach(E_{2})$ and $E_{2} \in \mathrm{SBSNNI}_{\wbis{\rm pw}}$ it follows that
$E'_{2} \setminus \cala_{\calh} \wbis{\rm pw} E'_{2} \, / \, \cala_{\calh}$, so that $E'_{2} \setminus
\cala_{\calh} \wbis{\rm pw} E'_{2} \, / \, \cala_{\calh} \wbis{\rm pw} E'_{1} \setminus \cala_{\calh}$,
i.e., $E'_{2} \setminus \cala_{\calh} \wbis{\rm pw} E'_{1} \setminus \cala_{\calh}$, as $\wbis{\rm pw}$ is
symmetric and transitive. As a consequence there exists $\sigma \in \sched(E'_{1} \setminus \cala_{\calh})$
such that $\proba_{\sigma}(E'_{1} \setminus \cala_{\calh}, \tau^{*} \, \hat{l} \, \tau^{*}, [E''_{2}
\setminus \cala_{\calh}]_{\wbis{\rm pw}}, \varepsilon_{E'_{1} \, \setminus \, \cala_{\calh}}) = 1$. Thus
there exists $\sigma' \in \sched(E'_{1} \setminus \cala_{\calh})$ such that $\proba_{\sigma'}(E'_{1}
\setminus \cala_{\calh}, \tau^{*} \, \hat{l} \, \tau^{*}, [((E''_{2} \pco{L} [1] F) \, / \, L) \setminus
\cala_{\calh}]_{\calb}, \varepsilon_{E'_{1} \, \setminus \, \cala_{\calh}}) = 1$ with $E''_{2} \in
\reach(E_{2})$ and $E''_{1} \setminus \cala_{\calh} \wbis{\rm pw} E''_{2} \, / \, \cala_{\calh}$ as $E_{2}
\in \mathrm{SBSNNI}_{\wbis{\rm pw}}$ for all $E''_{1} \in \reach(E_{1})$ such that $E''_{1} \setminus
\cala_{\calh} \wbis{\rm pw} E''_{2} \setminus \cala_{\calh}$.
 
\item Let $((E'_{2} \pco{L} F) \, / \, L) \setminus \cala_{\calh} \arrow{\tau}{\rm a} ((E''_{2} \pco{L} [1]
F) \, / \, L) \setminus \cala_{\calh}$ with $E'_{2} \arrow{\tau}{\rm a} E''_{2}$ so that $E'_{2} \setminus
\cala_{\calh} \linebreak \arrow{\tau}{\rm a} E''_{2} \setminus \cala_{\calh}$ as $\tau \notin
\cala_{\calh}$. The proof is like the one of the previous case.

\item If $((E'_{2} \pco{L} F) \, / \, L) \setminus \cala_{\calh} \arrow{\tau}{\rm a} (([1] E'_{2} \pco{L}
F') \, / \, L) \setminus \cala_{\calh}$ with $F \arrow{\tau}{\rm a} F'$, then trivially \linebreak $((([1]
E'_{2} \pco{L} F') \, / \, L) \setminus \cala_{\calh}, E'_{1} \setminus \cala_{\calh}) \in \calb$ as $E'_{1}
\setminus \cala_{\calh} \wbis{\rm pw} [1] E'_{2} \, / \, \cala_{\calh}$ -- because $[1] E'_{2} \wbis{\rm pw}
E'_{2}$ and hence $[1] E'_{2} \, / \, \cala_{\calh} \wbis{\rm pw} E'_{2} \, / \, \cala_{\calh}$ by
Lemma~\ref{lem:prob_bisim_congr}(4) -- with $\proba_{\sigma}(E'_{1} \setminus \cala_{\calh}, \tau^{*} \,
\hat{\tau} \, \tau^{*},$ \linebreak $[(([1] E'_{2} \pco{L} F') \, / \, L) \setminus \cala_{\calh}]_{\calb},
\varepsilon_{E'_{1} \, \setminus \, \cala_{\calh}}) = 1$ for $\sigma \in \sched(E'_{1} \setminus
\cala_{\calh})$ such that $\sigma(\varepsilon_{E'_{1} \, \setminus \, \cala_{\calh}}) \linebreak = \bot$.

\item Let $((E'_{2} \pco{L} F) \, / \, L) \setminus \cala_{\calh} \arrow{\tau}{\rm a} ((E''_{2} \pco{L} F')
\, / \, L) \setminus \cala_{\calh}$ with $E'_{2} \arrow{h}{\rm a} E''_{2}$ -- so that $E'_{2} \, / \,
\cala_{\calh} \linebreak \arrow{\tau}{\rm a} E''_{2} \, / \, \cala_{\calh}$ as $h \in \cala_{\calh}$ -- and
$F \arrow{h}{\rm a} F'$ for $h \in L$. We observe that from $E'_{2}, E''_{2} \in \reach(E_{2})$ and $E_{2}
\in \mathrm{SBSNNI}_{\wbis{\rm pw}}$ it follows that $E'_{2} \setminus \cala_{\calh} \wbis{\rm pw} E'_{2} \,
/ \, \cala_{\calh}$ and $E''_{2} \setminus \cala_{\calh} \linebreak \wbis{\rm pw} E''_{2} \, / \,
\cala_{\calh}$, so that $E'_{2} \setminus \cala_{\calh} \arrow{\tau}{\rm a} E''_{2} \setminus
\cala_{\calh}$,
as $E'_{2} \, / \, \cala_{\calh}
\arrow{\tau}{\rm a} E''_{2} \, / \, \cala_{\calh}$, and $E'_{2} \setminus \cala_{\calh} \linebreak \wbis{\rm
pw} E'_{2} \, / \, \cala_{\calh} \wbis{\rm pw} E'_{1} \setminus \cala_{\calh}$, i.e., $E'_{2} \setminus
\cala_{\calh} \wbis{\rm pw} E'_{1} \setminus \cala_{\calh}$, as $\wbis{\rm pw}$ is symmetric and transitive.
As a consequence there exists $\sigma \in \sched(E'_{1} \setminus \cala_{\calh})$ such that
$\proba_{\sigma}(E'_{1} \setminus \cala_{\calh}, \linebreak \tau^{*} \, \hat{\tau} \, \tau^{*}, [E''_{2}
\setminus \cala_{\calh}]_{\wbis{\rm pw}}, \varepsilon_{E'_{1} \, \setminus \, \cala_{\calh}}) = 1$. Thus
there exists $\sigma' \in \sched(E'_{1} \setminus \cala_{\calh})$ such that $\proba_{\sigma'}(E'_{1}
\setminus \cala_{\calh}, \tau^{*} \, \hat{\tau} \, \tau^{*}, [((E''_{2} \pco{L} F') \, / \, L) \setminus
\cala_{\calh}]_{\calb}, \varepsilon_{E'_{1} \, \setminus \, \cala_{\calh}}) = 1$ with $E''_{2} \in
\reach(E_{2})$ and $E''_{1} \setminus \cala_{\calh} \wbis{\rm pw} E''_{2} \, / \, \cala_{\calh}$ as $E_{2}
\in \mathrm{SBSNNI}_{\wbis{\rm pw}}$ for all $E''_{1} \in \reach(E_{1})$ such that $E''_{1} \setminus
\cala_{\calh} \wbis{\rm pw} E''_{2} \setminus \cala_{\calh}$.

				\end{itemize}

\noindent
As for probabilities, to avoid trivial cases let $E'_{1}, E'_{2}, F \in \procs_{\rm p}$ and consider an
equivalence class $C \in \procs / \calb$ that involves nondeterministic processes reachable from $E'_{1}
\setminus \cala_{\calh}$ and $((E'_{2} \pco{L} F) \, / \, L) \setminus \cala_{\calh}$, specifically $C = \{
G_{1, i} \setminus \cala_{\calh}, ((G_{2, j} \pco{L} H_{j}) \, / \, L) \setminus \cala_{\calh} \mid H_{j}
\in \procs \linebreak \textrm{having only actions in } \cala_{\calh} \land G_{k, \blue{\iota}} \in
\reach(E_{k}) \land G_{1, i} \setminus \cala_{\calh} \wbis{\rm pw} G_{2, j} \, / \, \cala_{\calh} \}$. If we
focus on a single probabilistic transition of $E'_{2}$, say $E'_{2} \arrow{p}{\rm p} E''_{2}$, then
$((E'_{2} \pco{L} F) \, / \, L) \setminus \cala_{\calh} \linebreak \arrow{p \cdot q_{j}}{\rm p} ((E''_{2}
\pco{L} F_{j}) \, / \, L) \setminus \cala_{\calh}$ for all $F \arrow{q_{j}}{\rm p} F_{j}$. From $\sum_{j \in
J} q_{j} = 1$ it follows that \linebreak $\proba(((E'_{2} \pco{L} F) \, / \, L) \setminus \cala_{\calh}, \{
((E''_{2} \pco{L} F_{j}) \, / \, L) \setminus \cala_{\calh} \mid j \in J \}) = p$, where all processes
$((E''_{2} \pco{L} F_{j}) \, / \, L)\setminus \cala_{\calh}$ belong to the same equivalence class of $\calb$
because each $F_{j}$ has only actions in $\cala_{\calh}$. Since the restriction and hiding operators do not
affect probabilistic transitions, we have that:
\cws{0}{\hspace*{-0.6cm}\begin{array}{rcl}
\proba(E'_{1} \setminus \cala_{\calh}, C) & \!\! = \!\! & \proba(E'_{1} \setminus \cala_{\calh}, \bar{C}) \\
\proba(((E'_{2} \pco{L} F) \, / \, L) \setminus \cala_{\calh}, C) & \!\! = \!\! & \proba(E'_{2} \, / \,
\cala_{\calh}, \bar{C}) \\
\end{array}}
where:
\cws{0}{\hspace*{-0.6cm}\begin{array}{rcl}
\bar{C} & \!\! = \!\! & \{ G_{1, i} \setminus \cala_{\calh} \in C \} \cup \{ G_{2, j} \, / \, \cala_{\calh}
\mid ((G_{2, j} \pco{L} H_{j}) \, / \, L) \setminus \cala_{\calh} \in C \} \\
\end{array}}
Since $E'_{1} \setminus \cala_{\calh} \wbis{\rm pw} E'_{2} \, / \, \cala_{\calh}$ and $\bar{C}$ is the union
of some $\wbis{\rm pw}$-equivalence classes, \linebreak we have that:
\cws{8}{\hspace*{-0.6cm}\begin{array}{rcl}
\proba(E'_{1} \setminus \cala_{\calh}, \bar{C}) & \!\! = \!\! & \proba(E'_{2} \, / \, \cala_{\calh},
\bar{C}) \\
\end{array}}

			\end{enumerate}

\noindent
\blue{Next we} prove the three results for the $\wbis{\rm pb}$-based properties:

			\begin{enumerate}

\item We proceed by induction on the number $n \in \natns$ of $\tau$- and probabilistic transitions along
$E' \, / \, \cala_{\calh} \warrow{}{} E'' \, / \, \cala_{\calh}$:

				\begin{itemize}

\item If $n = 0$ then the proof is like the one of the corresponding result for $\wbis{\rm pw}$.

\item Let $n > 0$ and $E'_{0} \, / \, \cala_{\calh} \warrow{}{} E'_{n - 1} \, / \, \cala_{\calh}
\arrow{\tau}{\rm a} E'_{n} \, / \, \cala_{\calh}$ or $E'_{0} \, / \, \cala_{\calh} \warrow{}{} E'_{n - 1} \,
/ \, \cala_{\calh} \arrow{p}{\rm p}$ \linebreak $E'_{n} \, / \, \cala_{\calh}$ where $E'_{0}$ is $E'$ and
$E'_{n}$ is $E''$. From the induction hypothesis it follows that $E' \setminus \cala_{\calh} \warrow{}{}
\hat{E}'_{n - 1} \setminus \cala_{\calh}$ with $E'_{n - 1} \setminus \cala_{\calh} \wbis{\rm pb} \hat{E}'_{n
- 1} \setminus \cala_{\calh}$. The rest of the proof is like the one of the corresponding result for
  $\wbis{\rm pw}$ with the following difference:

					\begin{itemize}

\item If $E'_{n - 1} \arrow{\tau}{\rm a} E'_{n}$ then $E'_{n - 1} \setminus \cala_{\calh} \arrow{\tau}{\rm
a} E'_{n} \setminus \cala_{\calh}$. Since $E'_{n - 1} \setminus \cala_{\calh} \wbis{\rm pb} \hat{E}'_{n - 1}
\setminus \cala_{\calh}$:

						\begin{itemize}

\item either $E'_{n} \setminus \cala_{\calh} \wbis{\rm pb} \hat{E}'_{n - 1} \setminus \cala_{\calh}$, in
which case $\hat{E}'_{n - 1} \setminus \cala_{\calh}$ stays idle and hence $E' \setminus \cala_{\calh}
\warrow{}{} \hat{E}'_{n - 1} \setminus \cala_{\calh}$ with $E'' \setminus \cala_{\calh} \wbis{\rm pb}
\hat{E}'_{n - 1} \setminus \cala_{\calh}$;

\item or there exists $\hat{E}'_{n - 1} \setminus \cala_{\calh} \warrow{}{} \bar{E}_{n - 1} \setminus
\cala_{\calh} \arrow{\tau}{\rm a} \hat{E}'_{n} \setminus \cala_{\calh}$ such that $E'_{n - 1} \setminus
\cala_{\calh} \wbis{\rm pb} \bar{E}_{n - 1} \setminus \cala_{\calh}$ and $E'_{n} \setminus \cala_{\calh}
\wbis{\rm pb} \hat{E}'_{n} \setminus \cala_{\calh}$, hence $E' \setminus \cala_{\calh} \warrow{}{}
\hat{E}'_{n} \setminus \cala_{\calh}$ with $E'' \setminus \cala_{\calh} \wbis{\rm pb} \hat{E}'_{n} \setminus
\cala_{\calh}$.

						\end{itemize}

					\end{itemize}

				\end{itemize}

\item Let $\calb$ be an equivalence relation containing all the pairs of processes mentioned at end of the
considered result that have to be shown to be $\wbis{\rm pb}$-equivalent under the constraints mentioned at
the beginning of the result itself. Starting from $(E_{1} \, / \, \cala_{\calh}, E_{2} \, / \,
\cala_{\calh}) \in \calb$, so that $E_{1} \setminus \cala_{\calh} \wbis{\rm pb} E_{2} \setminus
\cala_{\calh}$, there are three cases for action transitions based on the operational semantic rules in
Table~\ref{tab:prob_op_sem_nd}:

				\begin{itemize}

\item If $E_{1} \, / \, \cala_{\calh} \arrow{\tau}{\rm a} E'_{1} \, / \, \cala_{\calh}$ with $E_{1}
\arrow{h}{\rm a} E'_{1}$, then the proof is like the one of the corresponding result for $\wbis{\rm pw}$
with $E_{2} \, / \, \cala_{\calh}$ staying idle.

\item If $E_{1} \, / \, \cala_{\calh} \arrow{l}{\rm a} E'_{1} \, / \, \cala_{\calh}$ with $E_{1}
\arrow{l}{\rm a} E'_{1}$, then $E_{1} \setminus \cala_{\calh} \arrow{l}{\rm a} E'_{1} \setminus
\cala_{\calh}$ as $l \notin \cala_{\calh}$. From $E_{1} \setminus \cala_{\calh} \wbis{\rm pb} E_{2}
\setminus \cala_{\calh}$ it follows that there exists $E_{2} \setminus \cala_{\calh} \warrow{}{} \bar{E}_{2}
\setminus \cala_{\calh} \arrow{l}{\rm a} E'_{2} \setminus \cala_{\calh}$ such that $E_{1} \setminus
\cala_{\calh} \wbis{\rm pb} \bar{E}_{2} \setminus \cala_{\calh}$ and $E'_{1} \setminus \cala_{\calh}
\wbis{\rm pb} E'_{2} \setminus \cala_{\calh}$. Thus $E_{2} \, / \, \cala_{\calh} \warrow{}{} \bar{E}_{2} \,
/ \, \cala_{\calh} \arrow{l}{\rm a}$ \linebreak $E'_{2} \, / \, \cala_{\calh}$ as $l, \tau \notin
\cala_{\calh}$. Since $E_{1} \setminus \cala_{\calh} \wbis{\rm pb} \bar{E}_{2} \setminus \cala_{\calh}$ with
$E_{1}, \bar{E}_{2} \in \mathrm{SBNDC}_{\wbis{\rm pb}}$ and $E'_{1} \setminus \cala_{\calh} \wbis{\rm pb}
E'_{2} \setminus \cala_{\calh}$ with $E'_{1}, E'_{2} \in \mathrm{SBNDC}_{\wbis{\rm pb}}$, we have that
$(E_{1} \, / \, \cala_{\calh}, \bar{E}_{2} \, / \, \cala_{\calh}) \in \calb$ and $(E'_{1} \, / \,
\cala_{\calh}, E'_{2} \, / \, \cala_{\calh}) \in \calb$.

\item If $E_{1} \, / \, \cala_{\calh} \arrow{\tau}{\rm a} E'_{1} \, / \, \cala_{\calh}$ with $E_{1}
\arrow{\tau}{\rm a} E'_{1}$, then $E_{1} \setminus \cala_{\calh} \arrow{\tau}{\rm a} E'_{1} \setminus
\cala_{\calh}$ as $\tau \notin \cala_{\calh}$. There are two subcases:

					\begin{itemize}

\item If $E'_{1} \setminus \cala_{\calh} \wbis{\rm pb} E_{2} \setminus \cala_{\calh}$ then $E_{2} \setminus
\cala_{\calh}$ is allowed to stay idle with $(E'_{1} \, / \, \cala_{\calh}, E_{2} \, / \, \cala_{\calh}) \in
\calb$ because $E'_{1} \setminus \cala_{\calh} \wbis{\rm pb} E_{2} \setminus \cala_{\calh}$ and $E'_{1},
E_{2} \in \mathrm{SBNDC}_{\wbis{\rm pb}}$.

\item If $E'_{1} \setminus \cala_{\calh} \not\wbis{\rm pb} E_{2} \setminus \cala_{\calh}$ then the proof is
like the one of the previous case with $\arrow{\tau}{\rm a} \!$ used in place of $\arrow{l}{\rm a} \!$.

					\end{itemize}

				\end{itemize}

\noindent
As for probabilities, we reason like in the proof of the corresponding result for $\wbis{\rm pw}$.

\item Let $\calb$ be an equivalence relation containing all the pairs of processes mentioned at beginning of
the considered result that have to be shown to be $\wbis{\rm pb}$-equivalent under the constraints mentioned
at the end of the result itself. Starting from $E'_{1} \setminus \cala_{\calh}$ and $((E'_{2} \pco{L} F) \,
/ \, L) \setminus \cala_{\calh}$ related by $\calb$, so that $E'_{1} \setminus \cala_{\calh} \wbis{\rm pb}
E'_{2} \, / \, \cala_{\calh}$, there are six cases for action transitions based on the operational semantic
rules in Table~\ref{tab:prob_op_sem_nd}. In the first two cases, it is $E'_{1} \setminus \cala_{\calh}$ to
move first:

				\begin{itemize}

\item Let $E'_{1} \setminus \cala_{\calh} \arrow{l}{\rm a} E''_{1} \setminus \cala_{\calh}$. We observe that
from $E'_{2} \in \reach(E_{2})$ and $E_{2} \in \mathrm{SBSNNI}_{\wbis{\rm pb}}$ it follows that $E'_{2}
\setminus \cala_{\calh} \wbis{\rm pb} E'_{2} \, / \, \cala_{\calh}$, so that $E'_{1} \setminus \cala_{\calh}
\wbis{\rm pb} E'_{2} \, / \, \cala_{\calh} \wbis{\rm pb} E'_{2} \setminus \cala_{\calh}$, i.e., $E'_{1}
\setminus \cala_{\calh} \wbis{\rm pb} E'_{2} \setminus \cala_{\calh}$, as $\wbis{\rm pb}$ is symmetric and
transitive. As a consequence, since $l \neq \tau$ there exists $E'_{2} \setminus \cala_{\calh} \warrow{}{}
\bar{E}'_{2} \setminus \cala_{\calh} \arrow{l}{\rm a} E''_{2} \setminus \cala_{\calh}$ such that $E'_{1}
\setminus \cala_{\calh} \wbis{\rm pb} \bar{E}'_{2} \setminus \cala_{\calh}$ and $E''_{1} \setminus
\cala_{\calh} \wbis{\rm pb} E''_{2} \setminus \cala_{\calh}$. Thus $((E'_{2} \pco{L} F) \, / \, L) \setminus
\cala_{\calh} \warrow{}{} ((\bar{E}'_{2} \pco{L} F) \, / \, L) \setminus \cala_{\calh} \linebreak
\arrow{l}{\rm a} ((E''_{2} \pco{L} F) \, / \, L) \setminus \cala_{\calh}$ with $(E'_{1} \setminus
\cala_{\calh}, ((\bar{E}'_{2} \pco{L} F) \, / \, L) \setminus \cala_{\calh}) \in \calb$ -- because $E'_{1}
\in \reach(E_{1})$, $\bar{E}'_{2} \in \reach(E_{2})$, and $E'_{1} \setminus \cala_{\calh} \wbis{\rm pb}
\bar{E}'_{2} \, / \, \cala_{\calh}$ as $E_{2} \in \mathrm{SBSNNI}_{\wbis{\rm pb}}$ -- and $(E''_{1}
\setminus \cala_{\calh}, ((E''_{2} \pco{L} F) \, / \, L) \setminus \cala_{\calh}) \in \calb$ -- because
$E''_{1} \in \reach(E_{1})$, $E''_{2} \in \reach(E_{2})$, and $E''_{1} \setminus \cala_{\calh} \linebreak
\wbis{\rm pb} E''_{2} \, / \, \cala_{\calh}$ as $E_{2} \in \mathrm{SBSNNI}_{\wbis{\rm pb}}$ -- where the
right subprocess alternates between $F$ and $[1] F$ thus allowing the probabilistic transitions along
$E'_{2} \setminus \cala_{\calh} \warrow{}{} \bar{E}'_{2} \setminus \cala_{\calh}$ \linebreak to synchronize
with the only one of $[1] F$.

\item If $E'_{1} \setminus \cala_{\calh} \arrow{\tau}{\rm a} E''_{1} \setminus \cala_{\calh}$ there are two
subcases:

					\begin{itemize}

\item If $E''_{1} \setminus \cala_{\calh} \wbis{\rm pb} E'_{2} \, / \, \cala_{\calh}$ then $(E'_{2} \pco{L}
F) \, / \, L) \setminus \cala_{\calh}$ is allowed to stay idle with \linebreak $(E''_{1} \setminus
\cala_{\calh}, ((E'_{2} \pco{L} F) \, / \, L) \setminus \cala_{\calh}) \in \calb$ because $E''_{1} \in
\reach(E_{1})$ and $E'_{2} \in \reach(E_{2})$.

\item If $E''_{1} \setminus \cala_{\calh} \not\wbis{\rm pb} E'_{2} \, / \, \cala_{\calh}$ then the proof is
like the one of the previous case with $\arrow{\tau}{\rm a} \!$ used in place of $\arrow{l}{\rm a} \!$.

					\end{itemize}

				\end{itemize}

\noindent
In the other four cases, instead, it is $((E'_{2} \pco{L} F) \, / \, L) \setminus \cala_{\calh}$ to move
first:

				\begin{itemize}

\item Let $((E'_{2} \pco{L} F) \, / \, L) \setminus \cala_{\calh} \arrow{l}{\rm a} ((E''_{2} \pco{L} [1] F)
\, / \, L) \setminus \cala_{\calh}$ with $E'_{2} \arrow{l}{\rm a} E''_{2}$ so that $E'_{2} \setminus
\cala_{\calh} \linebreak \arrow{l}{\rm a} E''_{2} \setminus \cala_{\calh}$ as $l \notin \cala_{\calh}$. We
observe that from $E'_{2} \in \reach(E_{2})$ and $E_{2} \in \mathrm{SBSNNI}_{\wbis{\rm pb}}$ it follows that
$E'_{2} \setminus \cala_{\calh} \wbis{\rm pb} E'_{2} \, / \, \cala_{\calh}$, so that $E'_{2} \setminus
\cala_{\calh} \wbis{\rm pb} E'_{2} \, / \, \cala_{\calh} \wbis{\rm pb} E'_{1} \setminus \cala_{\calh}$,
i.e., $E'_{2} \setminus \cala_{\calh} \wbis{\rm pb} E'_{1} \setminus \cala_{\calh}$, as $\wbis{\rm pb}$ is
symmetric and transitive. As a consequence, since $l \neq \tau$ there exists $E'_{1} \setminus \cala_{\calh}
\warrow{}{} \bar{E}'_{1} \setminus \cala_{\calh} \arrow{l}{\rm a} E''_{1} \setminus \cala_{\calh}$ such that
$E'_{2} \setminus \cala_{\calh} \wbis{\rm pb} \bar{E}'_{1} \setminus \cala_{\calh}$ and $E''_{2} \setminus
\cala_{\calh} \wbis{\rm pb} E''_{1} \setminus \cala_{\calh}$. Thus $(((E'_{2} \pco{L} F) \, / \, L)
\setminus \cala_{\calh}, \bar{E}'_{1} \setminus \cala_{\calh}) \in \calb$ -- because $\bar{E}'_{1} \in
\reach(E_{1})$, $E'_{2} \in \reach(E_{2})$, and $\bar{E}'_{1} \setminus \cala_{\calh} \wbis{\rm pb} E'_{2}
\, / \, \cala_{\calh}$ as $E_{2} \in \mathrm{SBSNNI}_{\wbis{\rm pb}}$ -- and $(((E''_{2} \pco{L} [1] F) \, /
\, L) \setminus \cala_{\calh}, E''_{1} \setminus \cala_{\calh}) \in \calb$ -- because $E''_{1} \in
\reach(E_{1})$, $E''_{2} \in \reach(E_{2})$, and $E''_{1} \setminus \cala_{\calh} \wbis{\rm pb} E''_{2} \, /
\, \cala_{\calh}$ as $E_{2} \in \mathrm{SBSNNI}_{\wbis{\rm pb}}$.
 
\item If $((E'_{2} \pco{L} F) \, / \, L) \setminus \cala_{\calh} \arrow{\tau}{\rm a} ((E''_{2} \pco{L} [1]
F) \, / \, L) \setminus \cala_{\calh}$ with $E'_{2} \arrow{\tau}{\rm a} E''_{2}$ so that $E'_{2} \setminus
\cala_{\calh} \linebreak \arrow{\tau}{\rm a} E''_{2} \setminus \cala_{\calh}$ as $\tau \notin
\cala_{\calh}$, there are two subcases:

					\begin{itemize}

\item If $E''_{2} \setminus \cala_{\calh} \wbis{\rm pb} E'_{1} \setminus \cala_{\calh}$ then $E'_{1}
\setminus \cala_{\calh}$ is allowed to stay idle with $(((E''_{2} \pco{L} [1] F) \, / \, L) \setminus
\cala_{\calh}, E'_{1} \setminus \cala_{\calh}) \in \calb$ because $E'_{1} \in \reach(E_{1})$, $E''_{2} \in
\reach(E_{2})$, and $E'_{1} \setminus \cala_{\calh} \wbis{\rm pb} E''_{2} \, / \, \cala_{\calh}$ as $E_{2}
\in \mathrm{SBSNNI}_{\wbis{\rm pb}}$.

\item If $E''_{2} \setminus \cala_{\calh} \not\wbis{\rm pb} E'_{1} \setminus \cala_{\calh}$ then the proof
is like the one of the previous case with $\arrow{\tau}{\rm a} \!$ used in place of $\arrow{l}{\rm a} \!$.

					\end{itemize}

\item If $((E'_{2} \pco{L} F) \, / \, L) \setminus \cala_{\calh} \arrow{\tau}{\rm a} (([1] E'_{2} \pco{L}
F') \, / \, L) \setminus \cala_{\calh}$ with $F \arrow{\tau}{\rm a} F'$, then trivially \linebreak $((([1]
E'_{2} \pco{L} F') \, / \, L) \setminus \cala_{\calh}, E'_{1} \setminus \cala_{\calh}) \in \calb$ as $E'_{1}
\setminus \cala_{\calh} \wbis{\rm pb} [1] E'_{2} \, / \, \cala_{\calh}$ -- because $[1] E'_{2} \wbis{\rm pb}
E'_{2}$ and hence $[1] E'_{2} \, / \, \cala_{\calh} \wbis{\rm pb} E'_{2} \, / \, \cala_{\calh}$ by
Lemma~\ref{lem:prob_bisim_congr}(4) -- with $E'_{1} \setminus \cala_{\calh}$ staying idle.

\item Let $((E'_{2} \pco{L} F) \, / \, L) \setminus \cala_{\calh} \arrow{\tau}{\rm a} ((E''_{2} \pco{L} F')
\, / \, L) \setminus \cala_{\calh}$ with $E'_{2} \arrow{h}{\rm a} E''_{2}$ -- so that $E'_{2} \, / \,
\cala_{\calh} \linebreak \arrow{\tau}{\rm a} E''_{2} \, / \, \cala_{\calh}$ as $h \in \cala_{\calh}$ -- and
$F \arrow{h}{\rm a} F'$ for $h \in L$. We observe that from $E'_{2}, E''_{2} \in \reach(E_{2})$ and $E_{2}
\in \mathrm{SBSNNI}_{\wbis{\rm pb}}$ it follows that $E'_{2} \setminus \cala_{\calh} \wbis{\rm pb} E'_{2} \,
/ \, \cala_{\calh}$ and $E''_{2} \setminus \cala_{\calh} \wbis{\rm pb} E''_{2} \, / \, \cala_{\calh}$, so
that $E'_{2} \setminus \cala_{\calh} \arrow{\tau}{\rm a} E''_{2} \setminus \cala_{\calh}$, as $E'_{2} \, /
\, \cala_{\calh} \arrow{\tau}{\rm a} E''_{2} \, / \, \cala_{\calh}$, and $E'_{2} \setminus \cala_{\calh}
\wbis{\rm pb} E'_{2} \, / \, \cala_{\calh} \wbis{\rm pb} E'_{1} \setminus \cala_{\calh}$, i.e., $E'_{2}
\setminus \cala_{\calh} \wbis{\rm pb} E'_{1} \setminus \cala_{\calh}$, as $\wbis{\rm pb}$ is symmetric and
transitive. There are two subcases:

					\begin{itemize}

\item If $E''_{2} \setminus \cala_{\calh} \wbis{\rm pb} E'_{1} \setminus \cala_{\calh}$ then $E'_{1}
\setminus \cala_{\calh}$ is allowed to stay idle with $(((E''_{2} \pco{L} F') \, / \, L) \setminus
\cala_{\calh}, E'_{1} \setminus \cala_{\calh}) \in \calb$ because $E'_{1} \in \reach(E_{1})$, $E''_{2} \in
\reach(E_{2})$, and $E'_{1} \setminus \cala_{\calh} \wbis{\rm pb} E''_{2} \, / \, \cala_{\calh}$ as $E_{2}
\in \mathrm{SBSNNI}_{\wbis{\rm pb}}$.

\item If $E''_{2} \setminus \cala_{\calh} \not\wbis{\rm pb} E'_{1} \setminus \cala_{\calh}$ then there
exists $E'_{1} \setminus \cala_{\calh} \warrow{}{} \bar{E}'_{1} \setminus \cala_{\calh} \arrow{\tau}{\rm a}
E''_{1} \setminus \cala_{\calh}$ such that $E'_{2} \setminus \cala_{\calh} \wbis{\rm pb} \bar{E}'_{1}
\setminus \cala_{\calh}$ and $E''_{2} \setminus \cala_{\calh} \wbis{\rm pb} E''_{1} \setminus
\cala_{\calh}$. Thus $(((E'_{2} \pco{L} F) \, / \, L) \setminus \cala_{\calh}, \linebreak \bar{E}'_{1}
\setminus \cala_{\calh}) \in \calb$ -- because $\bar{E}'_{1} \in \reach(E_{1})$, $E'_{2} \in \reach(E_{2})$,
and $\bar{E}'_{1} \setminus \cala_{\calh} \wbis{\rm pb} E'_{2} \, / \, \cala_{\calh}$ as $E_{2} \in
\mathrm{SBSNNI}_{\wbis{\rm pb}}$ -- and $(((E''_{2} \pco{L} F') \, / \, L) \setminus \cala_{\calh}, E''_{1}
\setminus \cala_{\calh}) \in \calb$ -- because $E''_{1} \in \reach(E_{1})$, $E''_{2} \in \reach(E_{2})$, and
$E''_{1} \setminus \cala_{\calh} \wbis{\rm pb} E''_{2} \, / \, \cala_{\calh}$ as $E_{2} \in
\mathrm{SBSNNI}_{\wbis{\rm pb}}$.

					\end{itemize}

				\end{itemize}

\noindent
As for probabilities, we reason like in the proof of the corresponding result for $\wbis{\rm pw}$.
\qedhere 

			\end{enumerate}

		\end{proof}

	\end{lem}

	\begin{thm}\label{thm:prob_taxonomy_1}

Let $\wbis{} \: \in \{ \wbis{\rm pw}, \wbis{\rm pb} \}$. Then:
\cws{12}{\mathrm{SBNDC}_{\wbis{}} \subsetneq \mathrm{SBSNNI}_{\wbis{}} = \mathrm{P\_BNDC}_{\wbis{}}
\subsetneq \mathrm{BNDC}_{\wbis{}} \subsetneq \mathrm{BSNNI}_{\wbis{}}}

		\begin{proof}

\blue{First we} prove the relationships for the $\wbis{\rm pw}$-based properties. Let us examine each
relationship separately:

			\begin{itemize}

\item SBNDC$_{\wbis{\rm pw}}$ $\subsetneq$ SBSNNI$_{\wbis{\rm pw}}$.
Given $E \in \mathrm{SBNDC}_{\wbis{\rm pw}}$, the result follows by proving that the relation $\calb = \{
(E' \setminus \cala_{\calh}, E' \, / \, \cala_{\calh}) \mid E' \in \reach(E) \}$ is a weak probabilistic
bisimulation up to $\wbis{\rm pw}$. Starting from $(E' \setminus \cala_{\calh}, E' \, / \, \cala_{\calh})
\in \calb$, there are three cases for action transitions based on the operational semantic rules in
Table~\ref{tab:prob_op_sem_nd}. In the first case, \linebreak it is $E' \setminus \cala_{\calh}$ to move
first:

				\begin{itemize}

\item If $E' \setminus \cala_{\calh} \warrow{a}{} E'' \setminus \cala_{\calh}$ with $a \in \cala_{\call}
\cup \{ \tau \}$ \blue{due to some $\sigma_{1} \in \sched(E' \setminus \cala_{\calh})$}, then $E' \, / \,
\cala_{\calh} \warrow{\hat{a}}{} E'' \, / \, \cala_{\calh}$ as $a, \tau \notin \cala_{\calh}$, with $(E''
\setminus \cala_{\calh}, E'' \, / \, \cala_{\calh}) \in \calb$ as $E'' \in \reach(E)$. Thus $(E'' \setminus
\cala_{\calh}, E'' \, / \, \cala_{\calh}) \in (\calb \cup \calb^{-1} \cup \wbis{\rm pw})^{+}$ and there
exist\blue{s} $\sigma_{2} \in \sched(E' \, / \, \cala_{\calh})$ \linebreak such that $\proba_{\sigma_{1}}(E'
\setminus \cala_{\calh}, \tau^{*} \, a \, \tau^{*}, \blue{\{ E'' \setminus \cala_{\calh} \}},
\varepsilon_{E' \, \setminus \, \cala_{\calh}}) = \proba_{\sigma_{2}}(E' \, / \, \cala_{\calh}, \tau^{*} \,
\hat{a} \, \tau^{*}, \linebreak [E'' \setminus \cala_{\calh}]_{(\calb \cup \calb^{-1} \cup \wbis{\rm
pw})^{+}}, \varepsilon_{E' \, / \, \cala_{\calh}})$.

				\end{itemize}

\noindent
In the other two cases, instead, it is $E' \, / \, \cala_{\calh}$ to move first (note that possible
$\tau$-transitions along $\warrow{}{}$ arising from high actions in $E'$ can\blue{not} be executed \blue{by}
$E' \setminus \cala_{\calh}$, but for them we exploit $E \in \mathrm{SBNDC}_{\wbis{\rm pw}}$ and
Lemma~\ref{lem:prob_taxonomy}(1)):

				\begin{itemize}

\item If $E' \, / \, \cala_{\calh} \warrow{a}{} E'' \, / \, \cala_{\calh}$ with $a \in \cala_{\call} \cup \{
\tau \}$ \blue{due to some $\sigma_{1} \in \sched(E' \, / \, \cala_{\calh})$}, then there exist two
processes $\bar{E}', \bar{E}'' \in \reach(E')$ such that $E' \, / \, \cala_{\calh} \warrow{}{} \bar{E}' /
\cala_{\calh} \arrow{a}{\rm a} \bar{E}'' \, / \, \cala_{\calh} \warrow{}{}$ \linebreak $E'' \, / \,
\cala_{\calh}$. From $E' \, / \, \cala_{\calh} \warrow{}{} \bar{E}' \, / \, \cala_{\calh}$ and
Lemma~\ref{lem:prob_taxonomy}(1) it follows that $E' \setminus \cala_{\calh} \warrow{}{}$ \linebreak
$\hat{E}' \setminus \cala_{\calh}$ with $\bar{E}' \setminus \cala_{\calh} \wbis{\rm pw} \hat{E}' \setminus
\cala_{\calh}$. From $\bar{E}' \, / \, \cala_{\calh} \arrow{a}{\rm a} \bar{E}'' \, / \, \cala_{\calh}$ it
follows that \linebreak $\bar{E}' \setminus \cala_{\calh} \arrow{a}{\rm a} \bar{E}'' \setminus
\cala_{\calh}$ as $a \notin \cala_{\calh}$, hence $\hat{E}' \setminus \cala_{\calh} \warrow{\hat{a}}{}
\hat{E}'' \setminus \cala_{\calh}$ with $\bar{E}'' \setminus \cala_{\calh} \wbis{\rm pw} \hat{E}'' \setminus
\cala_{\calh}$ as $\bar{E}' \setminus \cala_{\calh} \wbis{\rm pw} \hat{E}' \setminus \cala_{\calh}$. From
$\bar{E}'' \, / \, \cala_{\calh} \warrow{}{} E'' / \cala_{\calh}$ and Lemma~\ref{lem:prob_taxonomy}(1) it
follows that \linebreak $\bar{E}'' \setminus \cala_{\calh} \warrow{}{} \hat{E}''' \setminus \cala_{\calh}$
with $E'' \setminus \cala_{\calh} \wbis{\rm pw} \hat{E}''' \setminus \cala_{\calh}$, hence \linebreak
$\hat{E}'' \setminus \cala_{\calh} \warrow{}{} \hat{E}'''' \setminus \cala_{\calh}$ with $\hat{E}'''
\setminus \cala_{\calh} \wbis{\rm pw} \hat{E}'''' \setminus \cala_{\calh}$ as $\bar{E}'' \setminus
\cala_{\calh} \wbis{\rm pw} \hat{E}'' \setminus \cala_{\calh}$. Note that $E'' \setminus \cala_{\calh}
\wbis{\rm pw} \hat{E}'''' \setminus \cala_{\calh}$ as $\wbis{\rm pw}$ is transitive. Summing up, we have
that \linebreak $E' \setminus \cala_{\calh} \warrow{\hat{a}}{} \hat{E}'''' \setminus \cala_{\calh}$ with
$E'' \, / \, \cala_{\calh} \; \calb^{-1} \, E'' \setminus \cala_{\calh} \wbis{\rm pw} \hat{E}'''' \setminus
\cala_{\calh}$, as $E'' \in \reach(E)$, and hence $(E'' \, / \, \cala_{\calh}, \hat{E}'''' \setminus
\cala_{\calh}) \in (\calb \cup \calb^{-1} \, \cup \wbis{\rm pw})^{+}$ and there exist\blue{s} $\sigma_{2}
\in \sched(E' \setminus \cala_{\calh})$ \linebreak such that $\proba_{\sigma_{1}}(E' \, / \, \cala_{\calh},
\tau^{*} \, a \, \tau^{*}, \blue{\{ E'' \, / \cala_{\calh} \}}, \varepsilon_{E' \, / \, \cala_{\calh}}) =
\proba_{\sigma_{2}}(E' \setminus \cala_{\calh}, \tau^{*} \, \hat{a} \, \tau^{*}, \linebreak [E'' \, / \,
\cala_{\calh}]_{(\calb \cup \calb^{-1} \cup \wbis{\rm pw})^{+}}, \varepsilon_{E' \, \setminus \,
\cala_{\calh}})$.

\item If $E' \, / \, \cala_{\calh} \warrow{\tau}{} E'' \, / \, \cala_{\calh}$ \blue{-- due to some
$\sigma_{1} \in \sched(E' \, / \, \cala_{\calh})$ --} stems from $\bar{E}' \arrow{h}{\rm a} \bar{E}''$ for
some $\bar{E}', \bar{E}'' \in \reach(E')$, then from Lemma~\ref{lem:prob_taxonomy}(1) it follows that $E'
\setminus \cala_{\calh} \warrow{}{}$ \linebreak $\hat{E}'' \setminus \cala_{\calh}$ with $E'' \setminus
\cala_{\calh} \wbis{\rm pw} \hat{E}'' \setminus \cala_{\calh}$. Since $E'' \, / \, \cala_{\calh} \;
\calb^{-1} \, E'' \setminus \cala_{\calh} \wbis{\rm pw} \hat{E}'' \setminus \cala_{\calh}$ as $E'' \in
\reach(E)$, we have that $(E'' \, / \, \cala_{\calh}, \hat{E}'' \setminus \cala_{\calh}) \in (\calb \cup
\calb^{-1} \cup \wbis{\rm pw})^{+}$ and there exist\blue{s} $\sigma_{2} \in \sched(E' \setminus
\cala_{\calh})$ such that $\proba_{\sigma_{1}}(E' \, / \, \cala_{\calh}, \tau^{*} \, a \, \tau^{*}, \blue{\{
E'' \, / \cala_{\calh} \}}, \varepsilon_{E' \, / \, \cala_{\calh}}) = \linebreak \proba_{\sigma_{2}}(E'
\setminus \cala_{\calh}, \tau^{*} \, \hat{a} \, \tau^{*}, [E'' \, / \, \cala_{\calh}]_{(\calb \cup
\calb^{-1} \cup \wbis{\rm pw})^{+}}, \varepsilon_{E' \, \setminus \, \cala_{\calh}})$.

				\end{itemize}

\noindent
As for probabilities, to avoid trivial cases let $E' \in \procs_{\rm p}$ and consider an equivalence class
$C \in \procs / (\calb \cup \calb^{-1} \, \cup \wbis{\rm pw})^{+}$ that involves nondeterministic processes
reachable from $E'$, specifically $C = \{ H \setminus \cala_{\calh}, H \, / \, \cala_{\calh} \mid H \in
\reach(E') \}$. Since the restriction and hiding operators do not affect probabilistic transitions, we have
that:
\cws{10}{\hspace*{-0.6cm}\begin{array}{rcl}
\proba(E_{h} \setminus \cala_{\calh}, C) & \!\! = \!\! & \proba(E_{h} \, / \, \cala_{\calh}, C) \\
\end{array}}

\item SBSNNI$_{\wbis{\rm pw}}$ = P\_BNDC$_{\wbis{\rm pw}}$.
$\mathrm{SBSNNI}_{\wbis{\rm pw}} \subseteq \mathrm{P\_BNDC}_{\wbis{\rm pw}}$ follows from
Lemma~\ref{lem:prob_taxonomy}(3) \linebreak by taking $E'_{1}$ identical to $E'_{2}$ and both reachable from
$E \in \mathrm{SBSNNI}_{\wbis{\rm pw}}$. \\
On the other hand, if $E \in \mathrm{P\_BNDC}_{\wbis{\rm pw}}$ then $E' \in \mathrm{BNDC}_{\wbis{\rm pw}}$
for every $E' \in \reach(E)$. Since $\mathrm{BNDC}_{\wbis{\rm pw}} \subsetneq \mathrm{BSNNI}_{\wbis{\rm
pw}}$ as will be shown in the last case of the proof of this part of the theorem, $E' \in
\mathrm{BSNNI}_{\wbis{\rm pw}}$ for every $E' \in \reach(E)$, i.e., $E \in \mathrm{SBSNNI}_{\wbis{\rm pw}}$.

\item SBSNNI$_{\wbis{\rm pw}}$ $\subsetneq$ BNDC$_{\wbis{\rm pw}}$.
If $E \in \mathrm{SBSNNI}_{\wbis{\rm pw}} = \mathrm{P\_BNDC}_{\wbis{\rm pw}}$ then it immediately follows
that $E \in \mathrm{BNDC}_{\wbis{\rm pw}}$.

\item BNDC$_{\wbis{\rm pw}}$ $\subsetneq$ BSNNI$_{\wbis{\rm pw}}$.
If $E \in \mathrm{BNDC}_{\wbis{\rm pw}}$, i.e., $E \setminus \cala_{\calh} \wbis{\rm pw} (E \pco{L} F) \, /
\, L) \setminus \cala_{\calh}$ for all $F \in \procs$ such that each of its actions belongs to
$\cala_{\calh}$ -- and $E, F \in \procs_{\rm n}$ or $E, F \in \procs_{\rm p}$ -- and for all $L \subseteq
\cala_{\calh}$, then we can consider in particular $\hat{F}$ capable of stepwise mimicking the high-level
behavior of $E$, in the sense that $\hat{F}$ is able to synchronize with all the high-level actions executed
by $E$ and its reachable processes, along with $\hat{L} = \cala_{\calh}$. As a consequence $(E \pco{\hat{L}}
\hat{F}) \, / \, \hat{L}) \setminus \cala_{\calh}$ is isomorphic to $E \, / \, \cala_{\calh}$, hence $E
\setminus \cala_{\calh} \wbis{\rm pw} (E \pco{\hat{L}} \hat{F}) \, / \, \hat{L}) \setminus \cala_{\calh}
\wbis{\rm pw} E \, / \, \cala_{\calh}$, i.e., $E \in \mathrm{BSNNI}_{\wbis{\rm pw}}$, as $\wbis{\rm pw}$ is
transitive.

			\end{itemize}

\noindent
\blue{Next we} prove the relationships for the $\wbis{\rm pb}$-based properties. Let us examine each
relationship separately:

			\begin{itemize}

\item SBNDC$_{\wbis{\rm pb}}$ $\subsetneq$ SBSNNI$_{\wbis{\rm pb}}$.
Given $E \in \mathrm{SBNDC}_{\wbis{\rm pb}}$, the result follows by proving that the relation $\calb = \{
(E' \setminus \cala_{\calh}, E' \, / \, \cala_{\calh}) \mid E' \in \reach(E) \}$ is a probabilistic
branching bisimulation up to $\wbis{\rm pb}$. Starting from $(E' \setminus \cala_{\calh}, E' \, / \,
\cala_{\calh}) \in \calb$, there are three cases for action transitions based on the operational semantic
rules in Table~\ref{tab:prob_op_sem_nd}. In the first case, \linebreak it is $E' \setminus \cala_{\calh}$ to
move first:

				\begin{itemize}

\item If $E' \setminus \cala_{\calh} \warrow{}{} \bar{E}' \setminus \cala_{\calh} \arrow{a}{\rm a} E''
\setminus \cala_{\calh}$ with $a \in \cala_{\call} \cup \{ \tau \}$, then $E' \, / \, \cala_{\calh}
\warrow{}{} \bar{E}' \, / \, \cala_{\calh} \arrow{a}{\rm a}$ \linebreak $E'' \, / \, \cala_{\calh}$ as $a,
\tau \notin \cala_{\calh}$, with $(\bar{E}' \setminus \cala_{\calh}, \bar{E}' \, / \, \cala_{\calh}) \in
\calb$ and $(E'' \setminus \cala_{\calh}, E'' \, / \, \cala_{\calh}) \in \calb$ as $\bar{E}', E'' \in
\reach(E)$. Thus $(\bar{E}' \setminus \cala_{\calh}, \bar{E}' \, / \, \cala_{\calh}) \in (\calb \cup
\calb^{-1} \, \cup \wbis{\rm pb})^{+}$ and $(E'' \setminus \cala_{\calh}, \linebreak E'' \, / \,
\cala_{\calh}) \in (\calb \cup \calb^{-1} \, \cup \wbis{\rm pb})^{+}$.

				\end{itemize}

\noindent
In the other two cases, instead, it is $E' \, / \, \cala_{\calh}$ to move first (note that possible
$\tau$-transitions along $\warrow{}{}$ arising from high actions in $E'$ can\blue{not} be executed \blue{by}
$E' \setminus \cala_{\calh}$, but for them we exploit $E \in \mathrm{SBNDC}_{\wbis{\rm pb}}$ and
Lemma~\ref{lem:prob_taxonomy}(1)):

				\begin{itemize}

\item Let $E' \, / \, \cala_{\calh} \warrow{}{} \bar{E}' \, / \, \cala_{\calh} \arrow{a}{\rm a} E'' \, / \,
\cala_{\calh}$ with $a \in \cala_{\call} \cup \{ \tau \}$. From $E' \, / \, \cala_{\calh} \warrow{}{}
\bar{E}' \, / \, \cala_{\calh}$ and Lemma~\ref{lem:prob_taxonomy}(1) it follows that $E' \setminus
\cala_{\calh} \warrow{}{} \hat{E}' \setminus \cala_{\calh}$ with $\bar{E}' \setminus \cala_{\calh} \wbis{\rm
pb} \hat{E}' \setminus \cala_{\calh}$. From $\bar{E}' \, / \, \cala_{\calh} \arrow{a}{\rm a} E'' \, / \,
\cala_{\calh}$ it follows that $\bar{E}' \setminus \cala_{\calh} \arrow{a}{\rm a} E'' \setminus
\cala_{\calh}$ as $a \notin \cala_{\calh}$. Since $\bar{E}' \setminus \cala_{\calh} \wbis{\rm pb} \hat{E}'
\setminus \cala_{\calh}$ there are two subcases:

					\begin{itemize}

\item If $a = \tau$ and $E'' \setminus \cala_{\calh} \wbis{\rm pb} \hat{E}' \setminus \cala_{\calh}$, then
$\bar{E}' \setminus \cala_{\calh} \wbis{\rm pb} E'' \setminus \cala_{\calh}$ as $\wbis{\rm pb}$ is symmetric
and transitive. From $\bar{E}', E'' \in \mathrm{SBNDC}_{\wbis{\rm pb}}$ and Lemma~\ref{lem:prob_taxonomy}(2)
it follows that $\bar{E}' \, / \, \cala_{\calh} \wbis{\rm pb} E'' \, / \, \cala_{\calh}$. Thus $E' \setminus
\cala_{\calh}$ is allowed to stay idle.

\item Otherwise there exists $\hat{E}' \setminus \cala_{\calh} \warrow{}{} \hat{E}'' \setminus \cala_{\calh}
\arrow{a}{\rm a} \hat{E}''' \setminus \cala_{\calh}$ such that $\bar{E}' \setminus \cala_{\calh} \wbis{\rm
pb} \hat{E}'' \setminus \cala_{\calh}$ and $E'' \setminus \cala_{\calh} \wbis{\rm pb} \hat{E}''' \setminus
\cala_{\calh}$. Summing up, we have that $E' \setminus \cala_{\calh} \warrow{}{} \hat{E}'' \setminus
\cala_{\calh}$ \linebreak $\arrow{a}{\rm a} \hat{E}''' \setminus \cala_{\calh}$ with $\bar{E'} \, / \,
\cala_{\calh} \; \calb^{-1} \, \bar{E}' \setminus \cala_{\calh} \wbis{\rm pb} \hat{E}'' \setminus
\cala_{\calh}$ and $E'' \, / \, \cala_{\calh} \; \calb^{-1} \, E'' \setminus \cala_{\calh} \wbis{\rm pb}
\hat{E}''' \setminus \cala_{\calh}$, as $\bar{E'}, E'' \in \reach(E)$, and hence $(\bar{E}' \, / \,
\cala_{\calh}, \hat{E}'' \setminus \cala_{\calh}) \in (\calb \cup \calb^{-1} \, \cup \wbis{\rm pb})^{+}$ and
$(E'' \, / \, \cala_{\calh}, \hat{E}''' \setminus \cala_{\calh}) \in (\calb \cup \calb^{-1} \, \cup
\wbis{\rm pb})^{+}$.

					\end{itemize}

\item Let $E' \, / \, \cala_{\calh} \warrow{}{} \bar{E}' \, / \, \cala_{\calh} \arrow{\tau}{\rm a} E'' \, /
\, \cala_{\calh}$ with $\bar{E}' \arrow{h}{\rm a} E''$. From $\bar{E}' \in \reach(E)$ and $E \in
\mathrm{SBNDC}_{\wbis{\rm pb}}$ it follows that $\bar{E}' \setminus \cala_{\calh} \wbis{\rm pb} E''
\setminus \cala_{\calh}$, hence $\bar{E}' \, / \, \cala_{\calh} \wbis{\rm pb} E'' \, / \, \cala_{\calh}$ by
virtue of Lemma~\ref{lem:prob_taxonomy}(2) as $\bar{E}', E'' \in \mathrm{SBNDC}_{\wbis{\rm pb}}$. Thus $E'
\setminus \cala_{\calh}$ is allowed to stay idle.

				\end{itemize}
				
\noindent
As for probabilities, the proof is like the one of the corresponding result for $\wbis{\rm pw}$.

\item SBSNNI$_{\wbis{\rm pb}}$ = P\_BNDC$_{\wbis{\rm pb}}$.
The proof is like the one of the corresponding result for $\wbis{\rm pw}$.

\item $\mathrm{SBSNNI}_{\wbis{\rm pb}}$ $\subsetneq$ $\mathrm{BNDC}_{\wbis{\rm pb}}$.
The proof is like the one of the corresponding result for $\wbis{\rm pw}$.

\item BNDC$_{\wbis{\rm pb}}$ $\subsetneq$ BSNNI$_{\wbis{\rm pb}}$.
The proof is like the one of the corresponding result for $\wbis{\rm pw}$.
\qedhere

			\end{itemize}

		\end{proof}

	\end{thm}

All the inclusions in the previous theorem are strict as shown by the following counterexamples (we omit
every $[1]$ for simplicity):

	\begin{itemize}
	
\item The process $\tau \, . \, l \, . \, \nil + l \, . \, l \, . \, \nil + h \, . \, l \, . \, \nil$ is
SBSNNI$_{\wbis{}}$ (resp.\ P\_BDNC$_{\wbis{}}$) because $(\tau \, . \, l \, . \, \nil + l \, . \, l \, . \,
\nil + h \, . \, l \, . \, \nil) \setminus \{ h \} \wbis{} (\tau \, . \, l \, . \, \nil + l \, . \, l \, .
\, \nil + h \, . \, l \, . \, \nil) \, / \, \{ h \}$ and action $h$ is enabled only at the beginning so any
other reachable process, i.e., $l \, . \, \nil$ and $\nil$, is BSNNI$_{\wbis{}}$ (resp.\ BNDC$_{\wbis{}}$).
It is not SBNDC$_{\wbis{}}$ because the low-level view of the process reached after action $h$, i.e., $(l \,
. \, \nil) \setminus \{ h \}$, is not $\wbis{}$-equivalent to $(\tau \, . \, l \, .  \, \nil + l \, . \, l
\, . \, \nil + h \, . \, l \, . \, \nil) \setminus \{ h \}$.

\item The process $l \, . \, \nil + l \, . \, l \, . \, \nil + l \, . \, h \, . \, l \, . \, \nil$ is
BNDC$_{\wbis{}}$ because, whether there are synchronizations with high-level actions or not, the overall
process can always perform either an $l$-action or a sequence of two $l$-actions. It is not
SBSNNI$_{\wbis{}}$ (resp.\ P\_BNDC$_{\wbis{}}$) because the reachable process $h \, . \, l \, . \, \nil$ is
not BSNNI$_{\wbis{}}$ (resp.\ BNDC$_{\wbis{}}$). 

\item The process $l \, . \, \nil + h \, . \, h \, . \, l \, . \, \nil$ is BSNNI$_{\wbis{}}$ due to $(l \, .
\, \nil + h \, . \, h \, . \, l \, . \, \nil) \setminus \{ h \} \wbis{} (l \, . \, \nil + h \, . \, h \, .
\, l \, . \, \nil) / \{ h \}$. It is not BNDC$_{\wbis{}}$ due to $(((l \, . \, \nil + h \, . \, h \, . \, l
\, . \, \nil) \pco{\{ h \}} (h \, . \, \nil)) / \{ h \}) \setminus \{ h \} \not\wbis{} (l \, . \, \nil + h
\, . \, h \, .\, l \, . \, \nil) \setminus \{ h \}$ because the former behaves as $l \, . \, \nil + \tau \,
. \, \nil$ while the latter behaves as $l \, . \, \nil$.

	\end{itemize}

We further observe that each of the $\wbis{\rm pb}$-based noninterference properties implies the
corresponding $\wbis{\rm pw}$-based one, due to the fact that $\wbis{\rm pb}$ is finer than $\wbis{\rm pw}$.

	\begin{thm}\label{thm:prob_taxonomy_2}

The following inclusions hold:

		\begin{enumerate}

\item $\mathrm{BSNNI}_{\wbis{\rm pb}} \subsetneq \mathrm{BSNNI}_{\wbis{\rm pw}}$.

\item $\mathrm{BNDC}_{\wbis{\rm pb}} \subsetneq \mathrm{BNDC}_{\wbis{\rm pw}}$.

\item $\mathrm{SBSNNI}_{\wbis{\rm pb}} \subsetneq \mathrm{SBSNNI}_{\wbis{\rm pw}}$.

\item $\mathrm{P\_BNDC}_{\wbis{\rm pb}} \subsetneq \mathrm{P\_BNDC}_{\wbis{\rm pw}}$.

\item $\mathrm{SBNDC}_{\wbis{\rm pb}} \subsetneq \mathrm{SBNDC}_{\wbis{\rm pw}}$.
\fullbox

		\end{enumerate}

	\end{thm}

All the inclusions above are strict by virtue of the following result; for an example of $E_{1}$ and $E_{2}$
below, see Figure~\ref{fig:prob_wb_brb_cex} with both systems extended with an identical action transition
at the beginning.

	 \begin{thm}\label{thm:prob_taxonomy_3}

Let $E_{1}, E_{2} \in \procs_{\rm n}$ be such that $E_{1} \wbis{\rm pw} E_{2}$ but $E_{1} \not\wbis{\rm pb}
E_{2}$. If no high-level actions occur in $E_{1}$ and $E_{2}$, then $F \in \{ E_{1} + h \, . \, [1] E_{2},
E_{2} + h \, . \, [1] E_{1} \}$ is such that:

		\begin{enumerate}

\item $F \in \mathrm{BSNNI}_{\wbis{\rm pw}}$ but $F \notin \mathrm{BSNNI}_{\wbis{\rm pb}}$.

\item $F \in \mathrm{BNDC}_{\wbis{\rm pw}}$ but $F \notin \mathrm{BNDC}_{\wbis{\rm pb}}$.

\item $F \in \mathrm{SBSNNI}_{\wbis{\rm pw}}$ but $F \notin \mathrm{SBSNNI}_{\wbis{\rm pb}}$.

\item $F \in \mathrm{P\_BNDC}_{\wbis{\rm pw}}$ but $F \notin \mathrm{P\_BNDC}_{\wbis{\rm pb}}$.

\item $F \in \mathrm{SBNDC}_{\wbis{\rm pw}}$ but $F \notin \mathrm{SBNDC}_{\wbis{\rm pb}}$.

		\end{enumerate}	 

		\begin{proof}

Let $F$ be $E_{1} + h \, . \, [1] E_{2}$ (the proof is similar for $F$ equal to $E_{2} + h \, . \, [1]
E_{1}$) and observe that no high-level actions occur in every process reachable from $F$ except $F$ itself:

			\begin{enumerate}

\item Since the only high-level action occurring in $F$ is $h$, in the proof of $F \in
\mathrm{BSNNI}_{\wbis{\rm pw}}$ the only interesting case is the transition $F \, / \, \cala_{\calh}
\arrow{\tau}{\rm a} ([1] E_{2}) \, / \, \cala_{\calh}$, to which $F \setminus \cala_{\calh}$ responds with a
scheduler that makes it stay idle because $([1] E_{2}) \, / \, \cala_{\calh} \wbis{\rm pw} [1] E_{2}
\wbis{\rm pw} E_{2} \wbis{\rm pw} E_{1} \wbis{\rm pw} F \setminus \cala_{\calh}$, i.e., $([1] E_{2}) \, / \,
\cala_{\calh} \wbis{\rm pw} F \setminus \cala_{\calh}$ as $\wbis{\rm pw}$ is symmetric and transitive. \\ On
the other hand, $F \notin \mathrm{BSNNI}_{\wbis{\rm pb}}$ because $E_{2} \not\wbis{\rm pb} E_{1}$ in the
same situation as before.

\item Since $F \in \mathrm{BSNNI}_{\wbis{\rm pw}}$ by the previous result and no high-level actions occur in
every process reachable from $F$ other than $F$, it holds that $F \in \mathrm{SBSNNI}_{\wbis{\rm pw}}$ and
hence $F \in \mathrm{BNDC}_{\wbis{\rm pw}}$ by virtue of Theorem~\ref{thm:prob_taxonomy_1}. \\ On the other
hand, from $F \notin \mathrm{BSNNI}_{\wbis{\rm pb}}$ by the previous result it follows that \linebreak $F
\notin \mathrm{BNDC}_{\wbis{\rm pb}}$ by virtue of Theorem~\ref{thm:prob_taxonomy_1}.

\item We already know from the proof of the previous result that $F \in \mathrm{SBSNNI}_{\wbis{\rm pw}}$. \\
On the other hand, from $F \notin \mathrm{BSNNI}_{\wbis{\rm pb}}$ by the first result it follows that $F
\notin \mathrm{SBSNNI}_{\wbis{\rm pb}}$ by virtue of Theorem~\ref{thm:prob_taxonomy_1}.

\item An immediate consequence of P\_BNDC$_{\wbis{\rm pw}}$ = SBSNNI$_{\wbis{\rm pw}}$ and
P\_BNDC$_{\wbis{\rm pb}}$ = SBSNNI$_{\wbis{\rm pb}}$ as established by Theorem~\ref{thm:prob_taxonomy_1}.

\item Since the only high-level action occurring in $F$ is $h$, in the proof of $F \in
\mathrm{SBNDC}_{\wbis{\rm pw}}$ the only interesting case is the transition $F \arrow{h}{\rm a} [1] E_{2}$,
for which it holds that $F \setminus \cala_{\calh} \linebreak \wbis{\rm pw} E_{1} \wbis{\rm pw} E_{2}
\wbis{\rm pw} [1] E_{2} \wbis{\rm pw} ([1] E_{2}) \setminus \cala_{\calh}$, i.e., $F \setminus \cala_{\calh}
\wbis{\rm pw} ([1] E_{2}) \setminus \cala_{\calh}$ as $\wbis{\rm pw}$ \linebreak is transitive. \\ On the
other hand, $F \notin \mathrm{SBNDC}_{\wbis{\rm pb}}$ because $E_{1} \not\wbis{\rm pb} E_{2}$ in the same
situation as before.
\qedhere

			\end{enumerate}

		\end{proof}

	\end{thm}

The diagram in Figure~\ref{fig:prob_taxonomy} summarizes the inclusions among the various noninterference
properties based on the results in Theorems~\ref{thm:prob_taxonomy_1} and~\ref{thm:prob_taxonomy_2}, where
$\calp \rightarrow \calq$ means that $\calp$ is strictly included in $\calq$. These inclusions follow the
same pattern as the nondeterministic setting~\cite{EABR25}. The arrows missing in the diagram, witnessing
incomparability, are justified by the following counterexamples (we omit every $[1]$ for simplicity):

	\begin{itemize}

\item SBNDC$_{\wbis{\rm pw}}$ vs.\ SBSNNI$_{\wbis{\rm pb}}$.
The process $\tau \, . \, l \, . \, \nil + l \, . \, l \, . \, \nil + h \, . \, l \, . \, \nil$ is
BSNNI$_{\wbis{\rm pb}}$ as $(\tau \, . \, l \, . \, \nil + l \, . \, l \, . \, \nil + h \, . \, l \, .  \,
\nil) \setminus \{ h \} \wbis{\rm pb} \tau \, . \, l \, . \, \nil + l \, . \, l \, . \, \nil \wbis{\rm pb}
\tau \, . \, l \, . \, \nil + l \, . \, l \, . \, \nil + \tau \, . \, l \, . \, \nil \wbis{\rm pb} (\tau \,
. \, l \, . \, \nil + l \, . \, l \, . \, \nil + h \, . \, l \, . \, \nil) \, / \, \{ h \}$. It is also
SBSNNI$_{\wbis{\rm pb}}$ because every reachable process does not enable further high-level actions.
However, it is not SBNDC$_{\wbis{\rm pw}}$ because after executing the high-level action $h$ it can perform
a single $l$-action, while the original process with the restriction on high-level actions can go along a
path where it can perform two $l$-actions. On the other hand, the process $F$ mentioned in
Theorem~\ref{thm:prob_taxonomy_3} is SBNDC$_{\wbis{\rm pw}}$ but neither BSNNI$_{\wbis{\rm pb}}$ nor
SBSNNI$_{\wbis{\rm pb}}$.

\item SBSNNI$_{\wbis{\rm pw}}$ vs.\ BNDC$_{\wbis{\rm pb}}$.
The process $l \, . \, h \, . \, l \, . \, \nil + l \, . \, \nil + l \, . \, l \, . \, \nil$ is
BSNNI$_{\wbis{\rm pb}}$ as $(l \, . \, h \, . \, l \, . \, \nil + l \, . \, \nil + l \, . \, l \, . \, \nil)
\setminus \{ h \} \wbis{\rm pb} l \, . \, \nil + l \, . \, \nil + l \, . \, l \, . \, \nil \wbis{\rm pb} l
\, . \, \tau \, . \, l \, . \, \nil + l \, . \, \nil + l \, . \, l \, . \, \nil \wbis{\rm pb} (l \, .  \, h
\, . \, l \, . \, \nil + l \, . \, \nil + l \, . \, l \, . \, \nil) \, / \, \{ h \}$. The same process is
BNDC$_{\wbis{\rm pb}}$ too as it includes only one high-level action, hence the only possible high-level
strategy coincides with the check conducted by BSNNI$_{\wbis{\rm pb}}$. However, it is not
SBSNNI$_{\wbis{\rm pw}}$ because of the reachable process $h \, . \, l \, . \, \nil$, which is not
BSNNI$_{\wbis{\rm pw}}$. On the other hand, the process $F$ mentioned in Theorem~\ref{thm:prob_taxonomy_3}
\linebreak is SBSNNI$_{\wbis{\rm pw}}$ but not BSNNI$_{\wbis{\rm pb}}$ and, therefore, not even
BNDC$_{\wbis{\rm pb}}$.

\item BNDC$_{\wbis{\rm pw}}$ vs.\ BSNNI$_{\wbis{\rm pb}}$.
The process $l \, . \, \nil + l \, . \, ([0.5] h_{1} \, . \, l_{1} \, . \, \nil \oplus [0.5] h_{2} \, . \,
l_{2} \, . \, \nil) + l \, . \, ([0.5] l_{1} \, . \, \nil \oplus [0.5] l_{2} \, . \, \nil)$ is
BSNNI$_{\wbis{\rm pb}}$ but not BNDC$_{\wbis{\rm pw}}$ as discussed in
Section~\ref{sec:prob_bisim_sec_prop}. In contrast, the process $F$ mentioned in
Theorem~\ref{thm:prob_taxonomy_3} is both BSNNI$_{\wbis{\rm pw}}$ and BNDC$_{\wbis{\rm pw}}$, but not
BSNNI$_{\wbis{\rm pb}}$.

	\end{itemize}

Like in the nondeterministic setting~\cite{EABR25}, the strongest property based on weak probabilistic
bisimilarity (SBNDC$_{\wbis{\rm pw}}$) and the weakest property based on probabilistic branching
bisimilarity (BSNNI$_{\wbis{\rm pb}}$) are incomparable too. The former is a very restrictive property
because it requires a local check every time a high-level action is performed, while the latter requires a
check only on the initial state. On the other hand, as shown in Theorem~\ref{thm:prob_taxonomy_3}, it is
very easy to construct processes that are secure under properties based on $\wbis{\rm pw}$ but not on
$\wbis{\rm pb}$, due to the minimal number of high-level actions in $F$.

	\begin{figure}[t]

\begin{center}
\begin{tikzpicture}[modal]

\node (bsnni)                                                         {BSNNI$_{\wbis{\rm pw}}$};
\node (bndc)    [above left = of bsnni]                               {BNDC$_{\wbis{\rm pw}}$};
\node (sbsnni)  [above left = of bndc, align=left, text width=1.8cm]  {SBSNNI$_{\wbis{\rm pw}}$ \\
								       {P\_BNDC}$_{\wbis{\rm pw}}$};
\node (sbndc)   [above left = of sbsnni]                              {SBNDC$_{\wbis{\rm pw}}$};

\node (brsnni)  [left = 1.1cm of bsnni]                               {BSNNI$_{\wbis{\rm pb}}$};
\node (brndc)   [left = 1.2cm of bndc]                                {BNDC$_{\wbis{\rm pb}}$};
\node (sbrsnni) [left = of sbsnni, left = 1cm of sbsnni, align=left,
						   text width=1.8cm]  {SBSNNI$_{\wbis{\rm pb}}$ \\
								       {P\_BNDC}$_{\wbis{\rm pb}}$};
\node (sbrndc)  [left = of sbndc]                             	      {SBNDC$_{\wbis{\rm pb}}$};

\path[<-] (bndc)    edge (sbsnni);
\path[<-] (bsnni)   edge (bndc);
\path[<-] (bsnni)   edge (brsnni);
\path[<-] (sbsnni)  edge (sbndc);
\path[<-] (sbsnni)  edge (sbrsnni);
\path[<-] (brndc)   edge (sbrsnni);
\path[<-] (brsnni)  edge (brndc);
\path[<-] (bndc)    edge (brndc);
\path[<-] (sbndc)   edge (sbrndc);
\path[<-] (sbrsnni) edge (sbrndc);

\end{tikzpicture}
\end{center}

\caption{Taxonomy of security properties based on probabilistic bisimilarities}
\label{fig:prob_taxonomy}

	\end{figure}

%
\subsection{Relating Nondeterministic and Probabilistic Taxonomies}
\label{sec:prob_taxonomy_rel}
%

Let us compare our probabilistic taxonomy with the nondeterministic one of~\cite{EABR25}. In the following,
we assume that $\wbis{\rm w}$ denotes the weak nondeterministic bisimilarity of~\cite{Mil89a} and $\wbis{\rm
b}$ denotes the nondeterministic branching bisimilarity of~\cite{GW96}. These can be obtained from the
corresponding definitions in Section~\ref{sec:prob_bisim} by restricting to nondeterministic states and
ignoring the clause involving the $\proba$ function. Since we are considering probabilistic choices as
internal, given a process $E \in \procs$ we can obtain its nondeterministic variant, denoted by $\nd(E)$, by
replacing every occurrence of $\bigoplus_{i \in I} [p_{i}] N_{i}$ with $\sum_{i \in I} \tau \, . \, N_{i}$.

The next proposition states that if two processes are equivalent according to any of the weak bisimilarities
in Section~\ref{sec:prob_bisim}, then their nondeterministic variants are equivalent according to the
corresponding nondeterministic weak bisimilarity. The inverse does not hold: e.g., processes $E_{1} = [0.5]
a_{1} \, . \, \nil \oplus [0.5] a_{2} \, . \, \nil$ and $E_{2} = [0.8] a_{1} \, . \, \nil \oplus [0.2] a_{2}
\, . \, \nil$ are such that $E_{1} \not\wbis{\rm pw} E_{2}$ and $E_{1} \not\wbis{\rm pb} E_{2}$, but their
nondeterministic counterparts coincide as both of them are equal to $\tau \, . \, a_{1} \, . \, \nil + \tau
\, . \, a_{2} \, . \, \nil$.

	\begin{prop}\label{prop:pr_implies_nd}

Let $E_{1}, E_{2} \in \procs$. Then:

		\begin{enumerate}

\item $E_{1} \wbis{\rm pw} E_{2} \Longrightarrow \nd(E_{1}) \wbis{\rm w} \nd(E_{2})$.

\item $E_{1} \wbis{\rm pb} E_{2} \Longrightarrow \nd(E_{1}) \wbis{\rm b} \nd(E_{2})$. 

		\end{enumerate}

		\begin{proof}

Let us denote by $\warrow{\hat{a}}{\rm a}$ the variant of $\warrow{\hat{a}}{}$ in which there are no
probabilistic transitions and by $\warrow{\tau^{*}}{\rm a}$ a possibly empty sequence of $\tau$-transitions:

			\begin{enumerate}

\item We need to prove that the symmetric relation $\calb = \{ (\nd(E_{1}), \nd(E_{2})) \mid E_{1} \wbis{\rm
pw} E_{2} \}$ is a weak bisimulation. We start by observing that from $E_{1} \wbis{\rm pw} E_{2}$ it follows
that for each $E_{1} \arrow{a}{\rm a} E'_{1}$ there exists $\sigma \in \sched(E_{2})$ such that
$\proba_{\sigma}(E_{2}, \tau^{*} \, \hat{a} \, \tau^{*}, [E'_{1}]_{\wbis{\rm pw}}, \varepsilon_{E_{2}}) =
1$. Since $\nd(E_{1})$ and $\nd(E_{2})$ are obtained by replacing each probabilistic transition with a
$\tau$-transition, for each $\nd(E_{1}) \arrow{a}{\rm a} \nd(E'_{1})$ there exists $\nd(E_{2})
\warrow{\hat{a}}{\rm a} \nd(E'_{2})$ such that $(\nd(E'_{1}), \nd(E'_{2})) \in \calb$.

\item We need to prove that the symmetric relation $\calb = \{ \nd(E_{1}), \nd(E_{2})) \mid E_{1} \wbis{\rm
pb} E_{2} \}$ is a branching bisimulation. We start by observing that from $E_{1} \wbis{\rm pb} E_{2}$ it
follows that for each $E_{1} \arrow{a}{\rm a} E'_{1}$ either $a = \tau$ and $E'_{1} \wbis{\rm pb} E_{2}$, or
there exists $E_{2} \warrow{}{} \bar{E}_{2} \arrow{a}{\rm a} E'_{2}$ such that $E_{1} \wbis{\rm pb}
\bar{E}_{2}$ and $E'_{1} \wbis{\rm pb} E'_{2}$. Since $\nd(E_{1})$ and $\nd(E_{2})$ are obtained by
replacing each probabilistic transition with a $\tau$-transition, for each $\nd(E_{1}) \arrow{a}{\rm a}
\nd(E'_{1})$ either $a = \tau$ and $(\nd(E'_{1}), \nd(E_{2})) \in \calb$, or there exists $\nd(E_{2})
\warrow{\tau^{*}}{\rm a} \nd(\bar{E}_{2}) \arrow{a}{\rm a} \nd(E'_{2})$ such that $(\nd(E_{1}),
\nd(\bar{E}_{2})) \in \calb$ and $(\nd(E'_{1}), \nd(E'_{2})) \in \calb$. 
\qedhere

			\end{enumerate}

		\end{proof}

	\end{prop}

An immediate consequence is that if a process is secure under any of the probabilistic noninterference
properties of Section~\ref{sec:prob_bisim_sec_prop}, then its nondeterministic variant is secure under the
corresponding nondeterministic property. The taxonomy of Figure~\ref{fig:prob_taxonomy} thus extends to the
left the one in~\cite{EABR25}, as each of the properties of Section~\ref{sec:prob_bisim_sec_prop} is finer
than its nondeterministic counterpart.

	\begin{cor}\label{cor:pr_implies_nd}

Let $\calp_{\rm pr} \in \{ \mathrm{BSNNI}_{\wbis{\rm pr}}, \mathrm{BNDC}_{\wbis{\rm pr}},
\mathrm{SBSNNI}_{\wbis{\rm pr}}, \mathrm{P\_BNDC}_{\wbis{\rm pr}}, \mathrm{SBNDC}_{\wbis{\rm pr}} \}$ and
$\calp_{\rm nd} \in \{ \mathrm{BSNNI}_{\wbis{\rm nd}}, \mathrm{BNDC}_{\wbis{\rm nd}},
\mathrm{SBSNNI}_{\wbis{\rm nd}}, \mathrm{P\_BNDC}_{\wbis{\rm nd}}, \mathrm{SBNDC}_{\wbis{\rm nd}} \}$ for
$\wbis{\rm pr} \: \in \{ \wbis{\rm pw}, \linebreak \wbis{\rm pb} \}$ and $\wbis{\rm nd} \: \in \{ \wbis{\rm
w}, \wbis{\rm b} \}$, where $\calp_{\rm nd}$ is meant to be the nondeterministic variant of $\calp_{\rm
pr}$. Then $E \in \calp_{\rm pr} \Longrightarrow \nd(E) \in \calp_{\rm nd}$ for all $E \in \procs$.

		\begin{proof}

The result directly follows from Proposition~\ref{prop:pr_implies_nd}.
\qedhere

		\end{proof}

	\end{cor}

%
%
\section{Reversibility via Weak Probabilistic Back-and-Forth Bisimilarity}
\label{sec:prob_branching_is_bf}
%
%

In~\cite{DMV90} it was shown that, over nodeterministic processes, weak back-and-forth bisimilarity
coincides with branching bisimilarity. In this section we extend that result so that probabilistic branching
bisimilarity can be employed in the noninterference analysis of reversible processes featuring
nondeterminism and probabilities.

A PLTS $(\cals, \cala_{\tau}, \! \arrow{}{} \!)$ represents a reversible process if each of its transitions
is seen as bidirectional. When going backward, it is of paramount importance to respect causality, i.e., the
last performed transition must be the first one to be undone. Following~\cite{DMV90} \linebreak we set up an
equivalence that enforces not only causality but also history preservation. This means that, when going
backward, a process can only move along the path representing the history that brought the process to the
current state even in the presence of concurrency. To accomplish this, the equivalence has to be defined
over computations, not over states, and the notion of transition has to be suitably revised. We start by
adapting the notation of the nondeterministic setting of~\cite{DMV90} to our strictly alternating
probabilistic setting. We use $\ell$ for a label in $\cala_{\tau} \cup \realns_{]0, 1]}$.

	\begin{defi}\label{def:prob_path}

A sequence $\xi = (s_{0}, \ell_{1}, s_{1}) (s_{1}, \ell_{2}, s_{2}) \dots (s_{n - 1}, \ell_{n}, s_{n}) \in
\: \arrow{}{}^{*}$ is a \emph{path} of length $n$ from state $s_{0}$; we let $\first(\xi) = s_{0}$ and
$\last(\xi) = s_{n}$. We denote by $\pt(s)$ the set of paths from $s$, which includes the empty path
$\varepsilon_{s}$ from $s$.
\fullbox

	\end{defi}

	\begin{defi}\label{def:prob_run}

A pair $\rho = (s, \xi)$ is called a \emph{run} from state $s$ iff $\xi \in \pt(s)$, in which case we let
$\pt(\rho) = \xi$, $\first(\rho) = \first(\xi) = s$, and $\last(\rho) = \last(\xi)$, with $\first(\rho) =
\last(\rho) = s$ when $\xi = \varepsilon_{s}$. We denote by $\rn(s)$ the set of runs from state $s$. Given
$\rho = (s, \xi) \in \rn(s)$ and $\rho' = (s', \xi') \in \rn(s')$, their composition $\rho \rho' = (s, \xi
\xi') \in \rn(s)$ is defined iff $\last(\rho) = \first(\rho') = s'$. We write $\rho \arrow{\ell}{} \rho'$
iff there exists $\bar{\rho} = (\bar{s}, (\bar{s}, \ell, s'))$ with $\bar{s} = \last(\rho)$ such that $\rho'
= \rho \bar{\rho}$; note that $\first(\rho) = \first(\rho')$. Moreover $\rho \warrow{}{} \rho'$, $\rho
\warrow{a}{} \rho'$, $\rho \warrow{\hat{a}}{} \rho'$, and $\proba(\rho, \rho')$ are defined like $s
\warrow{}{} s'$, $s \warrow{a}{} s'$, $s \warrow{\hat{a}}{} s'$, and $\proba(s, s')$ by considering
transitions between runs instead of transitions between states.
\fullbox

	\end{defi}

In the considered PLTS we work with the set $\calu$ of runs in lieu of $\cals$. Following~\cite{DMV90},
given a run $\rho$, during the weak bisimulation game we distinguish between \emph{outgoing} and
\emph{incoming} action transitions of $\rho$ \blue{for the forward direction and the backward direction,
respectively}. Like in~\cite{BM23a}, this does not apply to probabilistic transitions, which are thus
considered only in the forward direction. If the labels of incoming probabilistic transitions were taken
into account, then the nondeterministic state $a \, . \, \nil$ and the probabilistic state $[p] a \, . \,
\nil \oplus [1 - p] a \, . \, \nil$ would be told apart, because $a \, . \, \nil$ in the former state has no
incoming probabilistic transitions while $a \, . \, \nil$ in the latter state is reached with cumulative
probability~$1$. Unlike~\cite{BM23a}, where action execution and quantitative aspects are integrated in a
single transition relation, even a simpler clause requiring for any two related runs that they both have
incoming probabilistic transitions or neither has -- regardless of cumulative probabilities -- would
distinguish the two states exemplified before.

	\begin{defi}\label{def:weak_prob_bf_bisim}

Let $(\cals, \cala_{\tau}, \! \arrow{}{} \!)$ be a PLTS. We say that $\rho_{1}, \rho_{2} \in \calu$ (resp.\
$s_{1}, s_{2} \in \cals$) are \emph{weakly probabilistic back-and-forth bisimilar}, written $\rho_{1}
\wbis{\rm pbf} \rho_{2}$ (resp.\ $s_{1} \wbis{\rm pbf} s_{2}$), iff $(\rho_{1}, \rho_{2}) \in \calb$ (resp.\
$((s_{1}, \varepsilon_{s_{1}}), (s_{2}, \varepsilon_{s_{2}})) \in \calb)$ for some weak probabilistic
back-and-forth bisimulation $\calb$. An equivalence relation $\calb$ over $\calu$ is a \emph{weak
probabilistic back-and-forth bisimulation} iff, whenever $(\rho_{1}, \rho_{2}) \in \calb$, then:

		\begin{itemize}

\item For each outgoing action transition $\rho_{1} \arrow{a}{\rm a} \rho'_{1}$ there exists $\rho_{2}
\warrow{\hat{a}}{} \rho'_{2}$ such that $(\rho'_{1}, \rho'_{2}) \in \calb$.

\item For each incoming action transition $\rho'_{1} \arrow{a}{\rm a} \rho_{1}$ there exists $\rho'_{2}
\warrow{\hat{a}}{} \rho_{2}$ such that $(\rho'_{1}, \rho'_{2}) \in \calb$.

\item $\proba(\rho_{1}, C) = \proba(\rho_{2}, C)$ for all equivalence classes $C \in \calu / \calb$.
\fullbox

		\end{itemize}

	\end{defi}

We show that weak probabilistic back-and-forth bisimilarity over runs coincides with $\wbis{\rm pb}$, the
forward-only probabilistic branching bisimilarity over states. We proceed by adopting the proof strategy
followed in~\cite{DMV90} to show that their weak back-and-forth bisimilarity over runs coincides with the
forward-only branching bisimilarity over states of~\cite{GW96}. Therefore we start by proving that
$\wbis{\rm pbf}$ satisfies the \emph{cross property}. This means that, whenever two runs of two $\wbis{\rm
pbf}$-equivalent states can perform a sequence of finitely many $\tau$-transitions, alternating with
probabilistic transitions, such that each of the two target runs ends in a nondeterministic state and is
$\wbis{\rm pbf}$-equivalent to the source run of the other sequence, then the two target runs are $\wbis{\rm
pbf}$-equivalent to each other as well.

	\begin{lem}\label{lem:prob_cross_property}

Let $s_{1}, s_{2} \in \cals$ with $s_{1} \wbis{\rm pbf} s_{2}$. For all $\rho'_{1}, \rho''_{1} \in
\rn(s_{1})$ such that $\rho'_{1} \warrow{}{} \rho''_{1}$ with $\last(\rho''_{1}) \in \cals_{\rm n}$ and for
all $\rho'_{2}, \rho''_{2} \in \rn(s_{2})$ such that $\rho'_{2} \warrow{}{} \rho''_{2}$, with
$\last(\rho''_{2}) \in \cals_{\rm n}$, \linebreak if $\rho'_{1} \wbis{\rm pbf} \rho''_{2}$ and $\rho''_{1}
\wbis{\rm pbf} \rho'_{2}$ then $\rho''_{1} \wbis{\rm pbf} \rho''_{2}$.

		\begin{proof}

Given $s_{1}, s_{2} \in \cals$ with $s_{1} \wbis{\rm pbf} s_{2}$, consider the transitive closure
$\calb^{+}$ of the reflexive and \linebreak symmetric relation $\calb = \:\: \wbis{\rm pbf} \! \cup \: \{
(\rho''_{1}, \rho''_{2}), (\rho''_{2}, \rho''_{1}) \in (\rn(s_{1}) \times \rn(s_{2})) \cup (\rn(s_{2})
\times \rn(s_{1})) \mid \last(\rho''_{1}), \last(\rho''_{2}) \in \cals_{\tt n} \land \exists \rho'_{1} \in
\rn(s_{1}), \rho'_{2} \in \rn(s_{2}) \ldotp \rho'_{1} \warrow{}{} \rho''_{1} \land \rho'_{2} \warrow{}{}
\rho''_{2} \land \rho'_{1} \wbis{\rm pbf} \rho''_{2} \land \rho''_{1} \wbis{\rm pbf} \rho'_{2} \}$. The
result will follow by proving that $\calb^{+}$ is a weak probabilistic back-and-forth bisimulation, because
this implies that $\rho''_{1} \wbis{\rm pbf} \rho''_{2}$ for every additional pair -- i.e., $\calb^{+}$
satisfies the cross property -- as well as $\calb^{+} = \:\: \wbis{\rm pbf}$ -- hence $\wbis{\rm pbf}$
satisfies the cross property too. \\
Let $(\rho''_{1}, \rho''_{2}) \in \calb \, \setminus \! \wbis{\rm pbf}$ to avoid trivial cases. Then
$\last(\rho''_{1}), \last(\rho''_{2}) \in \cals_{\tt n}$ and there exist $\rho'_{1} \in \rn(s_{1})$ and
$\rho'_{2} \in \rn(s_{2})$ such that $\rho'_{1} \warrow{}{} \rho''_{1}$, $\rho'_{2} \warrow{}{} \rho''_{2}$,
$\rho'_{1} \wbis{\rm pbf} \rho''_{2}$, and $\rho''_{1} \wbis{\rm pbf} \rho'_{2}$. There are two cases for
action transitions:

			\begin{itemize}

\item In the forward case, assume that $\rho''_{1} \arrow{a}{\rm a} \rho'''_{1}$, from which we derive
$\rho'_{1} \warrow{}{} \rho''_{1} \arrow{a}{\rm a} \rho'''_{1}$. From $\rho'_{1} \wbis{\rm pbf} \rho''_{2}$
it follows that there exists $\rho''_{2} \warrow{}{} \rho'''_{2}$ if $a = \tau$ or $\rho''_{2} \warrow{}{}
\! \arrow{a}{\rm a} \! \warrow{}{} \rho'''_{2}$ \linebreak if $a \neq \tau$, such that $\rho'''_{1}
\wbis{\rm pbf} \rho'''_{2}$ and hence $(\rho'''_{1}, \rho'''_{2}) \in \calb$ as $\wbis{\rm pbf} \, \subseteq
\calb$. \\
When starting from $\rho''_{2} \arrow{a}{\rm a} \rho'''_{2}$, we exploit $\rho'_{2} \warrow{}{} \rho''_{2}$
and $\rho''_{1} \wbis{\rm pbf} \rho'_{2}$ instead.

\item In the backward case, assume that $\rho'''_{1} \arrow{a}{\rm a} \rho''_{1}$. From $\rho''_{1}
\wbis{\rm pbf} \rho'_{2}$ it follows that there exists $\rho'''_{2} \warrow{}{} \rho'_{2}$ if $a = \tau$, so
that $\rho'''_{2} \warrow{}{} \rho''_{2}$, or $\rho'''_{2} \warrow{}{} \! {\arrow{a}{\rm a}} \! \warrow{}{}
\rho'_{2}$ if $a \neq \tau$, so that $\rho'''_{2} \warrow{}{} \! \arrow{a}{\rm a} \! \warrow{}{}
\rho''_{2}$, such that $\rho'''_{1} \wbis{\rm pbf} \rho'''_{2}$ and hence $(\rho'''_{1}, \rho'''_{2}) \in
\calb$ as $\wbis{\rm pbf} \, \subseteq \calb$. \\
When starting from $\rho'''_{2} \arrow{a}{\rm a} \rho''_{2}$, we exploit $\rho'_{1} \wbis{\rm pbf}
\rho''_{2}$ and $\rho'_{1} \warrow{}{} \rho''_{1}$ instead.

			\end{itemize}

\noindent
As for probabilities, from $\last(\rho''_{1}), \last(\rho''_{2}) \in \cals_{\tt n}$ it follows that
$\proba(\rho''_{1}, \bar{C}) = 1 = \proba(\rho''_{2}, \bar{C})$ when $\bar{C}$ is the equivalence class with
respect to $\calb^{+}$ that contains $\rho''_{1}$ and $\rho''_{2}$, while $\proba(\rho''_{1}, C) = 0 =
\proba(\rho''_{2}, C)$ for any other equivalence class $C$.
\qedhere

		\end{proof}

	\end{lem}

	\begin{thm}\label{thm:prob_branching_is_bf}

Let $s_{1}, s_{2} \in \cals$. Then $s_{1} \wbis{\rm pbf} s_{2} \: \Longleftrightarrow \: s_{1} \wbis{\rm pb}
s_{2}$.

		\begin{proof}

The proof is divided into two parts:

			\begin{itemize}

\item Suppose that $s_{1} \wbis{\rm pbf} s_{2}$ and let $\calb$ be a weak probabilistic back-and-forth
bisimulation over $\calu$ such that $((s_{1}, \varepsilon_{s_{1}}), (s_{2}, \varepsilon_{s_{2}})) \in
\calb$. Assume that $\calb$ only contains all the pairs of $\wbis{\rm pbf}$-equivalent runs from $s_{1}$ and
$s_{2}$, so that Lemma~\ref{lem:prob_cross_property} is applicable to $\calb$. We show that $\calb' = \{
(\last(\rho_{1}), \last(\rho_{2})) \mid (\rho_{1}, \rho_{2}) \in \calb \}$ is a probabilistic branching
bisimulation over the states in $\cals$ reachable from $s_{1}$ and $s_{2}$, from which $s_{1} \wbis{\rm pb}
s_{2}$ will follow. Note that $\calb'$ is an equivalence relation because so is $\calb$. \\
Given $(\last(\rho_{1}), \last(\rho_{2})) \in \calb'$, by definition of $\calb'$ we have that $(\rho_{1},
\rho_{2}) \in \calb$. Let $r_{k} = \last(\rho_{k})$ for $k \in \{ 1, 2 \}$, so that $(r_{1}, r_{2}) \in
\calb'$. Suppose that $r_{1} \arrow{a}{\rm a} r'_{1}$, i.e., $\rho_{1} \arrow{a}{\rm a} \rho'_{1}$ where
$\last(\rho'_{1}) = r'_{1}$. There are two cases:

				\begin{itemize}

\item If $a = \tau$ then from $(\rho_{1}, \rho_{2}) \in \calb$ it follows that there exists $\rho_{2}
\warrow{}{} \rho'_{2}$ such that $(\rho'_{1}, \rho'_{2}) \in \calb$. This means that we have a sequence of
$n \ge 0$ transitions of the form $\rho_{2, i} \arrow{\tau}{\rm a} \rho_{2, i + 1}$ or $\rho_{2, i}
\arrow{p_{i}}{\rm p} \rho_{2, i + 1}$ for all $0 \le i \le n - 1$ -- with $\tau$-transitions and
probabilistic transitions alternating -- where $\rho_{2, 0}$ is $\rho_{2}$ while $\rho_{2, n}$ is
$\rho'_{2}$ so that $(\rho'_{1}, \rho_{2, n}) \in \calb$ as $(\rho'_{1}, \rho'_{2}) \in \calb$. \\
If $n = 0$ then we are done because $\rho'_{2}$ is $\rho_{2}$ and hence $(\rho'_{1}, \rho_{2}) \in \calb$ as
$(\rho'_{1}, \rho'_{2}) \in \calb$ -- thus $(r'_{1}, r_{2}) \in \calb'$ -- otherwise from $\rho_{2, n}$ we
go back to $\rho_{2, n - 1}$ via $\rho_{2, n - 1} \arrow{\tau}{\rm a} \rho_{2, n}$ or $\rho_{2, n - 1}
\arrow{p_{n - 1}}{\rm p} \rho_{2, n}$. Recalling that $(\rho'_{1}, \rho_{2, n}) \in \calb$, if it is a
$\tau$-transition and $\rho'_{1}$ can respond by staying idle, so that $(\rho'_{1}, \rho_{2, n - 1}) \in
\calb$, or it is a probabilistic transition with $(\rho'_{1}, \rho_{2, n - 1}) \in \calb$, and $n = 1$, then
we are done because $\rho_{2, n - 1}$ is $\rho_{2}$ and hence $(\rho'_{1}, \rho_{2}) \in \calb$ as
$(\rho'_{1}, \rho_{2, n - 1}) \in \calb$ -- thus $(r'_{1}, r_{2}) \in \calb'$ -- otherwise we go further
back to $\rho_{2, n - 2}$ via $\rho_{2, n - 2} \arrow{\tau}{\rm a} \rho_{2, n - 1}$ or $\rho_{2, n - 2}
\arrow{p_{n - 2}}{\rm p} \rho_{2, n - 1}$. If it is a $\tau$-transition and $\rho'_{1}$ can respond by
staying idle, so that $(\rho'_{1}, \rho_{2, n - 2}) \in \calb$, or it is a probabilistic transition with
$(\rho'_{1}, \rho_{2, n - 2}) \in \calb$, and $n = 2$, then we are done because $\rho_{2, n - 2}$ is
$\rho_{2}$ and hence $(\rho'_{1}, \rho_{2}) \in \calb$ as $(\rho'_{1}, \rho_{2, n - 2}) \in \calb$ -- thus
$(r'_{1}, r_{2}) \in \calb'$ -- otherwise we keep going backward. \\
By repeating this procedure, since $(\rho'_{1}, \rho_{2, n}) \in \calb$ either we get to $(\rho'_{1},
\rho_{2, n - n}) \in \calb$ and we are done because this implies that $(\rho'_{1}, \rho_{2}) \in \calb$ --
thus $(r'_{1}, r_{2}) \in \calb'$ -- or for some \linebreak $0 < m \le n$ such that $(\rho'_{1}, \rho_{2,
m}) \in \calb$ the incoming transition $\rho_{2, m - 1} \arrow{\tau}{\rm a} \rho_{2, m}$ is matched by
$\bar{\rho}_{1} \warrow{}{} \rho_{1} \arrow{\tau}{\rm a} \rho'_{1}$ with $(\bar{\rho}_{1}, \rho_{2, m - 1})
\in \calb$. In the latter case, since $\last(\rho_{1}), \last(\rho_{2, m - 1}) \linebreak \in \cals_{\tt
n}$, $\bar{\rho}_{1} \warrow{}{} \rho_{1}$, $\rho_{2} \warrow{}{} \rho_{2, m - 1}$, $(\bar{\rho}_{1},
\rho_{2, m - 1}) \in \calb$, and $(\rho_{1}, \rho_{2}) \in \calb$, from Lemma~\ref{lem:prob_cross_property}
we derive that $(\rho_{1}, \rho_{2, m - 1}) \! \in \! \calb$. Consequently $\rho_{2} \warrow{}{} \rho_{2, m
- 1} \arrow{\tau}{\rm a} \rho_{2, m}$ with $(\rho_{1}, \rho_{2, m - 1}) \! \in \! \calb$ \linebreak and
  $(\rho'_{1}, \rho_{2, m}) \in \calb$, thus $r_{2} \warrow{}{} \last(\rho_{2, m - 1}) \arrow{\tau}{\rm a}
\last(\rho_{2, m})$ with $(r_{1}, \last(\rho_{2, m - 1})) \in \calb'$ and $(r'_{1}, \last(\rho_{2, m})) \in
\calb'$.

\item If $a \neq \tau$ then from $(\rho_{1}, \rho_{2}) \in \calb$ it follows that there exists $\rho_{2}
\warrow{}{} \bar{\rho}_{2} \arrow{a}{\rm a} \bar{\rho}'_{2} \warrow{}{} \rho'_{2}$ such that $(\rho'_{1},
\rho'_{2}) \in \calb$. \\
From $(\rho'_{1}, \rho'_{2}) \! \in \! \calb$ and $\bar{\rho}'_{2} \warrow{}{} \rho'_{2}$ it follows that
there exists $\bar{\rho}'_{1} \warrow{}{} \rho'_{1}$ such that $(\bar{\rho}'_{1}, \bar{\rho}'_{2}) \! \in \!
\calb$. \linebreak Since $\rho_{1} \arrow{a}{\rm a} \rho'_{1}$ and hence the last transition in $\rho'_{1}$
is labeled with $a$, we derive that $\bar{\rho}'_{1}$ is $\rho'_{1}$ and hence $(\rho'_{1}, \bar{\rho}'_{2})
\in \calb$. \\
From $(\rho'_{1}, \bar{\rho}'_{2}) \in \calb$ and $\bar{\rho}_{2} \arrow{a}{\rm a} \bar{\rho}'_{2}$ it
follows that there exists $\bar{\rho}_{1} \warrow{}{} \rho_{1} \arrow{a}{\rm a} \rho'_{1}$ such that
$(\bar{\rho}_{1}, \bar{\rho}_{2}) \in \calb$. \\
Since $\last(\rho_{1}), \last(\bar{\rho}_{2}) \in \cals_{\tt n}$, $\bar{\rho}_{1} \warrow{}{} \rho_{1}$,
$\rho_{2} \warrow{}{} \bar{\rho}_{2}$, $(\bar{\rho}_{1}, \bar{\rho}_{2}) \in \calb$, and $(\rho_{1},
\rho_{2}) \in \calb$, from Lemma~\ref{lem:prob_cross_property} we derive that $(\rho_{1}, \bar{\rho}_{2})
\in \calb$. \\
Consequently $\rho_{2} \warrow{}{} \bar{\rho}_{2} \arrow{a}{\rm a} \bar{\rho}'_{2}$ with $(\rho_{1},
\bar{\rho}_{2}) \in \calb$ and $(\rho'_{1}, \bar{\rho}'_{2}) \in \calb$, thus $r_{2} \warrow{}{}
\last(\bar{\rho}_{2}) \linebreak \arrow{a}{\rm a} \last(\bar{\rho}'_{2})$ with $(r_{1},
\last(\bar{\rho}_{2})) \in \calb'$ and $(r'_{1}, \last(\bar{\rho}'_{2})) \in \calb'$.

				\end{itemize}

\noindent
As for probabilities, given $\rho \in \rn(s_{1}) \cup \rn(s_{2})$, the equivalence class $C'_{\rho}$ with
respect to~$\calb'$ is of the form $[\last(\rho)]_{\calb'} = \{ \last(\rho') \mid (\last(\rho),
\last(\rho')) \in \calb' \} = \last(\{ \rho' \mid (\rho, \rho') \in \calb \}) \linebreak =
\last([\rho]_{\calb})$, i.e., $C'_{\rho} = \last(C_{\rho})$ for some equivalence class $C_{\rho}$ with
respect to $\calb$, provided that function $\last$ is lifted from runs to sets of runs. Therefore
$\proba(r_{1}, C'_{\rho}) = \proba(\rho_{1}, C_{\rho}) = \proba(\rho_{2}, C_{\rho}) = \proba(r_{2},
C'_{\rho})$ for all equivalence classes $C'_{\rho}$ with respect to $\calb'$ such that $C'_{\rho} =
\last(C_{\rho})$ for some equivalence class $C_{\rho}$ with respect to $\calb$.

\item Suppose that $s_{1} \wbis{\rm pb} s_{2}$ and let $\calb$ be a probabilistic branching bisimulation
over $\cals$ such that $(s_{1}, s_{2}) \in \calb$. Assume that $\calb$ only contains all the pairs of
$\wbis{\rm pb}$-equivalent states reachable from $s_{1}$ and $s_{2}$. We show that the reflexive and
transitive closure $\calb'^{*}$ of $\calb' = \{ (\rho_{1}, \rho_{2}), (\rho_{2}, \rho_{1}) \in (\rn(s_{1})
\times \rn(s_{2})) \cup (\rn(s_{2}) \times \rn(s_{1})) \mid (\last(\rho_{1}), \last(\rho_{2})) \in \calb \}$
is a weak probabilistic back-and-forth bisimulation over the runs in $\calu$ from $s_{1}$ and $s_{2}$, from
which $(s_{1}, \varepsilon_{s_{1}}) \wbis{\rm pbf} (s_{2}, \varepsilon_{s_{2}})$, i.e., $s_{1} \wbis{\rm
pbf} s_{2}$, will follow. \\
Given $(\rho_{1}, \rho_{2}) \in \calb'$, by definition of $\calb'$ we have that $(\last(\rho_{1}),
\last(\rho_{2})) \in \calb$. Let $r_{k} = \last(\rho_{k})$ for $k \in \{ 1, 2 \}$, so that $(r_{1}, r_{2})
\in \calb$. There are two cases for action transitions:

				\begin{itemize}

\item In the forward case, if $\rho_{1} \arrow{a}{\rm a} \rho'_{1}$, i.e., $r_{1} \arrow{a}{\rm a} r'_{1}$
where $r'_{1} = \last(\rho'_{1})$, then either $a = \tau$ and $(r'_{1}, r'_{2}) \in \calb$ where $r'_{2} =
r_{2}$, or there exists $r_{2} \warrow{}{} \bar{r}_{2} \arrow{a}{\rm a} r'_{2}$ such that $(r_{1},
\bar{r}_{2}) \in \calb$ and $(r'_{1}, r'_{2}) \in \calb$. In both cases $\rho_{2} \warrow{\hat{a}}{}
\rho'_{2}$ where $\last(\rho'_{2}) = r'_{2}$, so that $(\rho'_{1}, \rho'_{2}) \in \calb'$.

\item In the backward case, if $\rho'_{1} \arrow{a}{\rm a} \rho_{1}$, i.e., $r'_{1} \arrow{a}{\rm a} r_{1}$
where $r'_{1} = \last(\rho'_{1})$, there are two subcases:

					\begin{itemize}

\item If $\rho'_{1}$ is $(s_{1}, \varepsilon_{s_{1}})$, i.e., $r'_{1} \arrow{a}{\rm a} r_{1}$ is $s_{1}
\arrow{a}{\rm a} r_{1}$ and $\last(\rho'_{1}) = s_{1}$, then from $(s_{1}, s_{2}) \in \calb$ \linebreak it
follows that either $a = \tau$ and $(r_{1}, r_{2}) \in \calb$ where $r_{2} = s_{2}$, or there exists $s_{2}
\warrow{}{} \bar{r}_{2} \arrow{a}{\rm a} r_{2}$ such that $(s_{1}, \bar{r}_{2}) \in \calb$ and $(r_{1},
r_{2}) \in \calb$. In both cases $\rho'_{2} \warrow{\hat{a}}{} \rho_{2}$ where $\last(\rho'_{2}) = s_{2}$,
so that $(\rho'_{1}, \rho'_{2}) \in \calb'$.

\item If $\rho'_{1}$ is not $(s_{1}, \varepsilon_{s_{1}})$ then from $(s_{1}, s_{2}) \in \calb$ it follows
that $s_{1}$ reaches $r'_{1}$ with a sequence of transitions that are $\calb$-compatible with those with
which $s_{2}$ reaches some $r'_{2}$ such that $(r'_{1}, r'_{2}) \in \calb$ as $\calb$ only contains all the
states reachable from $s_{1}$ and $s_{2}$. Therefore either $a = \tau$ and $(r_{1}, r'_{2}) \in \calb$ where
$r'_{2} = r_{2}$, or there exists $r'_{2} \warrow{}{} \bar{r}_{2} \arrow{a}{\rm a} r_{2}$ such that
$(r'_{1}, \bar{r}_{2}) \in \calb$ and $(r_{1}, r_{2}) \in \calb$. In both cases $\rho'_{2}
\warrow{\hat{a}}{} \rho_{2}$ where $\last(\rho'_{2}) = r'_{2}$, \linebreak so that $(\rho'_{1}, \rho'_{2})
\in \calb'$.

					\end{itemize}

				\end{itemize}

\noindent
As for probabilities, given $\rho \in \rn(s_{1}) \cup \rn(s_{2})$, the equivalence class $C'_{\rho}$ with
respect to $\calb'^{*}$ is of the form $[\rho]_{\calb'^{*}} = \{ \rho' \in \rn(s_{1}) \cup \rn(s_{2}) \mid
\last(\rho') \in [\last(\rho)]_{\calb} \}$, i.e., $C'_{\rho}$ corresponds to some equivalence class
$C_{\rho}$ with respect to $\calb$. Therefore $\proba(\rho_{1}, C'_{\rho}) = \proba(\last(\rho_{1}),
C_{\rho}) = \proba(\last(\rho_{2}), C_{\rho}) = \proba(\rho_{2}, C'_{\rho})$ for all equivalence classes
$C'_{\rho}$ with respect to $\calb'^{*}$.
\qedhere

			\end{itemize}
	
		\end{proof}

	\end{thm}

Therefore the properties $\mathrm{BSNNI}_{\wbis{\rm pb}}$, $\mathrm{BNDC}_{\wbis{\rm pb}}$,
$\mathrm{SBSNNI}_{\wbis{\rm pb}}$, $\mathrm{P\_BNDC}_{\wbis{\rm pb}}$, $\mathrm{SBNDC}_{\wbis{\rm pb}}$ do
not change if $\wbis{\rm pb}$ is replaced by $\wbis{\rm pbf}$. This allows us to study noninterference
properties for reversible systems featuring nondeterminism and probabilities by using $\wbis{\rm pb}$ in a
standard probabilistic process calculus like the one of Section~\ref{sec:prob_proc_lang}, without having to
resort to external memories~\cite{DK04}, communication keys~\cite{PU07a}, or fixed decorations~\cite{BR23}
for actions that have already been executed.

%
%
\section{Use Case: Probabilistic Smart Contract Lottery}
\label{sec:prob_example}
%
%

Probabilistic modeling~\cite{ARL20} and verification~\cite{SLHX24,KV24} of smart contracts for
blockchain-based, decentralized systems enable an in-depth analysis of potential vulnerabilities. This is
even more important if we consider probabilistic smart contracts for financial and gaming
applications~\cite{CGP19,QHLWLWZH23,PBPTKS21} that have recently emerged in modern systems. In fact, subtle
effects may be hidden in the implementation of randomness or in the inherent behavior of smart contracts.

As an example, in this section we employ our noninterference theory to analyze two vulnerabilities of a
lottery implemented with a probabilistic smart contract~\cite{CGP19} based on a public blockchain like,
e.g., Ethereum. The first vulnerability can only be revealed by considering the probabilistic behavior of
the smart contract, while the second one is intended to motivate the need to exploit the greater expressive
power of branching bisimulation semantics over weak bisimulation semantics.

In the lottery, initially anyone can buy a ticket by invoking a dedicated smart contract function that
allows one to pay a predefined amount for the ticket. When the lottery is closed, anyone can invoke another
smart contract function, called \textsf{draw()}, in which a random number $x$, between $1$ and the number of
sold tickets, is drawn and the entire amount of money is paid to the owner of ticket $x$ as soon as the win
is confirmed in a transaction of a new block of the blockchain.

The first critical issue that we consider is the randomization procedure of function \textsf{draw()}.  A
simple way to choose a seed for generating random numbers is picking one of the attributes of the block
containing a transaction, e.g., its hash or timestamp. However, as noticed in~\cite{CGP19}, this is a
vulnerable approach, because it gives advantage to potential malicious users. Indeed, consider a user who
succeeds in creating a new block containing the call to function \textsf{draw()}. If such a user is also a
lottery participant and the block timestamp is used as a seed, then the user could try to manipulate the
timestamp to win the lottery.

To model this scenario from a noninterference perspective, we follow an approach inspired by~\cite{FGM04},
where attacks to security conditions like, e.g., authentication and secrecy are modeled as interferences of
hostile environment activities toward good protocol behaviors. In our example, we assume that the
observable, low-level behaviors are those that are allowed by the smart contract, while the high-level
behaviors are those due to the hostile environment, e.g., an adversary user trying to corrupt the
randomization process. Hence, in this setting, the goal of the noninterference check is to verify potential
alterations of the protocol governing the lottery in the presence or in the absence of an adversary user. In
particular, in the following model, the high-level action $h$ denotes the intervention of such an adversary.

For simplicity, we assume that there are only two users purchasing one ticket each, where the malicious user
is the one buying ticket $1$ whilst the honest user buys ticket $2$. This scenario can be modeled in our
probabilistic framework as follows where we omit every $[1]$:
\cws{0}{\begin{array}{rcl}
\mathit{Lottery} & \!\! \eqdef \!\! & \tau \, . \, \mathit{draw} \, . \, ([0.5] \mathit{winner}_{1} \, . \,
\mathit{notif}_{1} \, . \, \nil \, \oplus [0.5] \mathit{winner}_{2} \, . \, \mathit{notif}_{2} \, . \, \nil)
\, + \\
& & \mathit{h} \, . \, \mathit{draw} \, . \, ([1 - \epsilon] \mathit{winner}_{1} \, . \, \mathit{notif}_{1}
\, . \, \nil \, \oplus [\epsilon] \mathit{winner}_{2} \, . \, \mathit{notif}_{2} \, . \, \nil) \\
\end{array}}
The invocation of function \textsf{draw()} shall lead to the probabilistic extraction of the ticket (action
$\mathit{draw}$), the determination of the winner (actions $\mathit{winner}_{i}$), and the notification of
the winner (actions $\mathit{notif}_{i}$). Instead of being fair, this procedure might be guided by an
adversary manipulating the random seed in such a way to pilot the extraction and, therefore, win the lottery
with a probability much higher than $1 / 2$. Notice that parameter $\epsilon \in \realns_{]0, 1[}$ can be
considered to be negligible but not equal to $0$ as, for computational reasons, the manipulation attempt may
fail. Since in a real-world scenario it is hard to predict who will be able to create a new block for the
blockchain, we employ a nondeterministic choice between action $\tau$ and the high-level action $h$ to model
the choice between a honest user and an adversary manipulating the random seed. Notice that, for mechanisms
like, e.g., the proof of stake of Ethereum, where the user chance of creating a new block is determined by
the amount of ethers staked by the user, the choice above could be modeled probabilistically, without any
significant impact upon the result of the analysis.

As far as nondeterministic noninterference analysis is concerned, process $\mathit{Lottery}$ does not leak
any information. More precisely, its nondeterministic variant satisfies all the security properties, for
both nondeterministic weak bisimilarity and branching bisimilarity. The reason is that if we abstract away
from probabilities, the behavior of the adversary (see the $h$-branch) is indistinguishable from the
behavior of the honest user (see the $\tau$-branch). However, $\mathit{Lottery}$ is not
$\mathrm{BSNNI}_{\wbis{}}$ for $\wbis{} \: \in \{ \wbis{\rm pw}, \wbis{\rm pb} \}$, hence both
bisimilarities can be used to capture the aforementioned interference in the probabilistic setting. Indeed,
the version of $\mathit{Lottery}$ with high-level actions hidden -- which includes both the branch with fair
extraction and the branch with unfair extraction -- and the version of $\mathit{Lottery}$ with high-level
actions restricted -- which includes only the fair branch -- cannot be $\wbis{}$-equivalent, because $[0.5]
N_{1} \oplus [0.5] N_{2} \not\wbis{} [1 - \epsilon] N_{1} \oplus [\epsilon] N_{2}$ for any pair of
$\wbis{}$-inequivalent nondeterministic processes $N_{1}$ and $N_{2}$ when $\epsilon \neq 0.5$.

Assuming that the seed for random number generation cannot be manipulated, the second critical issue
concerns another vulnerability that arises due to the rules of the blockchain. In fact, consider an
adversary who manages to create successfully a new block and realizes that the winning ticket belongs to
another participant. The adversary could then invalidate the lottery extraction by ignoring the transaction
related to the \textsf{draw()} call and forcing its rollback~\cite{CGP19}. We model such a behavior in the
following way:
\cws{0}{\begin{array}{rcl}
\mathit{Lottery}' & \!\! \eqdef \!\! & \mathit{draw} \, . \, ([0.5] \mathit{winner}_{1} \, . \,
\mathit{notif}_{1} \, . \, \tau \, . \, (\mathit{success} \, . \, \nil + \tau \, . \, \mathit{failure} \, .
\, \nil) \, \oplus \\
& & \hspace{35pt} [0.5] \mathit{winner}_{2} \, . \, \mathit{notif}_{2} \, . \, (\tau \, . \,
(\mathit{success} \, . \, \nil + \tau \, . \, \mathit{failure} \, . \, \nil) \, + \\
& & \hspace{141pt} h \, . \, (\tau \, . \, (\mathit{success} \, . \, \nil + \tau \, . \, \mathit{failure} \,
. \, \nil) \, + \\
& & \hspace{158pt} \mathit{failure} \, . \, \nil))) \\
\end{array}}
With respect to the previous scenario, the adversary cannot affect the probabilistic behavior of the smart
contract, i.e., the extraction procedure. However, the adversary can try to interfere if the outcome of the
extraction is different from ticket $1$.

On the one hand, consider the branch after action $\mathit{notif}_{1}$, which models the block creation
procedure. The first $\tau$-action expresses that the honest user is picked and creates the block. The
subsequent choice is between the successful creation (action $\mathit{success}$) and an event not depending
on the user (action $\tau$) that causes the failure of the process (action $\mathit{failure}$). Notice that
there might be several causes for this failure, such as a wrong transaction in the block, a fork in the
blockchain, and so on. On the other hand, in the branch after action $\mathit{notif}_{2}$, the adversary may
decide to participate actively in the procedure, as can be seen from the choice between the action $\tau$,
leading to the same behavior surveyed above, and the high-level action $h$. In the latter case, the race
between the adversary and the honest user is solved nondeterministically through a choice between action
$\tau$ and action $\mathit{failure}$. The former leads to the behavior of the honest user, while the latter
represents the behavior of the adversary, who decides to cause the immediate failure of the creation
procedure.

Formally, process $\mathit{Lottery}'$ is $\mathrm{SBNDC}_{\wbis{\rm pw}}$. Indeed, observing that we have
only one occurrence of the high-level action $h$, it holds that the subprocess $N_{1} = \tau \, . \,
(\mathit{success} \, . \, \nil + \tau \, . \, \mathit{failure} \, . \, \nil)$ -- denoting the low-level view
before executing $h$ -- is weakly probabilistic bisimilar to the subprocess $N_{2} = \tau \, . \,
(\mathit{success} \, . \, \nil + \tau \, . \, \mathit{failure} \, . \, \nil) + \mathit{failure} \, . \,
\nil$ -- denoting the low-level view after executing $h$. However, $\mathit{Lottery}'$ is not
$\mathrm{BSNNI}_{\wbis{\rm pb}}$ as can be seen by comparing the only part -- which is after action
$\mathit{notif}_{2}$ -- in which $\mathit{Lottery}' \setminus \{ h \}$ and $\mathit{Lottery}' \, / \, \{ h
\}$ differ, i.e., $N_{1}$ and $N_{1} + \tau \, . \, N_{2}$ respectively. In fact, $N_{1}$ is not
probabilistic branching bisimilar to $N_{1} + \tau \, . \, N_{2}$. This is because $N_{1} \not\wbis{\rm pb}
N_{2}$, while they are equated by~$\wbis{\rm pw}$. In essence, \linebreak $N_{1}$ cannot respond in
accordance with $\wbis{\rm pb}$ when $N_{2}$ immediately executes action $\mathit{failure}$.

By applying back-and-forth reasoning to $N_{2}$ -- which comes after action~$h$ -- undoing the rightmost
action \textit{failure} reveals that the failure has been forced by the adversary, while undoing the
leftmost action \textit{failure} reveals that the failure has been the consequence of a choice involving
also the action \textit{success}. This is sufficient to expose the behavior of the adversary, which would
not be detected by analyzing only the forward computations though. Intuitively, this interference is
observable in a setting in which the actions related to the blokchain management can be reversed, e.g., in
order to have the possibility of undoing efficiently activities that may cause failures and related side
effects. To conclude, the noninterference analysis based on the strongest $\wbis{\rm pw}$-based property of
Figure~\ref{fig:prob_taxonomy}, i.e., $\mathrm{SBNDC}_{\wbis{pw}}$, fails to reveal the covert channel
caused by the adversary, while the weakest $\wbis{\rm pb}$-based property of Figure~\ref{fig:prob_taxonomy},
i.e., $\mathrm{BSNNI}_{\wbis{pb}}$, can detect it.

%
%
\section{Conclusions}
\label{sec:concl}
%
%

In this paper we have investigated a taxonomy of noninterference properties for processes featuring both
nondeterminism and probabilities according to the strictly alternating model~\cite{HJ90}, along with
preservation and compositionality aspects of such properties. The two behavioral equivalences that we have
considered for those noninterference properties are the weak probabilistic bisimilarity of~\cite{PLS00} and
a probabilistic branching bisimilarity of~\cite{AGT12}.

Since we have shown that the latter coincides with a probabilistic variant of the weak back-and-forth
bisimilarity of~\cite{DMV90}, the noninterference properties based on the latter can be applied to
reversible probabilistic systems, thereby extending our previous results in~\cite{EABR25} for reversible
systems that are fully nondeterministic. Our work also extends the results of~\cite{ABG04}, where only
variants of BSNNI, BNDC, and SBNDC are considered over probabilistic systems of generative-reactive
type~\cite{GSS95} hence not allowing for full nondeterminism, in a way that avoids additional universal
quantifications over probabilistic parameters in the formalization of noninterference properties.

In the strictly alternating model of~\cite{HJ90} that we have used, states are divided into nondeterministic
and probabilistic. Each of the former may have only action transitions to probabilistic states, while each
of the latter may have only probabilistic transitions to nondeterministic states. In the non-strictly
alternating variant of~\cite{PLS00}, action transitions are admitted also between two nondeterministic
states. An alternative model is the non-alternating one given by Segala simple probabilistic
automata~\cite{Seg95a}, where every transition is labeled with an action and goes from a state to a
probability distribution over states. Regardless of the adopted model, in our theory some characteristics
seem to be independent from probabilities, as witnessed by almost all the counterexamples in
Section~\ref{sec:prob_bisim_sec_prop_char}.

Both the alternating model and the non-alternating one -- whose relationships have been studied
in~\cite{ST05} -- encompass nondeterministic models, generative models, and reactive models as special
cases. Since branching bisimulation semantics plays a fundamental role in reversible
systems~\cite{DMV90,BE23b}, in this paper we have adopted the alternating model because of the probabilistic
branching bisimulation congruence developed for it in~\cite{AGT12} along with equational and logical
characterizations and a polynomial-time decision procedure. In the non-alternating model, for which
branching bisimilarity has been just defined in~\cite{SL94}, weak variants of bisimulation semantics require
-- to achieve transitivity -- that a single transition be matched by a convex combination of several
transitions -- corresponding to the use of randomized schedulers -- which causes such equivalences to be
less manageable, although they can be decided in polynomial time~\cite{TH15}.

With respect to the earlier version of our study~\cite{EAB24}, the considered process language now supports
recursion. Like in~\cite{EABR25}, this has required us to develop a number of ancillary results and to
resort to the bisimulation-up-to technique~\cite{SM92}. As far as the latter is concerned, we have
introduced the corresponding definitions for $\wbis{\rm pw}$ and $\wbis{\rm pb}$ inspired
by~\cite{Mil89a,Gla93,HL97,BBG98b} and proven their correctness.

As for future work, we are planning to further extend the noninterference taxonomy so as to include
properties that take into account also stochastic aspects of process behavior like in~\cite{AB09,HMPR21}. In
those works actions are extended with rates expressing exponentially distributed durations, while following
the approach of the present paper we should consider action execution separated from stochastic time passing
like in~\cite{Her02}. The two different views, i.e., integrated time and orthogonal time, can be reconciled
as shown in~\cite{BCT16}. \linebreak In any case, it seems necessary to develop a notion of stochastic
branching bisimilarity and prove that it coincides with a stochastic variant of the weak back-and-forth
bisimilarity of~\cite{DMV90}, thus completing the work done for strong bisimilarity in~\cite{BM23a}.

\section*{Acknowledgment}
This research has been supported by the PRIN 2020 project \textit{NiRvAna -- Noninterference and
Reversibility Analysis in Private Blockchains}. We would like to thank the anonymous reviewers for their
thorough and constructive comments.

\bibliographystyle{alphaurl}
\bibliography{biblio}

\end{document}

%% file: Commands.tex
\newcommand{\ms}[1]
	{\null\ifmmode\mathord{\mathcode`-="702D\it #1\mathcode`\-="2200}
	\else$\mathord{\mathcode`-="702D\it #1\mathcode`\-="2200}$\fi}

\newcommand{\cws}[2]
	{\\ \centerline{$#2$} \\[-#1pt]}

\newcommand{\fullbox}
	{{\mbox{}\nolinebreak\hfill{$\rule{2mm}{2mm}$}}}

\newcommand{\bibtrick}[1]
	{}

\newcommand{\blue}[1]
	{{\color{blue}#1}}

\newcommand{\cala}
        {\mathcal{A}}

\newcommand{\calb}
        {\mathcal{B}}

\newcommand{\calh}
        {\mathcal{H}}

\newcommand{\cali}
        {\mathcal{I}}

\newcommand{\call}
        {\mathcal{L}}

\newcommand{\calp}
        {\mathcal{P}}

\newcommand{\calq}
        {\mathcal{Q}}

\newcommand{\cals}
        {\mathcal{S}}

\newcommand{\calu}
        {\mathcal{U}}

\newcommand{\natns}
	{\mathbb{N}}

\newcommand{\realns}
	{\mathbb{R}}

\newcommand{\procs}
	{\mathbb{P}}

\newcommand{\first}
	{\textit{first}}

\newcommand{\last}
	{\textit{last}}

\newcommand{\pt}
	{\textit{path}}

\newcommand{\rn}{\textit{run}}

\newcommand{\arrow}[2]
        {\, {\auxarrow\limits^{#1}}_{#2} \,}
\newcommand{\auxarrow}
	{\mathop{\longrightarrow}}

\newcommand{\warrow}[2]
        {\, {\wauxarrow\limits^{#1}}_{#2} \,}
\newcommand{\wauxarrow}
	{\mathop{= \!\!\!\! \Longrightarrow}}

\newcommand{\nil}
	{\underline 0}

\newcommand{\eqdef}
	{\triangleq}

\newcommand{\sbis}[1]
	{\sim_{#1}}

\newcommand{\wbis}[1]
        {\approx_{#1}}

\newcommand{\pco}[1]
	{\mathop{\Vert_{#1}}}

\renewcommand{\nd}
	{\textit{nd}}

\newcommand{\proba}
	{\textit{prob}}

\newcommand{\reach}
	{\textit{reach}}

\newcommand{\sched}
	{\textit{sched}}

\newcommand{\trace}
	{\textit{trace}}



%% file: biblio.bib
@article{Ald06,
Author = {A.~Aldini},
Title = {Classification of Security Properties in a {L}inda-Like Process Algebra},
Journal = {Science of Computer Programming},
Volume = {63},
Pages = {16--38},
Year = {2006}}

@inproceedings{AB09,
Author = {A.~Aldini and M.~Bernardo},
Title = {A General Framework for Nondeterministic, Probabilistic, and Stochastic Noninterference},
Booktitle = {Proc.\ of the 1st Joint Workshop on Automated Reasoning for Security Protocol Analysis and
	     Issues in the Theory of Security (ARSPA/WITS~2009)},
Publisher = {Springer},
Series = {LNCS},
Volume = {5511},
Pages = {18--33},
Year = {2009}}

@article{AB11,
Author = {A.~Aldini and M.~Bernardo},
Title = {Component-Oriented Verification of Noninterference},
Journal = {Journal of Systems Architecture},
Volume = {57},
Pages = {282--293},
Year = {2011}}

@article{ABG04,
Author = {A.~Aldini and M.~Bravetti and R.~Gorrieri},
Title = {A Process-Algebraic Approach for the Analysis of Probabilistic Noninterference},
Journal = {Journal of Computer Security},
Volume = {12},
Pages = {191--245},
Year = {2004}}

@article{AGT12,
Author = {S.~Andova and S.~Georgievska and N.~Trcka},
Title = {Branching Bisimulation Congruence for Probabilistic Systems},
Journal = {Theoretical Computer Science},
Volume = {413},
Pages = {58--72},
Year = {2012}}

@inproceedings{ARL20,
Author = {D.~Azzolini and F.~Riguzzi and E.~Lamma},
Title = {Modeling Smart Contracts with Probabilistic Logic Programming},
Booktitle = {Proc.\ of the 23rd Int.\ Business Information Systems Workshops (BIS~2020)},
Publisher = {Springer},
Series = {LNBIP},
Volume = {394},
Pages = {86--98},
Year = {2020}}

@book{BW90,
Author = {J.C.M.~Baeten and W.P.~Weijland},
Title = {Process Algebra},
Publisher = {Cambridge University Press},
Year = {1990}}

@inproceedings{BH97,
Author = {C.~Baier and H.~Hermanns},
Title = {Weak Bisimulation for Fully Probabilistic Processes},
Booktitle = {Proc.\ of the 9th Int.\ Conf.\ on Computer Aided Verification (CAV~1997)},
Publisher = {Springer},
Series = {LNCS},
Volume = {1254},
Pages = {119--130},
Year = {1997}}

@article{BT03,
Author = {R.~Barbuti and L.~Tesei},
Title = {A Decidable Notion of Timed Non-Interference},
Journal = {Fundamenta Informaticae},
Volume = {54},
Pages = {137--150},
Year = {2003}}

@article{Ben73,
Author = {C.H.~Bennett},
Title = {Logical Reversibility of Computation},
Journal = {IBM Journal of Research and Development},
Volume = {17},
Pages = {525--532},
Year = {1973}}

@article{BB03,
Author = {M.~Bernardo and M.~Bravetti},
Title = {Performance Measure Sensitive Congruences for {M}arkovian Process Algebras},
Journal = {Theoretical Computer Science},
Volume = {290},
Pages = {117--160},
Year = {2003}}

@article{BCT16,
Author = {M.~Bernardo and F.~Corradini and L.~Tesei},
Title = {Timed Process Calculi with Deterministic or Stochastic Delays: Commuting between Durational and
	 Durationless Actions},
Journal = {Theoretical Computer Science},
Volume = {629},
Pages = {2--39},
Year = {2016}}

@inproceedings{BE23b,
Author = {M.~Bernardo and A.~Esposito\bibtrick{b}},
Title = {Modal Logic Characterizations of Forward, Reverse, and Forward-Reverse Bisimilarities},
Booktitle = {Proc.\ of the 14th Int.\ Symp.\ on Games, Automata, Logics, and Formal Verification
	     (GANDALF~2023)},
Series = {EPTCS},
Volume = {390},
Pages = {67--81},
Year = {2023}}

@inproceedings{BLMMRS23,
Author = {M.~Bernardo and I.~Lanese and A.~Marin and C.A.~Mezzina and S.~Rossi and C.~Sacerdoti Coen},
Title = {Causal Reversibility Implies Time Reversibility},
Booktitle = {Proc.\ of the 20th Int.\ Conf.\ on the Quantitative Evaluation of Systems (QEST~2023)},
Publisher = {Springer},
Series = {LNCS},
Volume = {14287},
Pages = {270--287},
Year = {2023}}

@article{BM23a,
Author = {M.~Bernardo and C.A.~Mezzina},
Title = {Bridging Causal Reversibility and Time Reversibility: A Stochastic Process Algebraic Approach},
Journal = {Logical Methods in Computer Science},
Volume = {19(2)},
Pages = {6:1--6:27},
Year = {2023}}

@inproceedings{BR23,
Author = {M.~Bernardo and S.~Rossi},
Title = {Reverse Bisimilarity vs.\ Forward Bisimilarity},
Booktitle = {Proc.\ of the 26th Int.\ Conf.\ on Foundations of Software Science and Computation Structures
	     (FOSSACS~2023)},
Publisher = {Springer},
Series = {LNCS},
Volume = {13992},
Pages = {265--284},
Year = {2023}}

@inproceedings{BBG98b,
Author = {M.~Bravetti\bibtrick{b} and M.~Bernardo and R.~Gorrieri},
Title = {A Note on the Congruence Proof for Recursion in {M}arkovian Bisimulation Equivalence},
Booktitle = {Proc.\ of the 6th Int.\ Workshop on Process Algebra and Performance Modelling (PAPM~1998)},
Pages = {71--87},
Year = {1998}}

@article{BHR84,
Author = {S.D.~Brookes and C.A.R.~Hoare and A.W.~Roscoe},
Title = {A Theory of Communicating Sequential Processes},
Journal = {Journal of the ACM},
Volume = {31},
Pages = {560--599},
Year = {1984}}

@inproceedings{CGP19,
Author = {K.~Chatterjee and A.K.~Goharshady and A.~Pourdamghani},
Title = {Probabilistic Smart Contracts: Secure Randomness on the Blockchain}, 
Booktitle = {Proc.\ of the 1st IEEE Int.\ Conf.\ on Blockchain and Cryptocurrency (ICBC~2019)},
Publisher = {IEEE-CS Press},
Pages = {403--412},
Year = {2019}}

@inproceedings{DK04,
Author = {V.~Danos and J.~Krivine},
Title = {Reversible Communicating Systems},
Booktitle = {Proc.\ of the 15th Int.\ Conf.\ on Concurrency Theory (CONCUR~2004)},
Publisher = {Springer},
Series = {LNCS},
Volume = {3170},
Pages = {292--307},
Year = {2004}}

@inproceedings{DK05,
Author = {V.~Danos and J.~Krivine},
Title = {Transactions in {RCCS}},
Booktitle = {Proc.\ of the 16th Int.\ Conf.\ on Concurrency Theory (CONCUR~2005)},
Publisher = {Springer},
Series = {LNCS},
Volume = {3653},
Pages = {398--412},
Year = {2005}}

@inproceedings{DMV90,
Author = {R.~{De~Nicola} and U.~Montanari and F.~Vaandrager},
Title = {Back and Forth Bisimulations},
Booktitle = {Proc.\ of the 1st Int.\ Conf.\ on Concurrency Theory (CONCUR~1990)},
Publisher = {Springer},
Series = {LNCS},
Volume = {458},
Pages = {152--165},
Year = {1990}}

@inproceedings{EAB24,
Author = {A.~Esposito and A.~Aldini and M.~Bernardo},
Title = {Noninterference Analysis of Reversible Probabilistic Systems},
Booktitle = {Proc.\ of the 44th Int.\ Conf.\ on Formal Techniques for Distributed Objects, Components, and
	     Systems (FORTE~2024)},
Publisher = {Springer},
Series = {LNCS},
Volume = {14678},
Pages = {39--59},
Year = {2024}}

@article{EABR25,
Author = {A.~Esposito and A.~Aldini and M.~Bernardo and S.~Rossi},
Title = {Noninterference Analysis of Reversible Systems: An Approach Based on Branching Bisimilarity},
Journal = {Logical Methods in Computer Science},
Volume = {21(1)},
Pages = {6:1--6:28},
Year = {2025}}

@inproceedings{FG01,
Author = {R.~Focardi and R.~Gorrieri},
Title = {Classification of Security Properties},
Booktitle = {Proc.\ of the 1st Int.\ School on Foundations of Security Analysis and Design (FOSAD~2000)},
Publisher = {Springer},
Series = {LNCS},
Volume = {2171},
Pages = {331--396},
Year = {2001}}

@inproceedings{FGM04,
Author = {R.~Focardi and R.~Gorrieri and F.~Martinelli},
Title = {Classification of Security Properties ({Part II}: Network Security)},
Booktitle = {Proc.\ of the 2nd Int.\ School on Foundations of Security Analysis and Design (FOSAD~2001)},
Publisher = {Springer},
Series = {LNCS},
Volume = {2946},
Pages = {139--185},
Year = {2004}}

@inproceedings{FPR02,
Author = {R.~Focardi and C.~Piazza and S.~Rossi},
Title = {Proofs Methods for Bisimulation Based Information Flow Security},
Booktitle = {Proc.\ of the 3rd Int.\ Workshop on Verification, Model Checking, and Abstract Interpretation
	     (VMCAI~2002)},
Publisher = {Springer},
Series = {LNCS},
Volume = {2294},
Pages = {16--31},
Year = {2002}}

@article{FR06,
Author = {R.~Focardi and S.~Rossi},
Title = {Information flow security in dynamic contexts},
Journal = {Journal of Computer Security},
Volume = {14},
Pages = {65--110},
Year = {2006}}

@inproceedings{GLM14,
Author = {E.~Giachino and I.~Lanese and C.A.~Mezzina},
Title = {Causal-Consistent Reversible Debugging},
Booktitle = {Proc.\ of the 17th Int.\ Conf.\ on Fundamental Approaches to Software Engineering (FASE 2014)},
Publisher = {Springer},
Series = {LNCS},
Volume = {8411},
Pages = {370--384},
Year = {2014}}

@article{GM18,
Author = {R.~Giacobazzi and I.~Mastroeni},
Title = {Abstract Non-Interference: A Unifying Framework for Weakening Information-Flow},
Journal = {ACM Trans.\ on Privacy and Security},
Volume = {21(2)},
Pages = {9:1--9:31},
Year = {2018}}

@inproceedings{Gla93,
Author = {{R.J.~van}~Glabbeek},
Title = {A Complete Axiomatization for Branching Bisimulation Congruence of Finite-State Behaviours},
Booktitle = {Proc.\ of the 18th Int.\ Symp.\ on Mathematical Foundations of Computer Science (MFCS~1993)},
Publisher = {Springer},
Series = {LNCS},
Volume = {711},
Pages = {473--484},
Year = {1993}}

@inproceedings{Gla01,
Author = {{R.J.~van}~Glabbeek},
Title = {The Linear Time -- Branching Time Spectrum~{I}},
Booktitle = {Handbook of Process Algebra},
Publisher = {Elsevier},
Pages = {3--99},
Year = {2001}}

@article{GSS95,
Author = {{R.J.~van}~Glabbeek and S.A.~Smolka and B.~Steffen},
Title = {Reactive, Generative and Stratified Models of Probabilistic Processes},
Journal = {Information and Computation},
Pages = {59--80},
Volume = {121},
Year = {1995}}

@article{GW96,
Author = {{R.J.~van}~Glabbeek and W.P.~Weijland},
Title = {Branching Time and Abstraction in Bisimulation Semantics},
Journal = {Journal of the ACM},
Volume = {43},
Pages = {555--600},
Year = {1996}}

@inproceedings{GM82,
Author = {J.A.~Goguen and J.~Meseguer},
Title = {Security Policies and Security Models},
Booktitle = {Proc.\ of the 2nd IEEE Symp.\ on Security and Privacy (SSP~1982)},
Publisher = {IEEE-CS Press},
Pages = {11--20},
Year = {1982}}

@inproceedings{HJ90,
Author = {H.~Hansson and B.~Jonsson},
Title = {A Calculus for Communicating Systems with Time and Probabilities},
Booktitle = {Proc.\ of the 11th IEEE Real-Time Systems Symp.\ (RTSS~1990)},
Publisher = {IEEE-CS Press},
Pages = {278--287},
Year = {1990}}

@inproceedings{HS12,
Author = {D.~Hedin and A.~Sabelfeld},
Title = {A Perspective on Information-Flow Control},
Booktitle = {Software Safety and Security -- Tools for Analysis and Verification},
Publisher = {IOS Press},
Pages = {319--347},
Year = {2012}}

@book{Her02,
Author = {H.~Hermanns},
Title = {Interactive Markov Chains},
Publisher = {Springer},
Note = {Volume 2428 of LNCS},
Year = {2002}}

@inproceedings{HL97,
Author = {H.~Hermanns and M.~Lohrey}, 
Title = {Observation Congruence in a Stochastic Timed Calculus with Maximal Progress},
Publisher = {University of Erlangen, Technical Report IMMD VII-7/97},
Year = {1997}}

@book{Hil96,
Author = {J.~Hillston},
Title = {A Compositional Approach to Performance Modelling},
Publisher = {Cambridge University Press},
Year = {1996}}

@article{HMPR21,
Author = {J.~Hillston and A.~Marin and C.~Piazza and S.~Rossi},
Title = {Persistent Stochastic Non-Interference},
Journal = {Fundamenta Informaticae},
Volume = {181},
Pages = {1--35},
Year = {2021}}

@article{Kel76,
Author = {R.M.~Keller},
Title = {Formal Verification of Parallel Programs},
Journal = {Communications of the ACM},
Volume = {19},
Pages = {371--384},
Year = {1976}}

@article{KV24,
Author = {L.V.~Kovalchuk and A.A.~Vykhlo}, 
Title = {Estimation of the Probability of Success of a Frontrunning Attack on Smart Contracts},
Journal = {Cybernetics and Systems Analysis},
Volume = {60},
Pages = {881--890},
Year = {2024}}

@article{Lan61,
Author = {R.~Landauer},
Title = {Irreversibility and Heat Generation in the Computing Process},
Journal = {IBM Journal of Research and Development},
Volume = {5},
Pages = {183--191},
Year = {1961}}

@inproceedings{LLMSS13,
Author = {I.~Lanese and M.~Lienhardt and C.A.~Mezzina and A.~Schmitt and J.-B.~Stefani},
Title = {Concurrent Flexible Reversibility},
Booktitle = {Proc.\ of the 22nd European Symp.\ on Programming (ESOP~2013)},
Publisher = {Springer},
Series = {LNCS},
Volume = {7792},
Pages = {370--390},
Year = {2013}}

@inproceedings{LNPV18a,
Author = {I.~Lanese and N.~Nishida and A.~Palacios and G.~Vidal},
Title = {Cau{DE}r: A Causal-Consistent Reversible Debugger for {E}rlang},
Booktitle = {Proc.\ of the 14th Int.\ Symp.\ on Functional and Logic Programming (FLOPS~2018)},
Publisher = {Springer},
Series = {LNCS},
Volume = {10818},
Pages = {247--263},
Year = {2018}}

@article{LS91,
Author = {K.G.~Larsen and A.~Skou},
Title = {Bisimulation Through Probabilistic Testing},
Journal = {Information and Computation},
Volume = {94},
Pages = {1--28},
Year = {1991}}

@article{LES18,
Author = {J.S.~Laursen and L.-P.~Ellekilde and U.P.~Schultz},
Title = {Modelling Reversible Execution of Robotic Assembly},
Journal = {Robotica},
Volume = {36},
Pages = {625--654},
Year = {2018}}

@inproceedings{Man11,
Author = {H.~Mantel},
Title = {Information Flow and Noninterference},
Booktitle = {Encyclopedia of Cryptography and Security},
Publisher = {Springer},
Pages = {605--607},
Year = {2011}}

@inproceedings{Mar98,
Author = {F.~Martinelli},
Title = {Partial Model Checking and Theorem Proving for Ensuring Security Properties},
Booktitle = {Proc.\ of the 11th IEEE Computer Security Foundations Workshop (CSFW~1998)},
Publisher = {IEEE-CS Press},
Pages = {44--52},
Year = {1998}}

@article{Mar03,
Author = {F.~Martinelli},
Title = {Analysis of Security Protocols as Open Systems},
Journal = {Theoretical Computer Science},
Volume = {290},
Pages = {1057--1106},
Year = {2003}}

@book{Mil89a,
Author = {R.~Milner},
Title = {Communication and Concurrency},
Publisher = {Prentice Hall},
Year = {1989}}

@inproceedings{Par81,
Author = {D.~Park},
Title = {Concurrency and Automata on Infinite Sequences},
Booktitle = {Proc.\ of the 5th GI Conf.\ on Theoretical Computer Science},
Publisher = {Springer},
Series = {LNCS},
Volume = {104},
Pages = {167--183},
Year = {1981}}

@article{PBPTKS21,
Author = {N.S.~Patel and P.~Bhattacharya and S.B.~Patel and S.~Tanwar and N.~Kumar and H.~Song},
Title = {Blockchain-Envisioned Trusted Random Oracles for {IoT}-Enabled Probabilistic Smart Contracts}, 
Journal = {IEEE Internet of Things Journal}, 
Volume = {8},
Pages = {14797--14809},
Year = {2021}}

@article{PP14,
Author = {K.S.~Perumalla and A.J.~Park},
Title = {Reverse Computation for Rollback-Based Fault Tolerance in Large Parallel Systems -- {E}valuating
	 the Potential Gains and Systems Effects},
Journal = {Cluster Computing},
Volume = {17},
Pages = {303--313},
Year = {2014}}

@inproceedings{PLS00,
Author = {A.~Philippou and I.~Lee and O.~Sokolsky},
Title = {Weak Bisimulation for Probabilistic Systems},
Booktitle = {Proc.\ of the 11th Int.\ Conf.\ on Concurrency Theory (CONCUR~2000)},
Publisher = {Springer},
Series = {LNCS},
Volume = {1877},
Pages = {334--349},
Year = {2000}}

@article{PU07a,
Author = {I.~Phillips and I.~Ulidowski\bibtrick{a}},
Title = {Reversing Algebraic Process Calculi},
Journal = {Journal of Logic and Algebraic Programming},
Volume = {73},
Pages = {70--96},
Year = {2007}}

@inproceedings{PUY12,
Author = {I.~Phillips and I.~Ulidowski\bibtrick{b} and S.~Yuen},
Title = {A Reversible Process Calculus and the Modelling of the {ERK} Signalling Pathway},
Booktitle = {Proc.\ of the 4th Int.\ Workshop on Reversible Computation (RC~2012)},
Publisher = {Springer},
Series = {LNCS},
Volume = {7581},
Pages = {218--232},
Year = {2012}}

@inproceedings{Pin17,
Author = {G.M.~Pinna},
Title = {Reversing Steps in Membrane Systems Computations},
Booktitle = {Proc.\ of the 18th Int.\ Conf.\ on Membrane Computing (CMC~2017)},
Publisher = {Springer},
Series = {LNCS},
Volume = {10725},
Pages = {245--261},
Year = {2017}}

@article{QHLWLWZH23,
Author = {P.~Qian and J.~He and L.~Lu and S.~Wu and Z.~Lu and L.~Wu and Y.~Zhou and Q.~He},
Title ={Demystifying Random Number in {E}thereum Smart Contract: Taxonomy, Vulnerability Identification, and
	Attack Detection}, 
Journal = {IEEE Trans.\ on Software Engineering}, 
Volume = {49},
Pages = {3793--3810},
Year = {2023}}

@inproceedings{SabSan00,
Author = {A.~Sabelfeld and D.~Sands},
Title = {Probabilistic Noninterference for Multi-Threaded Programs}, 
Booktitle = {Proc.\ of the 13th IEEE Computer Security Foundations Workshop (CSFW~2000)}, 
Publisher = {IEEE-CS Press},
Pages = {200--214},
Year = {2000}}

@inproceedings{SM92,
Author = {D.~Sangiorgi and R.~Milner},
Title = {The Problem of ``Weak Bisimulation up to''},
Booktitle = {Proc.\ of the 3rd Int.\ Conf.\ on Concurrency Theory (CONCUR~1992)},
Publisher = {Springer},
Series = {LNCS},
Volume = {630},
Pages = {32--46},
Year = {1992}}

@article{SOJB18,
Author = {M.~Schordan and T.~Oppelstrup and D.R.~Jefferson and P.D.~{Barnes Jr.}},
Title = {Generation of Reversible {C++} Code for Optimistic Parallel Discrete Event Simulation},
Journal = {New Generation Computing},
Volume = {36},
Pages = {257--280},
Year = {2018}}

@book{Seg95a,
Author = {R.~Segala},
Title = {Modeling and Verification of Randomized Distributed Real-Time Systems},
Publisher = {PhD Thesis},
Year = {1995}}

@inproceedings{SL94,
Author = {R.~Segala\bibtrick{a} and N.A.~Lynch},
Title = {Probabilistic Simulations for Probabilistic Processes},
Booktitle = {Proc.\ of the 5th Int.\ Conf.\ on Concurrency Theory (CONCUR~1994)},
Publisher = {Springer},
Series = {LNCS},
Volume = {836},
Pages = {481--496},
Year = {1994}}

@inproceedings{ST05,
Author = {R.~Segala\bibtrick{a} and A.~Turrini},
Title = {Comparative Analysis of Bisimulation Relations on Alternating and Non-Alternating Probabilistic
	 Models},
Booktitle = {Proc.\ of the 2nd Int.\ Conf.\ on the Quantitative Evaluation of Systems (QEST~2005)},
Publisher = {IEEE-CS Press},
Pages = {44--53},
Year = {2005}}

@inproceedings{SLHX24,
Author = {S.~Semujju and F.~Liu and H.~Huang and Y.~Xiang},
Title = {Enhancing Fault Detection in Smart Contract Loops through Adaptive Probabilistic Sampling},
Booktitle = {Proc.\ of the 26th Genetic and Evolutionary Computation Conf.\ (GECCO~2024)},
Publisher = {ACM Press},
Pages = {731--734},
Year = {2024}}

@article{SPP19,
Author = {H.~Siljak and K.~Psara and A.~Philippou},
Title = {Distributed Antenna Selection for Massive {MIMO} Using Reversing {P}etri Nets},
Journal = {IEEE Wireless Communication Letters},
Volume = {8},
Pages = {1427--1430},
Year = {2019}}

@article{TH15,
Author = {A.~Turrini and H.~Hermanns},
Title = {Polynomial time decision algorithms for probabilistic automata},
Journal = {Information and Computation},
Volume = {244},
Pages = {134--171},
Year = {2015}}

@inproceedings{VS18,
Author = {M.~Vassor and J.-B.~Stefani},
Title = {Checkpoint/Rollback vs Causally-Consistent Reversibility},
Booktitle = {Proc.\ of the 10th Int.\ Conf.\ on Reversible Computation (RC~2018)},
Publisher = {Springer},
Series = {LNCS},
Volume = {11106},
Pages = {286--303},
Year = {2018}}

@inproceedings{VS98,
Author = {D.~Volpano and G.~Smith},
Title = {Probabilistic Noninterference in a Concurrent Language},
Booktitle = {Proc.\ of the 11th IEEE Computer Security Foundations Workshop (CSFW~1998)},
Publisher = {IEEE-CS Press},
Pages = {34--43},
Year = {1998}}

@inproceedings{VKH10,
Author = {{E.~de}~Vries and V.~Koutavas and M.~Hennessy},
Title = {Communicating Transactions},
Booktitle = {Proc.\ of the 21st Int.\ Conf.\ on Concurrency Theory (CONCUR~2010)},
Publisher = {Springer},
Series = {LNCS},
Volume = {6269},
Pages = {569--583},
Year = {2010}}

@article{Yca93,
Author = {B.~Ycart},
Title = {The Philosophers' Process: An Ergodic Reversible Nearest Particle System},
Journal = {Annals of Applied Probability},
Volume = {3},
Pages = {356--363},
Year = {1993}}

@inproceedings{ZM04,
Author = {L.~Zheng and A.~Myers},
Title = {Dynamic Security Labels and Noninterference},
Booktitle = {Proc.\ of the 2nd IFIP Workshop on Formal Aspects in Security and Trust (FAST~2004)},
Publisher = {Springer},
Series = {IFIP AICT},
Volume = {173},
Pages = {27--40},
Year = {2004}}
